\documentclass[10pt,twocolumn,letterpaper]{article}

\usepackage[pagenumbers]{cvpr} 
\usepackage{amsmath,bm}
\usepackage{multicol}
\usepackage{multirow}
\usepackage{wrapfig}
\usepackage{fontawesome5}
\usepackage{standalone}
\usepackage{tikz}
\usepackage{graphicx}
\usepackage{subcaption}
\usepackage{makecell}
\usepackage{tabularx}
\usepackage{booktabs} 
\usepackage{algorithm}
\usepackage{algpseudocode}
\usepackage[switch]{lineno}

\newcommand{\nonl}{\renewcommand{\nl}{\let\nl\oldnl}}
\newlength\savedwidth

\newcommand\blfootnote[1]{%
\begingroup
\renewcommand\thefootnote{}\footnote{#1}%
\addtocounter{footnote}{-1}%
\endgroup
}

\definecolor{cvprblue}{rgb}{0.21,0.49,0.74}
\usepackage{hyperref}
\hypersetup{pagebackref,breaklinks,colorlinks,citecolor=cvprblue}

\title{Gaussian Splashing: Unified Particles for Versatile Motion Synthesis and
Rendering}

\author{
    Yutao Feng$^{1,2\ast}$\ \ \ \ \ 
    Xiang Feng$^{1,2\ast}$\ \ \ \ \ 
    Yintong Shang$^{1}$\ \ \ \ \ \ 
    Ying Jiang$^{3}$\ \ \ \ \ 
    Chang Yu$^{3}$\ \ \ \ \ 
    Zeshun Zong$^{3}$
    \\
    Tianjia Shao$^{2}$\ \ \ \ \ 
    Hongzhi Wu$^{2}$\ \ \ \ \ 
    Kun Zhou$^{2}$\ \ \ \ \ 
    Chenfanfu Jiang$^{3}$\ \ \ \ \ 
    Yin Yang$^{1}$
    \\
    $^{1}$University of Utah, USA\ \ \ \ \ 
    $^{2}$Zhejiang University, China\ \ \ \ \ 
    $^{3}$UCLA, USA
}

\begin{document}

\twocolumn[{%
\renewcommand\twocolumn[1][]{#1}%
\maketitle
\begin{center}
    \centering
    \begin{minipage}{\textwidth}
        \centering
        \begin{minipage}[b]{.23\linewidth}
            \centering
            \includegraphics[width=\linewidth]{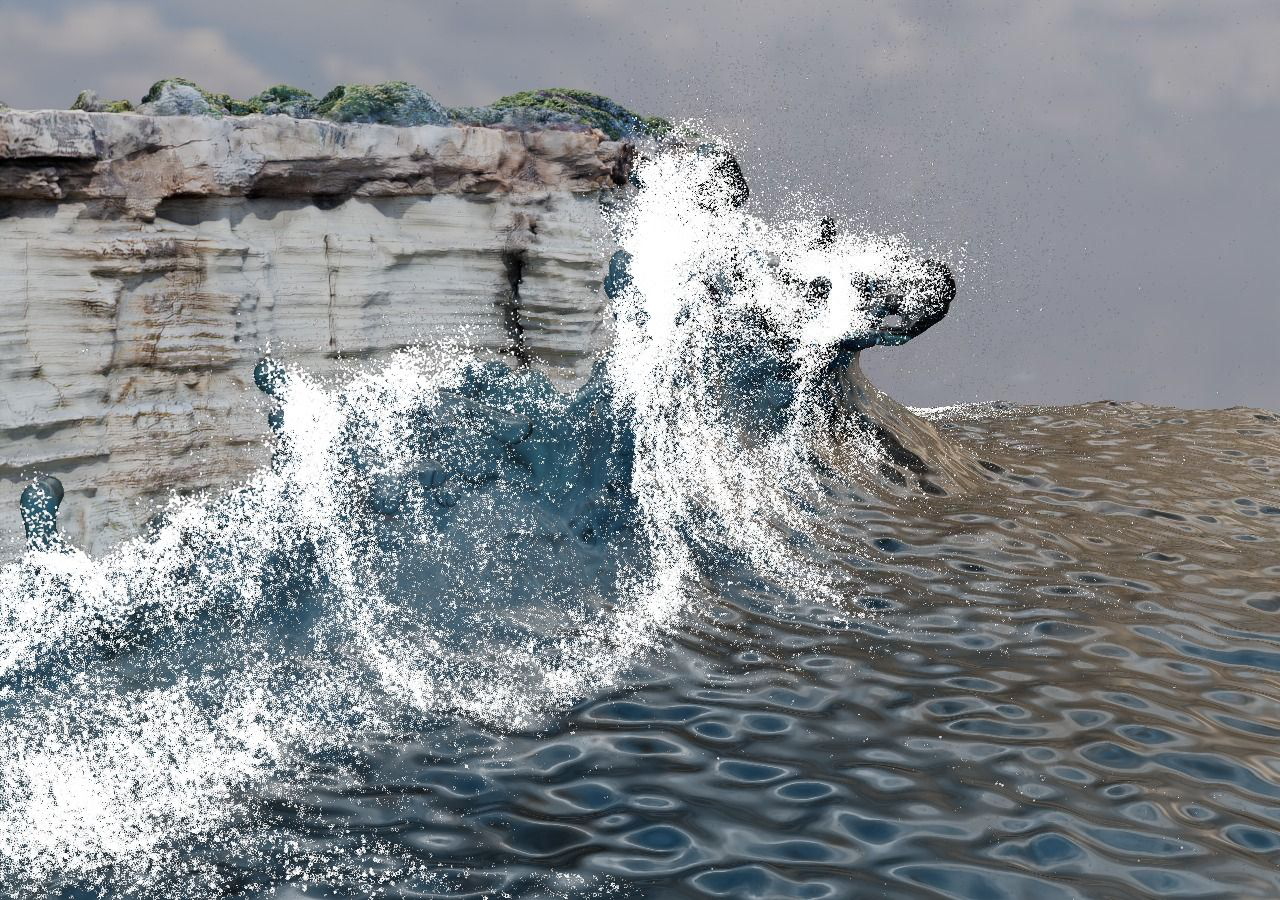}
            \put(-11,4){\scalebox{.9}{\color{white} (a)}}\\
        \end{minipage}
        \begin{minipage}[b]{.23\linewidth}
            \centering
            \includegraphics[width=\linewidth]{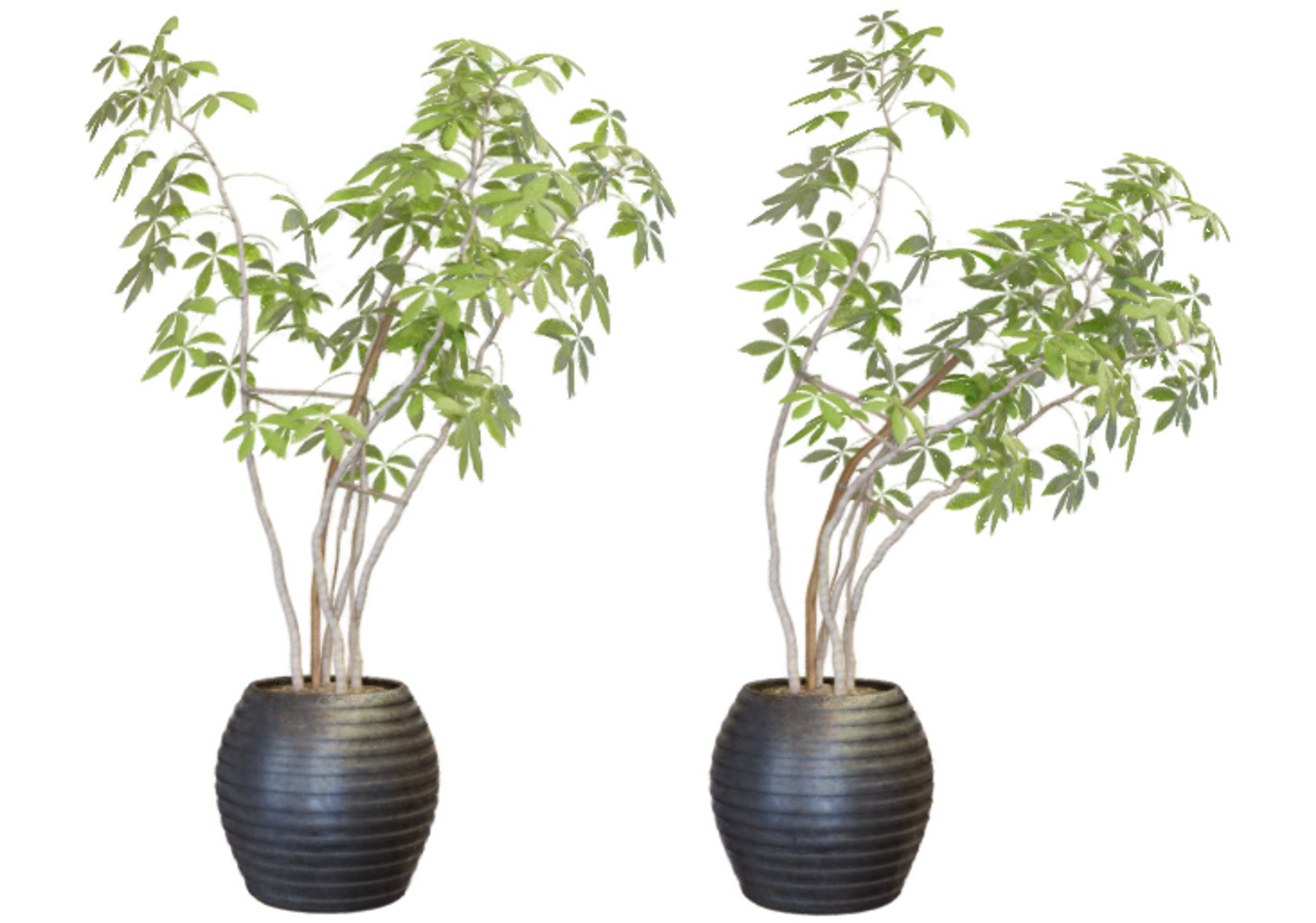}
            \put(-11,4){\scalebox{.9}{\color{black} (b)}}\\
        \end{minipage}
        \begin{minipage}[b]{0.265\linewidth} 
            \centering
            \includegraphics[width=\linewidth,trim={1cm 5cm 1cm 5cm},clip]{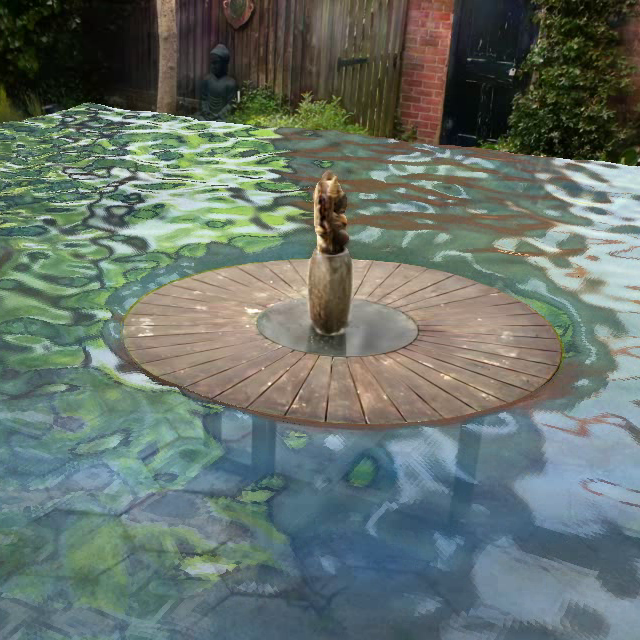}
            \put(-11,4){\scalebox{.9}{\color{white} (c)}}\\
        \end{minipage}
        \begin{minipage}[b]{.2515\linewidth}
            \centering
            \includegraphics[width=\linewidth,trim={4.75cm 0 4.75cm 0},clip]{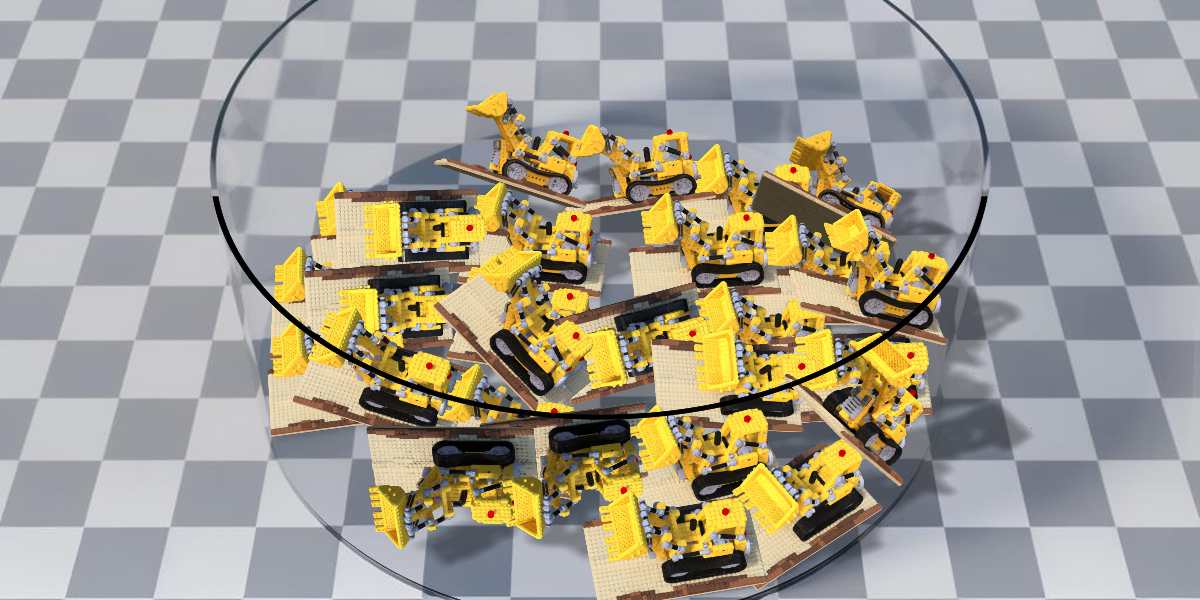}
            \put(-11,4){\scalebox{.9}{\color{white} (d)}}\\
        \end{minipage}
    \end{minipage}
    \captionsetup{type=figure}
    \captionof{figure}{\textbf{Versatile Motion Synthesis and Rendering.}~~Gaussian Splashing (GSP) is a unified framework combining 3D Gaussian Splatting (3DGS) and position-based dynamics. By leveraging their coherent point-based representations, GSP delivers high-quality rendering for novel dynamic views involving interacting deformable bodies, rigid objects, and fluids. GSP enables a variety of compelling effects and new human-computer interaction modalities not available with existing NeRF/3DGS-based systems. The teaser figure showcases a cliff battered by waves (a), a deformable ficus plant (b), flooding garden (c) and piled and scattered rigid lego bulldozers in a box (d). The reconstructed Gaussians not only capture the nonlinear dynamics of fluids and solids but can also be rasterized to realistically render with both diffuse and specular shadings. GSP re-engineers several state-of-the-art techniques from neural surface reconstruction, specular-aware Gaussian shading, position-based surface tension, and AI inpainting to ensure the quality of both simulation and rendering with 3DGS.}
    \label{fig:teaser}
\end{center}
}]

\begin{abstract}
\blfootnote{$\ast$ indicates equal contributions.} 
We demonstrate the feasibility of integrating physics-based animations of solids and fluids with 3D Gaussian Splatting (3DGS) to create novel effects in virtual scenes reconstructed using 3DGS. Leveraging the coherence of the Gaussian Splatting and Position-Based Dynamics (PBD) in the underlying representation, we manage rendering, view synthesis, and the dynamics of solids and fluids in a cohesive manner. Similar to GaussianShader, we enhance each Gaussian kernel with an added normal, aligning the kernel's orientation with the surface normal to refine the PBD simulation. This approach effectively eliminates spiky noises that arise from rotational deformation in solids. It also allows us to integrate physically based rendering to augment the dynamic surface reflections on fluids. Consequently, our framework is capable of realistically reproducing surface highlights on dynamic fluids and facilitating interactions between scene objects and fluids from new views.
For more information, please visit our project page at \url{https://gaussiansplashing.github.io/}.
\end{abstract}

\section{Introduction}\label{sec:intro}
Visualization and reconstruction of 3D scenes have been the core of 3D graphics and vision. Recent development of learning-based techniques such as the neural radiance fields (NeRFs)~\cite{Mildenhall20eccv_nerf} sheds new light on this topic. NeRF casts the reconstruction pipeline as a training procedure and delivers state-of-the-art results by encapsulating the color, texture, and geometry of the 3D scene into an implicit MLP net. Its superior convenience and efficacy inspire many follow-ups, e.g., with improved visual quality~\cite{liu2020neural}, faster performance~\cite{yu2021plenoctrees,garbin2021fastnerf}, and sparser inputs~\cite{jain2021putting,yuan2022neural}. NeRF is computationally expensive. Image synthesis with NeRF has to follow the path integral, which is less suitable for real-time or interactive applications unless dedicated compression or acceleration methods are employed e.g., with NGP encoding~\cite{Muller2022ngp}. 3D Gaussian Splatting (3DGS)~\cite{kerbl2023gaussians} provides an elegant alternative. As the name suggests, 3DGS learns a collection of Gaussian kernels from the input images.Apart from NeRF, a novel view of the scene from an unseen camera pose is generated using rasterization with the tile-splatting technique. Therefore, fast rendering with 3DGS is feasible. 

It is noted that Gaussian kernels not only serve as a good rendering agent but also explicitly encode rich information of the scene geometry. This feature suggests 3DGS a good candidate for dynamic scenes~\cite{wu20234d,yang2023real,duisterhof2023md}, animated avatars~\cite{moreau2023human,zielonka2023drivable}, or simulated physics~\cite{xie2023physgaussian}. We expand on this intuition, enhancing the current 3DGS ecosystem by injecting physics-based fluid and solid interactions into a 3DGS scene. This appears straightforward at first sight. Since 3DGS kernels are essentially a collection of ellipsoids, they can be used for the discretization of the fluid and solid dynamics just as position-based dynamics~\cite{macklin2016xpbd}, oriented particles~\cite{muller2011solid} or other particle-based simulation techniques. Unfortunately, a simple combination of those techniques does not yield the results expected. Large rotational deformation of the solid objects affects the splatting results with sharp and spiky noises. During fluid motion, fluid particles undergo substantial positional shifts, moving from the inside to the outside or vice versa. Fluids are both translucent and specular. The vanilla 3DGS simplifies the composition of the light field without well-defined appearance properties. This limitation makes fluid rendering cumbersome with 3DGS.

This paper presents a system namely Gaussian Splashing (GSP), a 3DGS-based framework that enables realistic interactions between solid objects and fluids in a physically meaningful way, and thus generates two-way coupled fluids-solids dynamics in novel views. GSP integrates Lagrangian fluid and 3DGS scenes through a unified framework of position-based dynamics (PBD)~\cite{muller2007pbd,macklin2016xpbd}. We follow a recent contribution of GaussianShader~\cite{jiang2023gaussianshader} to augment Gaussian kernels with additional environmental information so that specular shading can be dynamically synthesized along with the fluid's movement. For solid objects, GSP uses an anisotropy loss to cap the stretching ratio during 3DGS training and mitigate the rendering artifacts induced by rotational deformation. We approximate the normal of a fluid kernel based on the surface tension if it is near the fluid surface. For scattered fluid droplets, we resort to a depth volume rendered via the current camera pose to estimate the normal information~\cite{van2009screenspace}. GSP is versatile, due to the flexibility of PBD. It handles deformable bodies, rigid objects, and fluid dynamics in a unified way. While it is possible to incorporate more complicated constitutional models as in~\cite{feng2023pienerf} and \cite{li2023pacnerf}. We found that PBD-based simulation suffices in many situations. We further augment GSP with an image-space segmentation module to select objects of interest from the 3DGS scene. We exploit the latest generative AI to fill the missing pixels to enable interesting physics-based scene editing. 

In a nutshell, GSP leverages a unified, volumetric, particle-based representation for rendering, 3D reconstruction, view synthesis, and dynamic simulation. It contributes a novel 3D graphics/vision system that allows natural and realistic solid-fluid interactions in real-world 3DGS scenes. This is achieved by carefully engineering the pipeline to overcome the limitations of the vanilla 3DGS. GSP could enable a variety of intriguing effects and new human-computer interaction modalities in a diverse range of applications. For instance, one can pour water to flood the scene, floating the objects within, or directly liquefy an object, just as in science fiction. Fig.~\ref{fig:teaser} showcases the dynamic interaction between a LEGO excavator and the splashing waves. There are $334$,$815$ solid Gaussian kernels and $280$,$000$ fluid Gaussian kernels. Through the two-way coupling dynamics, the excavator is animated to surf on the splashing waves.


\section{related work}
\label{sec:related}

\paragraph{Dynamic neural radiance field}
Dynamic neural radiance fields generalize the original NeRF system to capture time-varying scenes e.g., by decomposing time-dependent neural fields into an inverse displacement field and canonical time-invariant neural fields~\cite{park2021nerfies,park2021hypernerf,tretschk2021non,weng2022humannerf}, or estimating the time-continuous 3D motion field~\cite{pumarola2021d,du2021neural,li2021neural,xian2021space,gao2021dynamic,liu2022devrf,guo2023forward} with an added temporal dimension. Existing arts enable direct edits of NeRF reconstructions~\cite{bao2023sine,dong2023vica,jiang2022nerffaceediting,Li2023ClimateNeRF}. In dynamic scenes, the rendering process needs to trace deformed sample points back to rest-shape space to correctly retrieve the color/texture information~\cite{yuan2022nerfediting,qiao2023dynamic,xu2022deforming,peng2022cagenerf}. They often extract a mesh/grid from the NeRF volume. It is also possible to integrate physical simulation with NeRF using meshless methods~\cite{li2023pacnerf, feng2023pienerf}. Point NeRF~\cite{xu2022pointnerf} and 
3DGS~\cite{kerbl2023gaussians} offer a different perspective to scene representation explicitly using points/Gaussian kernels to encode the scene. The success of 3DGS has inspired many studies to transplant techniques for dynamic NeRF to 3DGS~\cite{wu20234d,liang2023gaufre,xu2023gsheadavatar}. They incorporate learning-based deformation and editing techniques to reconstruct or generate dynamics of NeRF scenes. It is noteworthy that a recent work from \citet{xie2023physgaussian} integrates physical simulation with 3DGS, leveraging the unified proxy for both simulating and rendering.

\paragraph{Lagrangian fluid simulation}
Lagrangian fluid simulation tracks fluid motion using individual particles as they traverse the simulation domain. A seminal approach within this domain is smoothed particle hydrodynamics (SPH)~\cite{monaghan1992sph}, which solves fluid dynamics equations by assessing the influence of neighboring particles. Despite its efficacy, SPH, particularly in its standard and weakly compressible forms (WCSPH)~\cite{Becker2007wcsph}, suffers from parameter sensitivity, e.g., kernel radius and time-step size for stiff equations. To relax the time-step restriction, predictive-corrective incompressible SPH (PCISPH)~\cite{solenthaler2009pcisph} iteratively corrects pressure based on the density error. Similarly, position-based dynamics~\cite{muller2007pbd} provides a robust method of solving a system of non-linear constraints using Gauss-Seidel iterations by updating particle positions directly, which can also be employed in fluid simulation~\cite{macklin2013pbf} with improved stability. Furthermore, surface tension can also be generated~\cite{xing2022pbtf} using the position-based iterative solver by tracking surface particles and solving constraints on them to minimize the surface area.

\paragraph{Reflective object rendering}
Achieving precise rendering of reflective surfaces relies on accurately estimating scene illumination, such as environmental light, and material properties like bidirectional reflectance distribution function (BRDF). This task falls under the domain of inverse rendering~\cite{barron2014shape,nimier2019mitsuba}. Some NeRF-related methodologies disentangle the visual appearance into lighting and material properties, which can jointly optimize environmental illumination, surface geometry, and material~\cite{bi2020neural,boss2021nerd,boss2021neural,srinivasan2021nerv,zhang2021physg,zhang2021nerfactor}. Other NeRF studies~\cite{liang2022spidr,liang2023envidr,liu2023nero,kang2021view} aim to enhance the accuracy of the normal estimation in physically based rendering (PBR). Nevertheless, these efforts face challenges such as time-consuming training and slow rendering speed. On the contrary, 3DGS naturally offers a good normal estimation as the shortest axis of the Gaussian kernel~\cite{guedon2023sugar,jiang2023gaussianshader}. Following this idea, it is possible to achieve high-quality rendering of reflective objects and training efficiency simultaneously~\cite{jiang2023gaussianshader}.

\paragraph{Point-based rendering}
Point-based rendering has been an active topic in computer graphics since the 1980s~\cite{levoy1985point}. The simplest method~\cite{grossman1998point} renders a set of points with a fixed size.  It suffers from holes and relies on post-processing steps such as hole-filling~\cite{grossman1998point,pintus2011real} to correct the resulting rendering. An improvement is to use ellipsoids instead of volume-less points. This strategy is usually referred to as \emph{point splatting}~\cite{zwicker2001splatting}. The ellipsoids are rendered with a Gaussian alpha-mask to eliminate visible artifacts between neighboring splats and then combined by a normalizing blend function~\cite{zwicker2001splatting,alexa2004point}. Point-based rendering well synergizes with  Lagrangian fluid rendering, enabling the calculation of fluid thickness and depth through splatting. This approach~\cite{van2009screenspace} achieves fluid rendering at an impressive real-time speed. Further extensions to splatting aim to automatically compute the shape and color of ellipsoids, for example, auto splats~\cite{childs2012auto}. With the development of deep learning in recent years, learning-based approaches improve the image quality of splatting~\cite{bui2018point,yang2020surfelgan}. 3DGS~\cite{kerbl2023gaussians} has introduced this technique into 3D reconstruction, enabling high-quality real-time novel view synthesis for reconstructed scenes. A natural idea is to combine 3DGS with fluid rendering, enabling interaction between reconstructed scenes and dynamic fluids.


\begin{figure*}
    \centering
    \includegraphics[width=\textwidth]{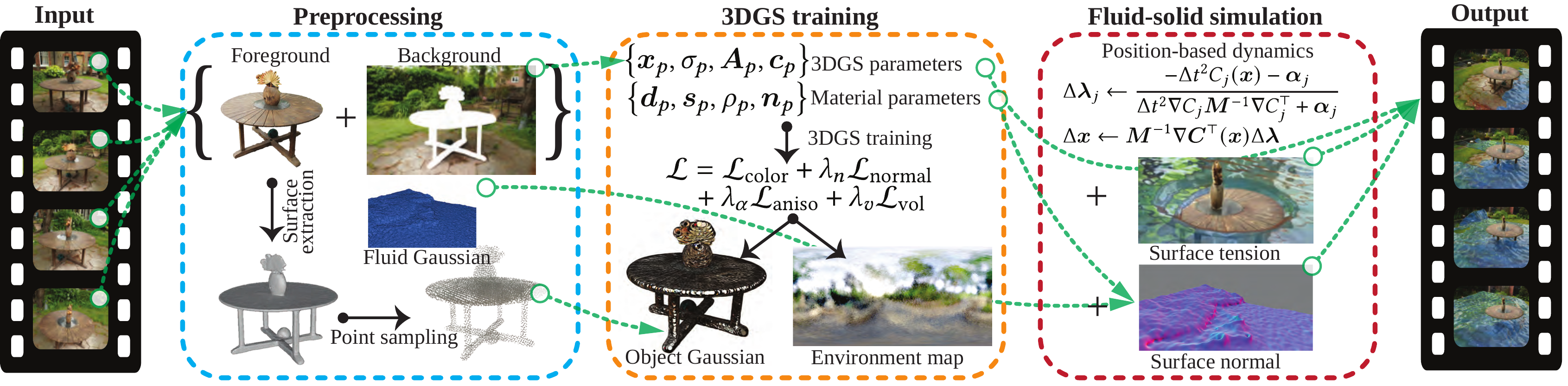}
    \caption{\textbf{An overview of GSP pipeline.}~~The input to our system comprises multi-view images that capture a 3D scene. During the preprocessing stage, foreground objects are isolated and reconstructed. This is followed by point sampling to facilitate scene discretization for PBD simulation and Gaussian rendering. We train the Gaussian kernels using differentiable 3DGS, which takes into account appearance materials and lighting conditions. These kernels are animated using PBD, in conjunction with fluid particles, to tackle the dynamics of both solids and fluids within the scene. Finally, the dynamic scene is rendered into images. This rendering process includes detailed modeling of specular reflections, thereby providing visually accurate representations of the simulated interactions between solids and fluids.}
    \label{fig:overview}
\end{figure*}

\section{Background}\label{sec:background}
To make our paper self-contained, we start with a brief review of PBD and 3DGS on which our pipeline is built. More detailed discussions are available in the supplementary material and relevant literature e.g.,~\cite{macklin2016xpbd,muller2007pbd,kerbl2023gaussians,jiang2023gaussianshader}.

\subsection{Position-Based Dynamics}
PBD/XPBD treats a dynamic system as a set of $N$ vertices and $M$ constraints. This perspective offers an easy and efficient simulation modality, converting the variational optimization to the so-called constraint projections. Specifically, XBPD considers the total system potential $U$ as a quadratic form of all the constraints $\bm{C}(\bm{x}) = [C_1(\bm{x}), C_2(\bm{x}),...,C_M(\bm{x})]^\top$ such that $
U = \frac{1}{2}\bm{C}^\top(\bm{x}) \bm{\alpha}^{-1} \bm{C}(\bm{x})$. Here, $\bm{x}$ represents the position of vertices and $\bm{\alpha}$ is the compliance matrix, i.e., the inverse of the constraint stiffness. XBPD estimates an update of constraint force (i.e., the multiplier) $\Delta \bm{\lambda}$ by solving:
\begin{equation}
    \left[\Delta t^2 \nabla \bm{C}(\bm{x})\bm{M}^{-1}\nabla\bm{C}^\top(\bm{x}) + \bm{\alpha} \right]\Delta \bm{\lambda}= -\Delta t^2 \bm{C}(\bm{x}) - \bm{\alpha} \bm{\lambda},
\end{equation}
where $\Delta t$ is the time step size, and $\bm{M}$ is the lumped mass matrix. The update of the primal variable $\Delta \bm{x}$ can then be computed as:
\begin{equation}
    \Delta \bm{x} = \bm{M}^{-1}\nabla\bm{C}^\top(\bm{x})\Delta\bm{\lambda}.
\end{equation}

We note that such constraint-projection-based simulation naturally synergizes with 3DGS. It is also versatile and can deal with a wide range of physical problems such as fluid~\cite{macklin2013pbf,xing2022pbtf}, rigid bodies~\cite{muller2020detailed}, or hyperelastic objects~\cite{macklin2021constraint}.

\subsection{3D Gaussian Splatting}
3D Gaussian Splatting (3DGS) is a learning-based rasterization technique for 3D scene reconstruction and novel view synthesis. 3DGS encodes a radiance field using a set of Gaussian kernels $\mathcal{P}$ with trainable parameters $\mathbf{x}_p, \sigma_p, \mathbf{A}_p, \mathbf{c}_p$ for $p\in \mathcal{P}$, 
where $\bm{x}_p$, $\sigma_p$, $\bm{A}_p$ and $\bm{c}_p$ represent the center, opacity, covariance matrix, and color function of each kernel. To generate a scene render, 3DGS projects these kernels onto the imaging plane according to the viewing matrix and blends their colors based on the opacity and depth. The final color of the $i$-th pixel is computed as:
\begin{equation}\label{eq:GaussianSlatting}
\bm{c}_i=\sum_{k} G_k(i) \sigma_k \bm{c}_k (\bm{r}_i) \prod_{j=1}^{k-1}\left(1 - G_j(i) \sigma_j \right).
\end{equation}
Here, all the kernels are re-ordered based on the z-values at kernel centers under the current view. $G_k(i)$ denotes the 2D Gaussian weight of the $k$-th kernel at pixel $i$, and $\bm{r}_i$ is the view direction from camera to pixel $i$. The color functions only depend on the viewing direction. 

GaussianShader~\cite{jiang2023gaussianshader} further enhances 3DGS by incorporating additional trainable material parameters for kernel $p$ such as diffuse $\bm{d}_p$, specular $\bm{s}_p$, roughness $\rho_p$, and normal $\bm{n}_p$, along with a global environment map. It fuses more information into the kernel's color:
\begin{equation}\label{eq:GaussianShader}
\bm{c}_p(\bm{r}_i) = \bm{d}_p + \bm{s}_p \odot L_{s}(\bm{r}_i, \bm{n}_p, \rho_p),
\end{equation}
where $L_{s}(\bm{r}_i, \bm{n}_p, \rho_p)$ is the specular light for the kernel along $\bm{r}_i$ given the normal and roughness of the kernel. It can be pre-filtered into multiple mip maps. The symbol $\odot$ denotes the element-wise multiplication.

\section{Method}\label{sec:method}
As shown in Fig.~\ref{fig:overview}, the input of our system is a collection of images of a given 3D scene taken from different viewpoints. We separate foreground objects and the image background for all the inputs and extract the surface of masked objects. We apply an anisotropy loss to mitigate undesired splatting render to prevent over-stretching Gaussian kernels when training 3DGS for the solid object. Doing so mitigates the rendering artifacts near the surface of solid models. We decouple solid simulation and rendering by utilizing a separate set of sampled particles for simulation, and interpolating deformations of these particles onto trained Gaussian kernels during rendering. This approach ensures high-quality and robust results in both simulation and rendering. On the other hand, fluids use a unified set of Gaussian kernels (for rendering) or particles (for simulation). We track the fluids surface during simulation and use surface normal to properly synthesize specular effects by augmenting them with a decomposed environment map. Under the PBD framework, both fluids and solids are made of particles, and can be animated in a unified way based on local constraint projection.

\subsection{Training}
Our training process is generally similar to traditional Gaussian Splatting. However, due to our use of Physically-Based Rendering (PBR) modeling, the visual attributes or material parameters (e.g., diffuse, specular, roughness) of Gaussian kernels need to be determined. Without them, high-quality rendering results are not possible. Similar to \cite{jiang2023gaussianshader}, we leverage the differentiable 3DGS pipeline to optimize the appearance of every Gaussian kernel. All Gaussian kernels are first shaded with their corresponding material parameters (Eq.~\eqref{eq:GaussianShader}) and are then splatted to a rendered image. Comparing it with the training view gives a loss back-propagated to update the corresponding parameters of Gaussian kernels. Specifically, the trainable parameters for each kernel $p$ are position $\bm{x}_p$, opacity $\sigma_p$, covariance $\bm{A}_p$ and material $\{\bm{d}_p, s_p, \rho_p, \bm{n}_p\}$. The loss is defined as:
\begin{equation}\label{eq:Loss}
    \mathcal{L}=\mathcal{L}_{\rm color}+\lambda_n\mathcal{L}_{\rm normal}+\lambda_a\mathcal{L}_{\rm aniso},
\end{equation}
where $\mathcal{L}_{\rm color}$ is the color loss between render and training image; $\mathcal{L}_{\rm normal}$ is normal consistency regularization adopted from GaussianShader~\cite{jiang2023gaussianshader}. The anisotropy loss $\mathcal{L}_{\rm aniso}$ is designed to prevent Gaussian kernels from becoming excessively elongated or compressed and potentially producing artifacts under large deformations. It is defined as:
\begin{equation}\label{eq:AnisotropyLoss}
\mathcal{L}_{\rm aniso}=\frac{1}{|\mathcal{P}|} \sum\limits_{p\in\mathcal{P}}{\max}\left\{\frac{\bm{S}_p^1}{\bm{S}_p^2}-a,0\right\},
\end{equation}
where $a$ is a ratio threshold and $\bm{S}_p=\left\{\bm{S}_p^1,\bm{S}_p^2,\bm{S}_p^3\right\}$ are the scalings of Gaussian kernels with $\bm{S}_p^1$ being the largest scaling and $\bm{S}_p^3$ being the smallest scaling. Note that as the shading normal is based on the minimal axis of Gaussian kernels, we do not constrain the minimal axis in the anisotropy loss. Otherwise, a spherical Gaussian kernel will result in normal ambiguity.
We set $a=1.1$, $\lambda_n=0.2$, and $\lambda_a=10$ in our experiments.

\subsection{Solid Simulation and Interpolation}
In many dynamic Gaussian Splatting applications~\cite{xie2023physgaussian, qian2023gaussianavatars}, simulation and rendering are often coupled, utilizing the Gaussian kernels for both processes. This approach presents several issues. First, Gaussian kernels tend to distribute primarily on the surface of objects, necessitating the addition of internal kernels that, unless they contribute to effects like fracturing or breakage, are predominantly unrendered. Additionally, an uneven distribution of surface kernels can compromise the quality of the simulation, while a distribution that is too uniform can detrimentally affect rendering quality. Therefore, decoupling simulation and rendering is necessary to maintain quality in both aspects. In our pipeline, we leverage reconstructed 3DGS kernels for solid rendering and utilize a distinct set of particles for simulation.

To properly handle the object-object interaction or fluid-solid coupling, the particles are required to have an accurate description of object boundaries and interiors. To this end, we use the Poisson disk sampling~\cite{bridson2007fast} to place simulation particles within the surface mesh, which is explicitly reconstructed for segmented foreground model using NeuS~\cite{wang2021neus}. NeuS extracts the zero-level set of a signed distance function corresponding to the foreground object. It is important to note that existing frameworks that incorporate physics with 3DGS~\cite{xie2023physgaussian} also require spatial discretization over the model to numerically solve the dynamics equation. Those prior approaches sample the model based on the 
trained 3DGS kernels. This method can result in sparsely sampled regions, especially for objects with thin parts, potentially affecting simulation quality (see Fig.~\ref{fig:boundary_comparison} in supplementary material).

Once the simulation particles are sampled and in place, we perform PBD on them to animate frames of dynamic motion. Each frame, deformation gradients and displacements are interpolated from the simulation particles to the trained Gaussian kernels using generalized moving least squares (GMLS)~\cite{martin2010unified}, which produces smooth and robust results.

\subsection{Position-Based Fluids}
We employ the Position-Based Fluids (PBF)~\cite{macklin2013pbf} as our Lagrangian fluid synthesizer. To enforce the fluid incompressibility, PBF imposes a density constraint $C^\rho_i$ on each particle, maintaining the integrated density $\rho_i$ computed by the SPH kernel as:

\begin{equation}\label{eq:FluidsConstraint_paper}
C_i^\rho =\frac{\rho_i}{\rho_0} - 1=\sum_j\frac{m_j}{\rho_0}W(\bm{p}_i-\bm{p}_j) - 1,
\end{equation}
where $m_j$ is the mass of particle $j$. $\bm{p}_i$ is the position of particle $i$, and $W$ is the SPH kernel function. Each constraint can be straightforwardly parallelized using GPUs.

GSP incorporates a position-based tension model~\cite{xing2022pbtf} to more accurately simulate fluid surface dynamics. Surface detection for a particle is performed through occlusion estimation, where each particle is enclosed by a spherical screen. Neighboring particles project onto this screen, and a particle is classified as on the surface if the cumulative projection area is below a specified threshold, reflecting its partial exposure. Given the natural tendency of tensions to reduce surface area, PBF enforces an area constraint on each surface particle to minimize local surface area. This process begins with the calculation of the normal $\bm{n}_i$ for surface particle $i$ as:
\begin{equation}\label{eq:surfacenormal_paper}
    \bm{n}_i = \text{normalize}(-\nabla_{\bm{p}_i}C_{i}^\rho),
\end{equation}
where $C_i^\rho = 0$ indicates the particle is inside the fluid, and $C_i^\rho = -1$ indicates it is outside. After that, we project the neighboring surface particles onto a plane perpendicular to $\bm{n}_i$ and triangularize the plane. The area constraint can then be built as:
\begin{equation}\label{eq:AreaConstraint_paper}
C_i^A=\sum_{t\in T(i)}\frac{1}{2}\left\|(\bm{p}_{t^2}-\bm{p}_{t^1})\times (\bm{p}_{t^3}-\bm{p}_{t^1})\right\|
\end{equation}
where $T(i)$ is the set of neighboring triangles for particle $i$. To promote a more uniform particle distribution, additional distance constraints are introduced to push apart particles that are too close to each other:
\begin{equation}\label{eq:DistanceConstraint_paper}
C_{ij}^D={\min}\left\{ 0, \left\|\bm{p}_i-\bm{p}_j\right\|-d_0 \right\},
\end{equation}
where $d_0$ is the distance threshold. The original version was parallelized on CPUs; we have enhanced it for GPU parallelization. Calculations for $\bm{n}_i$, $C_i^A$ and $C_{ij}^D$ are efficiently parallelizable on GPUs. Considering that the number of surface particles surrounding each particle is typically low, local triangulation for each particle is conducted independently on separate threads.

\subsection{Rendering}
The rendering of the dynamic scene reuses the existing 3DGS pipeline. For dynamic solids, 
we first transform each solid Gaussian kernel from rest positions $\bm{x}_p$ to deformed positions $\bm{x}^t_p$ where $t$ indicates the time step index. We directly place these kernels at deformed positions. For a kernel with deformation gradient $\bm{F}_p$, its covariance $\bm{A}_p^t$ and normal $\bm{n}_p^t$ after deformation is updated by:
\begin{equation}\label{eq:deform_gs_update}
    \bm{A}_p^t=\bm{F}_p\bm{A}_p\bm{F}_p^\top, \;\text{and} \; \bm{n}_p^t=\frac{\bm{F}_p^{-\top}\bm{n}_p}{\left\|\bm{F}_p^{-\top}\bm{n}_p\right\|}.
\end{equation}
We then shade the deformed Gaussian kernels $\{\bm{x}^t_p,\sigma_p,\bm{A}_p^t, \bm{n}_p^t\}$ with material $\{\bm{d}_p,s_p,\rho_p\}$, i.e., Eq.~\eqref{eq:GaussianShader} and splat them into an image $\bm{c}^{\rm bg}$. As shadows are important to visual outcomes in dynamic scenes, we further re-engineer nearly-soft shadows~\cite{donnelly2006variance} into our system to enhance the realism.

\begin{figure}
    \centering
    \includegraphics[width=\linewidth]{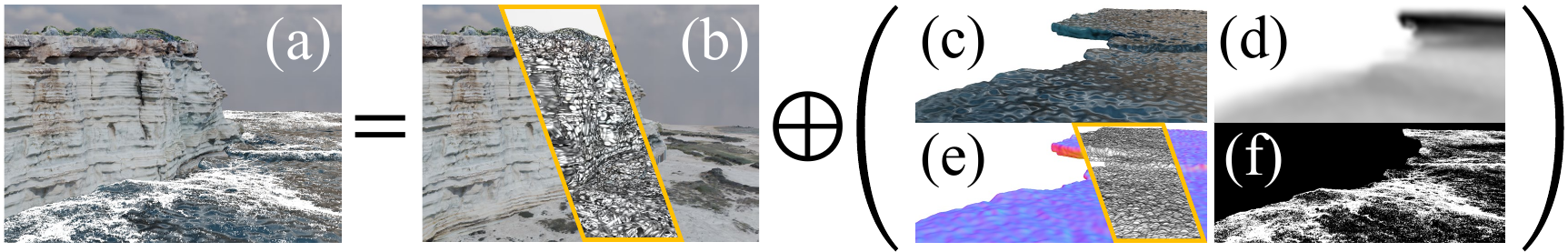}
    \caption{\textbf{GSP rendering.}~~GSP synthesizes high-quality images corresponding to dynamically interacting fluids and solids. (a) The final rendered image combining rendered solids, fluids, and foams. (b) The rendering result of solids. (c) The rendering result of fluids. (d) The fluid thickness by additive splatting, where the darker color indicates the higher thickness. (e) The normal of fluids. (f) The intensity of foam particles. The insets in (b) and (e) represent solid and fluid Gaussian kernels, respectively.}
    \label{fig:fluid_rendering_intermediate_results}
\end{figure}

The dynamic particle-based fluids are rendered with ellipsoids splatting~\cite{macklin2013pbf}, which is inherently compatible with the existing 3DGS pipeline. We begin by generating spherical fluid Gaussian kernels at each fluid particle. The initial covariance $\bm{A}_p$ of each kernel is determined by the particle radius. The normal $\bm{n}_p$ adopts the surface normal of the nearest surface fluid particle from PBF simulation. 
We proceed to set up the appearance material for each fluid Gaussian kernel, which requires careful engineering due to the fluid's strong reflection and refraction effects. We employ the current PBR workflow to model the reflection, which adopts the same formula of Eq.~\eqref{eq:GaussianShader}. We set a specular material ($s_p=1$, $\rho_p=0.05$ in our experiments) for all fluid kernels to imitate reflective behavior. The refraction needs careful treatment, which we model into a thickness-dependent diffuse color $\bm{d}_p$. As the fluid thickness increases, the absorption of light within the fluid intensifies, resulting in reduced visibility for objects behind. Conversely, when there is less fluid present, it exhibits a more transparent appearance. The fluid thickness $\tau$ comes from the modified splatting pipeline, with the alpha blending replaced by additive blending. The refraction color $\bm{d}_p$ is then represented by Beer's Law~\cite{swinehart1962beer}:
\begin{equation}\label{eq:refraction_color}
    \bm{d}_p=e^{-k\tau_p}\bm{c}_p^{\rm bg}.
\end{equation}
Here, the absorption coefficient $k$ is defined differently for each color channel, $\tau_p$,$\bm{c}_p^{\rm bg}$ is the fluid thickness and background back-projected to each Gaussian kernel respectively. Note that for background back-projection, a distortion $\beta\bm{n}_p$ is added to mimic the change of light path due to refraction. Opacity $\sigma_p$ is set to $1$ as most of transmission and refraction has already been modeled into $\bm{d}_p$. We finally shade all fluid Gaussian kernels $\{\bm{x}^t,\sigma_p,\bm{A}_p^t, \bm{n}_p^t\, \bm{d}_p,s_p,\rho_p\}$ and splat them to the fluid rendering result $\bm{c}^{\rm fluid}$ with Gaussian Splatting. To enhance the realism of fluids, foam particles are synthesized~\cite{ihmsen2012unified} and rendered~\cite{akinci2013screen}. The 3DGS pipeline is re-engineered to incorporate additive splatting with distinct kernels for foam, bubble, and spray particles. The final rendering result is achieved by combining the $\bm{c}^{\rm bg}$ and $\bm{c}^{\rm fluid}$, as shown in Fig.~\ref{fig:fluid_rendering_intermediate_results}. 
Please refer to the supplementary material for more details on the rendering part.

\subsection{Inpainting}
Displaced object exposes unseen areas that were originally covered to the camera. Since they are not present in the input image, 3DGS is unable to recover the color and texture information of these areas, leading to black smudges and dirty textures in the result. GSP remedies this issue using an inpainting trick. First, we remove all the Gaussian kernels of the object that may be displaced. We then use LaMa~\cite{suvorov2022lama} to inpaint the rendered results, noting the new colors of pixels originally in spots and assigning them to the first Gaussian kernel encountered by the rays emitted from these pixels. We average the recorded colors on all noted Gaussian kernels for their diffuse color, set their opacity to $1$, and their specular color to $\bm{0}$ to prevent highlights.


\section{Experiments}\label{sec:Experiments}

We implemented Gaussian Splashing pipeline using \texttt{Python}, \texttt{C++} and \texttt{CUDA} on a desktop PC equipped with a 12-core \texttt{Intel} \texttt{i7-12700F} CPU and an \texttt{NVIDIA} \texttt{RTX} \texttt{3090} GPU. Specifically, for the rendering part, we ported the published implementation of GaussianShader~\cite{jiang2023gaussianshader} and integrated our fluid rendering using \texttt{PyTorch}~\cite{imambi2021pytorch}. PBD/PBF engine was implemented with \texttt{CUDA}, and we also group independent constraints to efficiently parallelize the constraint projections on the GPU. Please refer to the supplementary material for more details of the implementation.

\subsection{Ablation}

\begin{figure}
    \centering
    \includegraphics[width=\linewidth]{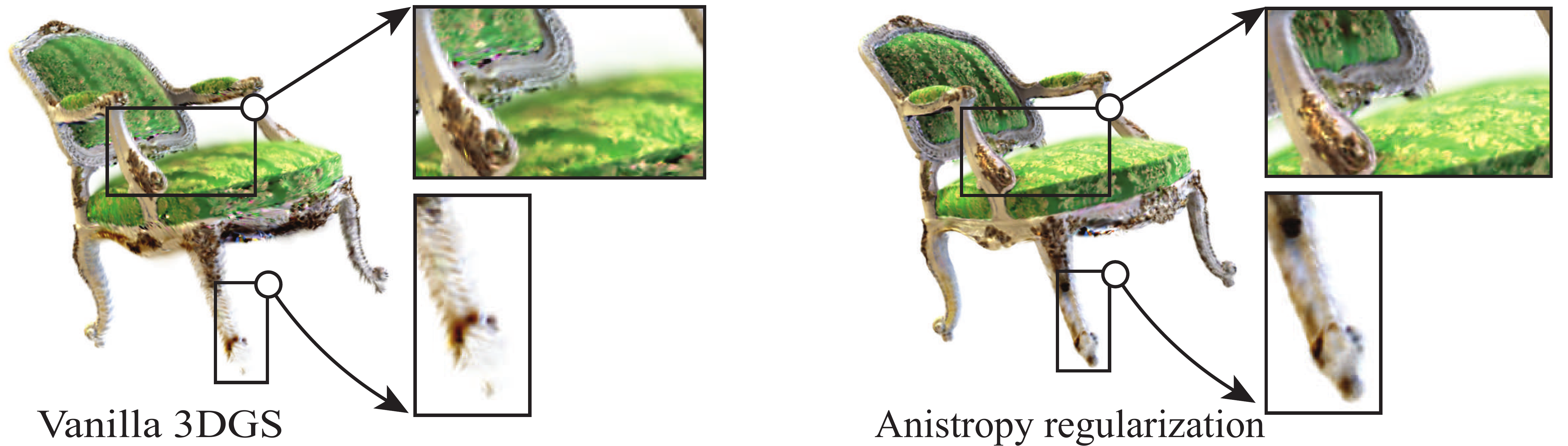}
    \caption{\textbf{Anistropy regularization.}~~Anistropy regularization effectively maintains rendering quality under large deformations. Without the regularization term, 3DGS tends to generate fuzzy and spiky artifacts, especially near the surface of the model (left). When the regularization is applied, image quality is greatly improved with correct specular effects.}   
    \label{fig:anisotropy_regularization}
\end{figure}

\paragraph{Anisotropy Regularization} 3DGS is originally designed for view synthesis. 3DGS obtained from a static scene produces low-quality renders when Gaussian kernels undergo large rotational deformations. The anisotropy regularization is effective against this limitation as shown Fig.~\ref{fig:anisotropy_regularization}. Detailed statistics regarding the experiment settings and timings are reported in Tab.~\ref{tab:time}. Most experiments can also be found in the supplementary video.

\begin{figure}
    \centering
    \begin{minipage}{\linewidth}
        \centering
        \begin{minipage}{.48\linewidth}
            \centering
            \includegraphics[width=\linewidth,trim={1cm 3cm 1cm 3cm},clip]{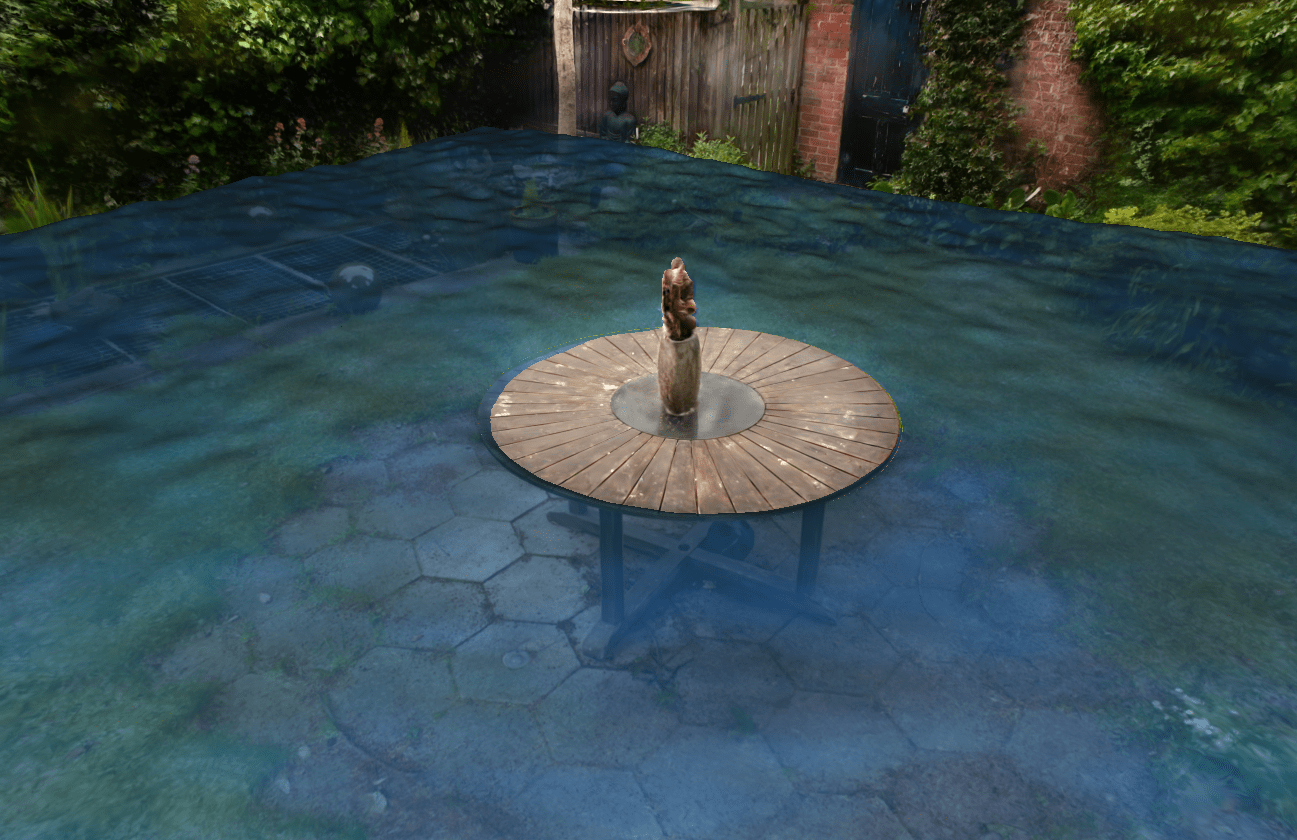}
        \end{minipage}
        \begin{minipage}{.48\linewidth}
            \centering
            \includegraphics[width=\linewidth,trim={1cm 3cm 1cm 3cm},clip]{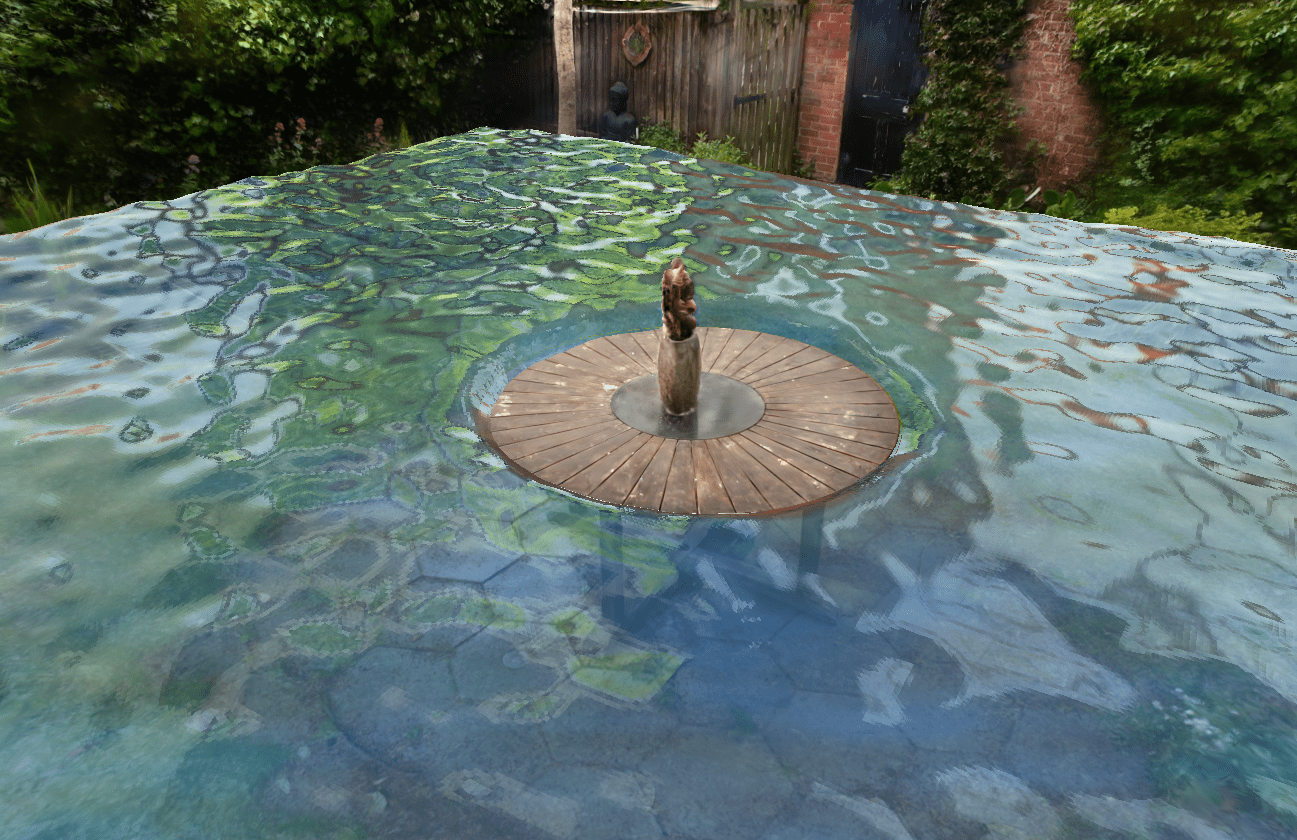}
        \end{minipage}
    \end{minipage}
    \caption{\textbf{Ablation study of specular.}~~We demonstrate the impact of specular highlights on the quality of rendering. On the left is a fluid rendered with diffuse color only. On the right, surface reflective specular are added, which exhibits a more realistic and dynamic fluid. }
    \label{fig:specular}
\end{figure}

\paragraph{PBR Material} To show the importance of specular highlights in fluid rendering, we show a side-by-side comparison in a 3DGS scene. As shown in Fig.~\ref{fig:specular}, we compare fluids rendered using purely diffuse materials with those that incorporated specular reflections. The fluids without specular reflection appear almost smoke-like, while the inclusion of specular term significantly enhances the realism of the fluids. It should be noted that incorporating PBR in 3DGS is key to improving the render quality of largely deformed fluids. This is in contrast to previous work~\cite{xie2023physgaussian}, which baked lighting into spherical harmonics and failed to produce realistic rendering when fluids underwent significant motion.

\begin{figure}
    \centering
    \includegraphics[width=\linewidth]{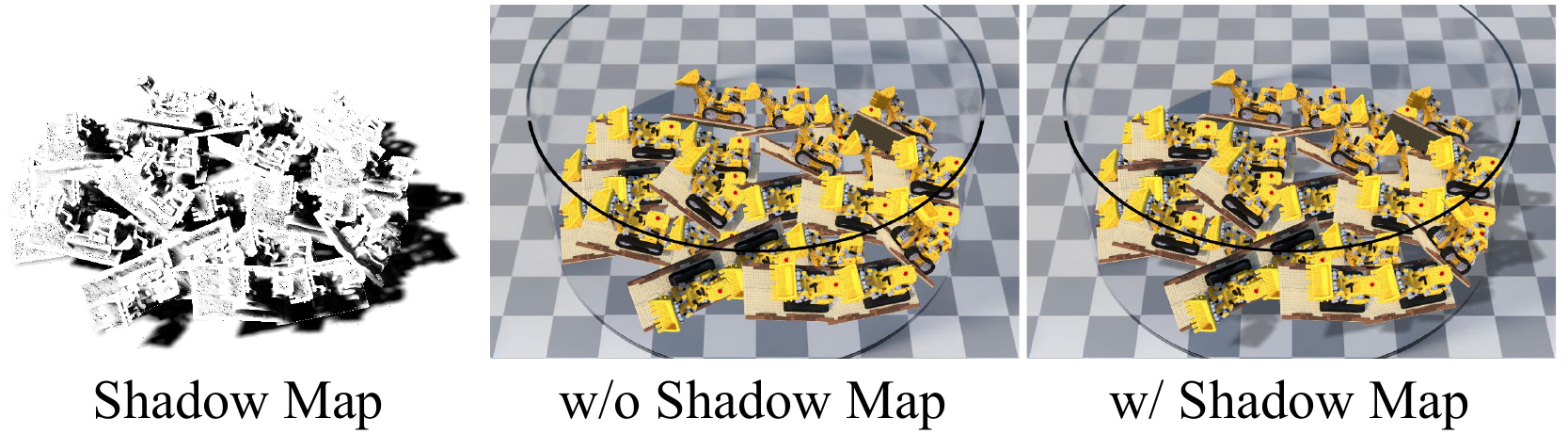}
    \caption{\textbf{Shadow map.}~GSP incorporates dynamic shadows into the rendering pipeline to enhance visual realism. We re-engineer variance shadow mapping~\cite{donnelly2006variance} within the existing 3DGS pipeline to produce nearly-soft shadows.}
    \label{fig:shadow}
\end{figure}

\paragraph{Shadow Map} In Fig.~\ref{fig:shadow}, we demonstrate that GSP, with the addition of nearly soft shadows, significantly improves visual realism compared to vanilla 3DGS. Without these shadows, objects appear like flat layers pasted onto the background, lacking a sense of depth.

\begin{figure}
    \centering
    \includegraphics[width=\linewidth]{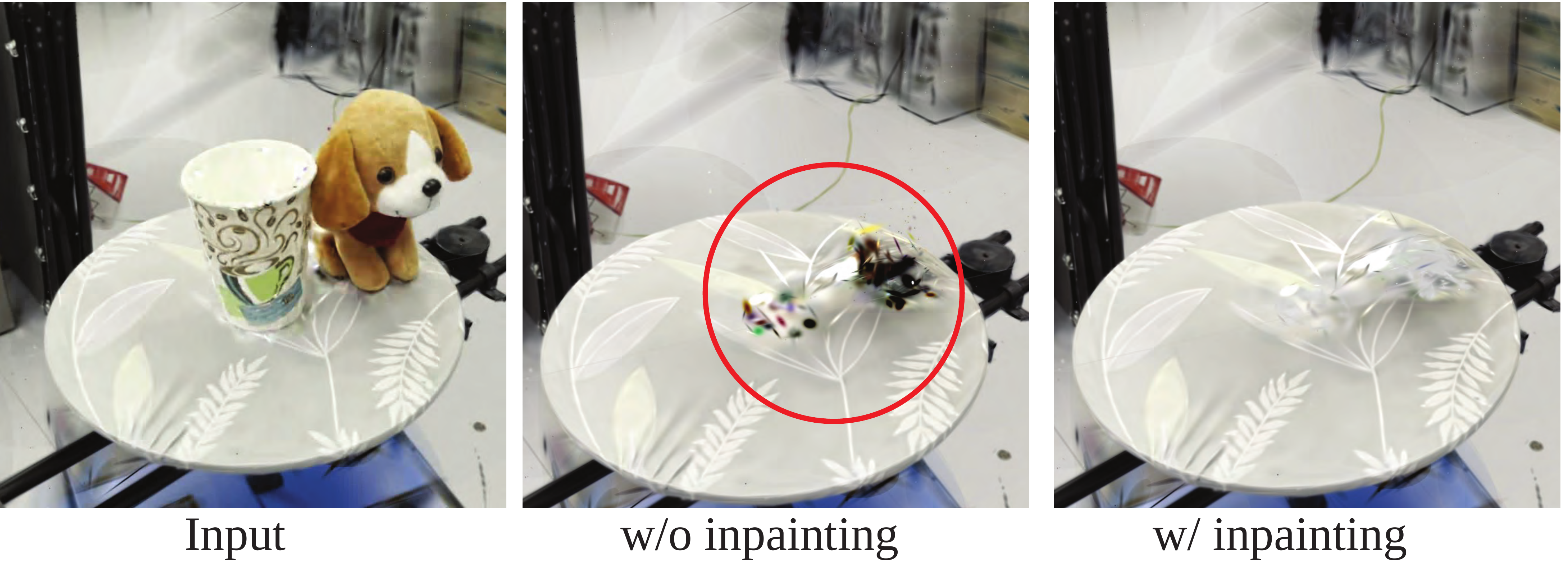}
    \caption{\textbf{3DGS inpainting.}~~In this indoor scene, both the paper cup and the stuffed toy dog are segmented from the input image (left). 3DGS leaves empty spots and dirty textures blended from irrelevant kernels, as highlighted in the middle figure. Applying the inpainting with generative AI~\cite{suvorov2022lama} ameliorates this issue (right).}  
    \label{fig:inpainting}
\end{figure}

\paragraph{Inpainting} We conduct an ablation experiment on inpainting as shown in Fig.~\ref{fig:inpainting}. In this indoor scene, displacing cup and dog with vanilla 3DGS results in the occurance of black smudges and dirty textures, as the hidden area by the object could not be reconstructed properly due to missing information in the input images. GSP addresses this by using LaMa~\cite{suvorov2022lama} for inpainting.

\subsection{Evaluation}

\begin{figure}
    \centering

   \begin{minipage}{\linewidth}
        \centering
        \begin{minipage}{\linewidth}
            \centering
            \begin{minipage}{0.49\linewidth} 
                \centering
                \includegraphics[width=\linewidth]{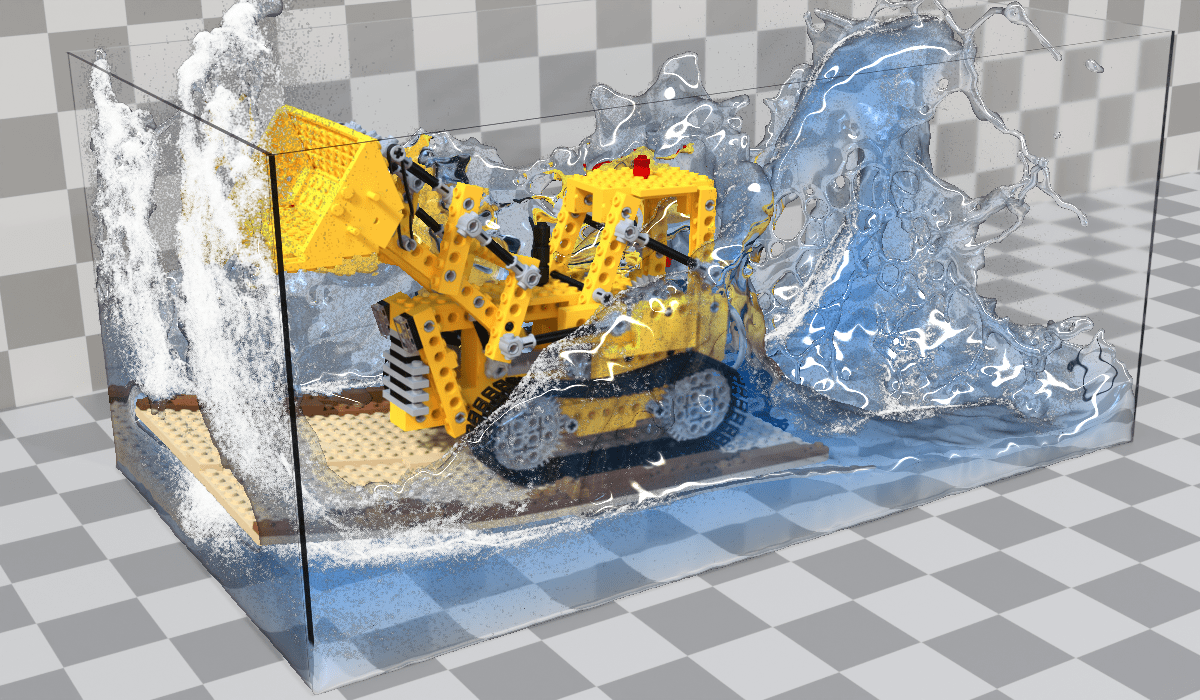}
            \end{minipage}
            \begin{minipage}{0.49\linewidth} 
                \centering
                \includegraphics[width=\linewidth]{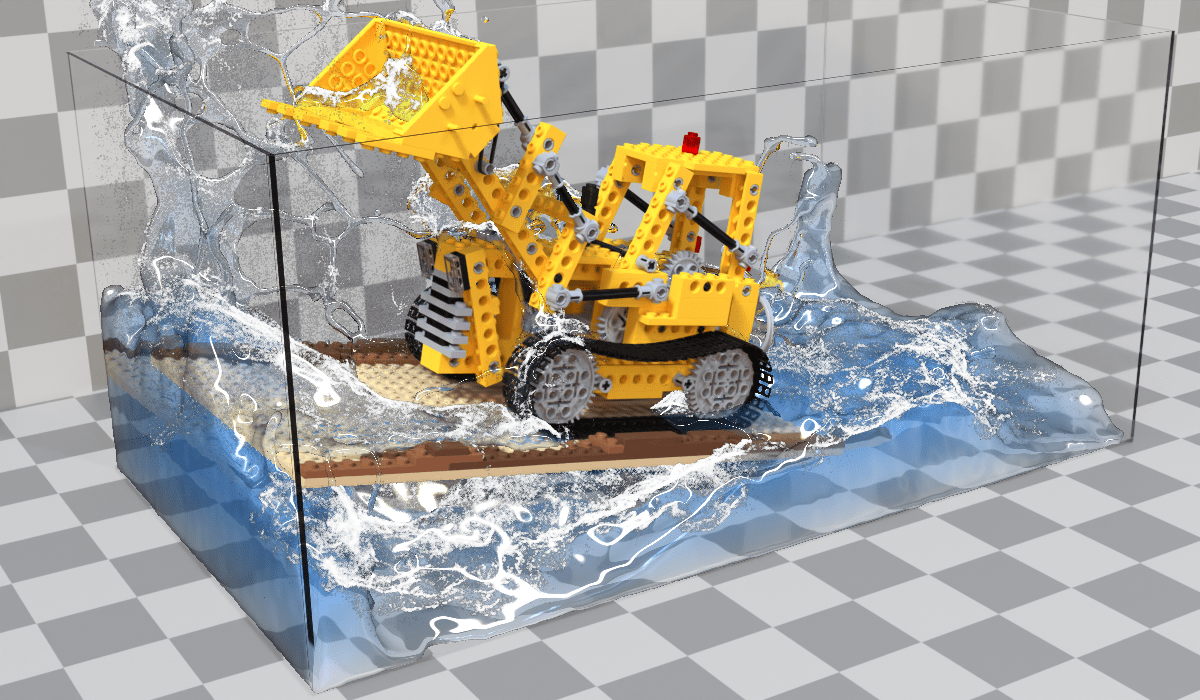}
            \end{minipage}
        \end{minipage}
    \end{minipage}

    \caption{\textbf{Splashing LEGO.~ Through the two-way coupling dynamics, the LEGO bulldozer is animated to surf on the splashing waves.}}
    \label{fig:splashing_brief}
\end{figure}

We evaluate GSP through a diverse set of experiments, covering the dynamics of deformable bodies, rigid bodies, and fluids, as well as two-way coupling between solids and fluids. For additional frames and more detailed results, please refer to the supplementary material. All experiments are also available in the supplementary video. A preliminary test is depicted in Fig.~\ref{fig:chair}, where a soft chair from the NeRF synthetic datasets \cite{Mildenhall20eccv_nerf} falls into a pool, demonstrating the two-way coupling between deformable solids and fluids. The chair deforms, floats due to buoyancy, and generates fluid ripples. Fig.~\ref{fig:wave} illustrates another dynamic fluid scene featuring a coastal cliff and waves. The waves continuously push towards the shore from a distance, and upon colliding with the cliff, they splash out foam and spray. This accurately models the complex interactions between the fluid and the solid cliff face. Fig.~\ref{fig:garden} showcases another fluid-solid interaction test. The garden scene is sourced from the Mip-NeRF 360~\cite{barron2022mipnerf360} dataset. The foreground objects consist of a fixed table and a potted plant. We pour water into the garden slowly. The water rises up and eventually sinks the table, and sweeps the plant. In Fig.~\ref{fig:splashing_brief} and Fig.~\ref{fig:splashing}, a lego bulldozer is surfing on the splashing waves. Through the two-way coupling, the baseplate and the bucket deform and vibrate under the impact of the waves.

GSP has a semantic segmentation module. Therefore, the user is able to freely manipulate the models in the scene. Furthermore, since everything is represented as Gaussian particles, GSP allows the user to transform the state of the model. An example is shown in Fig.~\ref{fig:cup_and_dog}, 
the scene includes a round white table, on which a paper cup and a stuffed toy dog are placed. Water is poured into the cup. The state of the toy dog and the cup are changed to water, and splashes on the desk. As mentioned, we use LaMA to inpaint the table texture so that the user does not observe rendering artifacts when the liquefied cup and toy dog splash away.

\begin{table}
\caption{\textbf{Time performance.}~~We present detailed time statistics for the experiments reported in the paper. All time-related evaluations are expressed in seconds. From left to right, (1) \textbf{\# Kernels} indicates the number of Gaussian kernels, while \textbf{\# Solids}, \textbf{\# Fluids}, and \textbf{\# BG} denote the foreground solid, the fluid, and the background, respectively. In some experiments (e.g., \textbf{Cup \& Dog} and \textbf{Headset}), the number of fluid particles varies over time. We report the maximum number of fluid particles. (2) \textbf{Sim. Time} provides time statistics for simulations, with \textbf{Overall} representing the total simulation time per time step and \textbf{Tension} indicating the time taken for each surface tension constraint projection. (3) \textbf{Render Time} details the time statistics for rendering, with \textbf{Solids}, \textbf{Fluids}, and \textbf{BG} denoting the time spent on rendering the solids, fluids, and background, respectively.}\label{tab:time}
{\huge
\begin{center}
\vspace{-10pt}
\resizebox{\linewidth}{!}{
\begin{tabular}{c|ccc|cc|ccc}
\multirow{2}*{\textbf{Scene (Fig.)}} &
\multicolumn{3}{c|}{\textbf{\# Kernels}} &
\multicolumn{2}{c|}{\textbf{Sim. Time (s)}} &
\multicolumn{3}{c}{\textbf{Render Time ($\times10^{-2}$ s)}} \\

 &
\textbf{\# Solids} &
\textbf{\# Fluids} &
\textbf{\# BG} &
\textbf{Overall} &
\textbf{Tension} &
\textbf{Solids} &
\textbf{Fluids} &
\textbf{BG}
\\

\hline

\textbf{\makecell{Chair (Fig.~\ref{fig:chair})}} & 315K & 300K & 0 & 5.4 & 1.1 & 3.9 & 5.9  & 0 \\
\textbf{\makecell{Waves (Fig.~\ref{fig:wave})}} & 420K & 817K & 0 & 8.1 & 3.1  & 8.2 & 11.6 & 0 \\
\textbf{\makecell{Garden (Fig.~\ref{fig:garden})}} & 450K & 614K & 2.27M & 7.3 & 1.6  & 6.9 & 10.3 & 2.3 \\
\textbf{\makecell{Lego (Fig.~\ref{fig:splashing_brief}\&\ref{fig:splashing})}} & 330K & 280K & 290K & 3.8 & 1.0 & 2.9 & 4.6 & 1.9 \\
\textbf{\makecell{Cup \& dog (Fig.~\ref{fig:cup_and_dog})}} & 156K & 160K & 310K & 2.1 & 0.6 & 2.4 & 4.1 & 1.9 \\
\textbf{\makecell{Headset (Fig.~\ref{fig:headset_brief}\&\ref{fig:headset}})} & 357K & 64K & 2.22M & 1.9 & 1.8 & 1.5 & 3.8 & 2.7 \\
\textbf{\makecell{Can (Fig.~\ref{fig:can})}} & 390K & 254K & 1.19M & 1.6 & 0.8 & 2.5 & 4.7 & 2.0 \\
\textbf{\makecell{Astronaut (Fig.~\ref{fig:astronaut})}} & 0 & 145K & 0 & 0.8 & 0.6 & 0 & 4.0 & 0 \\
\textbf{\makecell{Ficus (Fig.~\ref{fig:ficus})}} & 204K & 0 & 0 & 2.3 & 0 & 2.6 & 0 & 0 \\

\textbf{\makecell{Bulldozers (Fig.~\ref{fig:piled_bulldozers})}} & 6.67M & 0 & 350K & 4.5 & 0 & 24.6 & 0 & 2.1

\end{tabular}

}
\end{center}

}
\end{table}

\begin{figure}
    \centering
   \begin{minipage}{\linewidth}
        \centering
        \begin{minipage}{\linewidth}
            \centering
            \begin{minipage}{0.495\linewidth} 
                \centering
                \includegraphics[width=\linewidth]{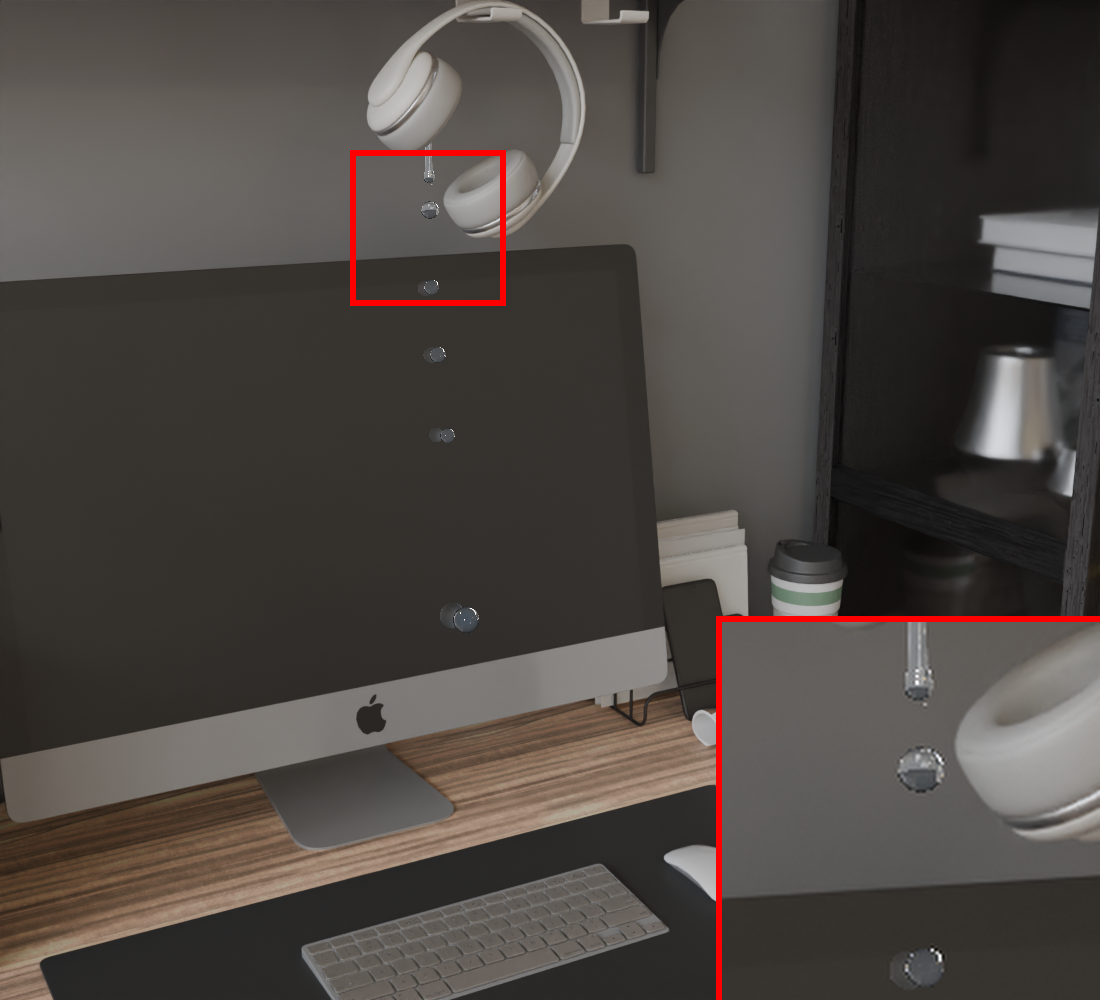}
            \end{minipage}
            \begin{minipage}{0.495\linewidth} 
                \centering
                \includegraphics[width=\linewidth]{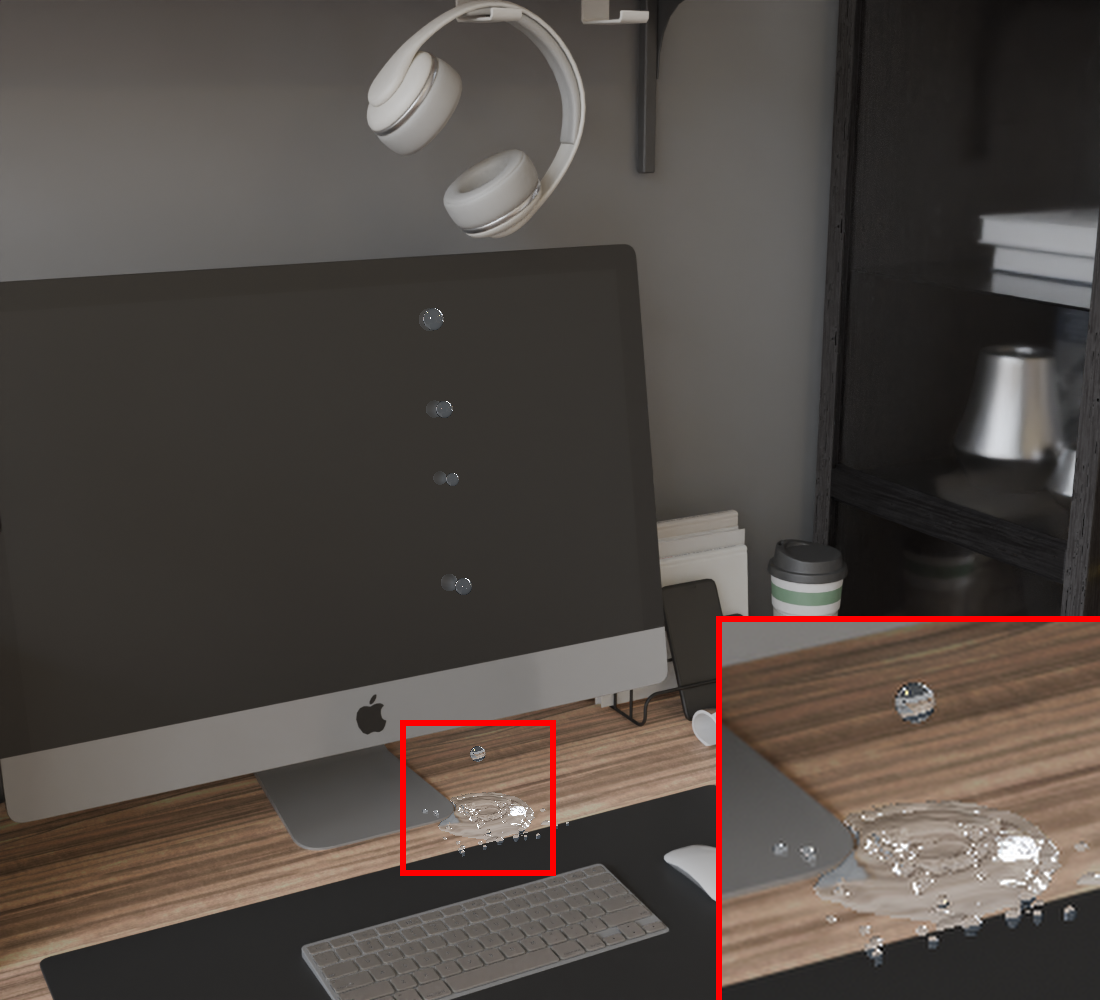}
            \end{minipage}
        \end{minipage}
    \end{minipage}

    \caption{\textbf{Headset droplets}. Water flows from a headset hanging above an office desk, resembling a faucet. Due to surface tension, the water forms droplets, sliding down the computer screen and splashing onto the desk, creating a puddle.}
    \label{fig:headset_brief}
\end{figure}

Our fluid simulator can also capture the surface tension within the PBD framework. This enables realistic low-volume fluid-solid interaction. As shown in Fig.~\ref{fig:headset_brief} and Fig.~\ref{fig:headset}, water flows from a headset hanging above an office desk, resembling a faucet. Due to surface tension, the water forms droplets as it falls, sliding down the computer screen and splashing onto the desk, creating a puddle. In Fig.~\ref{fig:can}, droplets of water continuously drip onto the top of the can until they reaches the capacity and overflow. The surface tension of the liquid causes the droplets to gradually aggregate at the surface. As more droplets fall onto the liquid surface, they rise above the edge of the can. Eventually, the accumulated water exceeds the limit of surface tension and spills over. Fig.~\ref{fig:astronaut} showcases another interesting use of GSP, where the user applies Trisolarans' black magic on an astronaut. The astronaut is strucked by the magic and is transformed to water. It finally collapses into a water ball in a zero-gravity space due to the presence of surface tension. 

GSP is a versatile system and supports the manipulation of both rigid objects and deformable bodies. As shown in Fig.~\ref{fig:ficus}, a deformable ficus is waving at you. Due to the external force applied to it, the plant undergoes continuous deformation. In Fig.~\ref{fig:piled_bulldozers}, bulldozers are piled and dropped in a bowl. They collide and contact with each other and eventually scattered within the box.


\section{Conclusion}
GSP is a novel pipeline combining versatile position-based dynamics with 3DGS. The principle design philosophy of Gaussian Splashing is to harness the consistency of volume particle-based discretization to enable integrated processing of various 3D graphics and vision tasks, such as 3D reconstruction, deformable simulation, fluid dynamics, rigid contacts, and rendering from novel viewpoints. While the concept is straightforward, building Gaussian Splashing involves significant research and engineering efforts. The presence of fluid complicates the 3DGS processing due to the specular highlights at the fluid surface. Fluid-solid coupling resorts to accurate surface information; Large deformation on the solid object generates defective rendering; Displaced models also leave empty regions that the input images fail to capture. We overcome those difficulties by systematically integrating and adapting a collection of state-of-the-art technologies into the framework. As a result, Gaussian Splashing enables realistic view synthesis not only under novel camera poses but also with novel physically-based dynamics for various deformable, rigid, and fluid objects, or even novel object state transform. It should be noted that incorporating physically-based fluid dynamics in NeRF/3DGS has not been explored previously. The primary contribution of this work is to showcase the feasibility of building a unified framework for integrated physics and learning-based 3D reconstruction. Gaussian Splashing still has many limitations. For instance, PBD is known to be less physically accurate. It may be worth generalizing PBD with other meshless simulation methods. The fluid rendering in Gaussian Splashing in its current form is far from perfect -- ellipsoid splatting is an ideal candidate for position-based fluid but does not physically handle real world light transport, e.g. refraction.

\clearpage

{
    \small
    \bibliographystyle{ieeenat_fullname}
    \bibliography{main}
}

\clearpage
\setcounter{page}{1}
\maketitlesupplementary

\section{Simulation Details}

\subsection{Position-Based Dynamics}

PBD/XPBD treats a dynamic system as a set of $N$ vertices, i.e., $\bm{x} = [\bm{x_0}, \bm{x_1},...,\bm{x_N}]^\top$  and $M$ constraints, i.e., $\bm{C}(\bm{x}) = [C_1(\bm{x}), C_2(\bm{x}),...,C_M(\bm{x})]^\top$. Here, $\bm{x}$ represents the position of vertices and $\bm{C}(\bm{x})$ represents the set of constraints.  Specifically, the total system potential $U$ is defined as a quadratic form of all the constraints such that $U = \frac{1}{2}\bm{C}^\top(\bm{x}) \bm{\alpha}^{-1} \bm{C}(\bm{x})$. Here, $\bm{\alpha}$ is the compliance matrix, i.e., the inverse of the constraint stiffness. For example, if there are only two vertices and they form a mass-spring system, constraint and compliance matrix could be written as $\bm{C}(\bm{x})=[\|\bm{x_0}-\bm{x_1}\|-d_0]^\top$ and $\bm{\alpha} = [k]$, where $d_0$ is the rest length of the spring and $k$ is the stiffness of the spring.

Motion at each time step can be solved by minimizing the system energy. However, PBD/XPBD offers an easy and efficient simulation modality, converting the variational optimization to the so-called constraint projections.

XBPD estimates an update of constraint force (i.e., the multiplier) $\Delta \bm{\lambda}$ by solving:
\begin{equation}
    \left[\Delta t^2 \nabla \bm{C}(\bm{x})\bm{M}^{-1}\nabla\bm{C}^\top(\bm{x}) + \bm{\alpha} \right]\Delta \bm{\lambda}= -\Delta t^2 \bm{C}(\bm{x}) - \bm{\alpha} \bm{\lambda},
\end{equation}
where $\Delta t$ is the time step size, and $\bm{M}$ is the lumped mass matrix. The update of the primal variable $\Delta \bm{x}$ can then be computed as:
\begin{equation}
    \Delta \bm{x} = \bm{M}^{-1}\nabla\bm{C}^\top(\bm{x})\Delta\bm{\lambda}.
    \label{eq:compute_delta_x}
\end{equation}
The parallelization of XPBD is enabled with a Gauss-Seidel-like scheme, which computes $\Delta \bm{\lambda}_j$ at each constraint $C_j$ independently:
\begin{equation}
    \label{eq:compute_delta_lambda}
    \Delta \bm{\lambda}_j \leftarrow \frac{-\Delta t^2 C_j(\bm{x}) - \bm{\alpha}_j}{\Delta t^2\nabla C_j \bm{M}^{-1} \nabla C_j^\top + \bm{\alpha}_j}.
\end{equation}
A typical XPBD simulation loop is shown in Algorithm~\ref{algo:xpbd}.
\begin{algorithm}
\caption{XPBD simulation loop for time step $n+1$}
\label{algo:xpbd}
\begin{algorithmic}[1]
    \State {predict position $\tilde{\bm{x}}\Leftarrow \bm{x}^n + \Delta t\bm{v}^n + \Delta t^2\bm{M}^{-1}\bm{f}_{ext}(\bm{x})^n$}
    \State
    \State initialize solve $\bm{x}_0\Leftarrow\title{\bm{x}}$
    \State initialize multipliers $\bm{\lambda}_0\Leftarrow{\bm{0}}$
    \While{$i < {\rm solverIterations}$}
        \ForAll{constraints}
            \State compute $\Delta\lambda$ using Eq.~\ref{eq:compute_delta_lambda}
            \State compute $\Delta\bm{x}$ using Eq.~\ref{eq:compute_delta_x}
            \State update $\lambda_{i+1}\Leftarrow \lambda_i+\Delta\lambda$
            \State update $\bm{x}_{i+1}\Leftarrow\bm{x}_i+\Delta\bm{x}$
        \EndFor
    \EndWhile
    \State update positions $\bm{x}^{n+1}\Leftarrow\bm{x}_i$
    \State update velocities $\bm{v}^{n+1}\Leftarrow\frac{1}{\Delta t}(\bm{x}^{n+1}-\bm{x}^n)$
\end{algorithmic}
\end{algorithm}

\subsection{Position-Based Fluids}
We employ the Position-Based Fluids (PBF)~\cite{macklin2013pbf} as our Lagrangian fluid synthesizer. PBF is based on PBD, which means it also use constraint projections to simulate fluid behaviour. In PBF, fluid is composed of a large amount of particles. To enforce the fluid incompressibility, PBF imposes a density constraint $C^\rho_i$ on each particle, maintaining the integrated density $\rho_i$ computed by the SPH kernel as:

\begin{equation}\label{eq:FluidsConstraint}
C_i^\rho =\frac{\rho_i}{\rho_0} - 1=\sum_j\frac{m_j}{\rho_0}W(\bm{p}_i-\bm{p}_j, r) - 1,
\end{equation}
where $m_j$ is the mass of particle $j$. $\bm{p}_i$ is the position of particle $i$, $W$ is the SPH kernel function and $r$ is the kernel radius. Intuitively, projecting this constraint to $0$ ensures that the density at the current time remains consistent with the initial state. We use the following cubic SPH kernel:
\begin{equation}\label{eq:cubic_sph_kernel}
    W(\bm{p}, r)=
    \begin{cases}
        \frac{8}{\pi r^3}(6q^2(q-1)+1), & 0\leq q\leq0.5\\
        \frac{16}{\pi r^3}(1-q)^3, & 0.5<q\leq1\\
        0, &{\rm otherwise}
    \end{cases}
\end{equation}
where $q=\frac{\|\bm{p}\|}{r}$. The Jacobian of constraint is computed as:
\begin{equation}\label{eq:JacobianFluidsConstraint}
\nabla_{\bm{p}_k} C_i^\rho = 
\begin{cases}
    \displaystyle \sum_j\frac{m_j}{\rho_0}\nabla_{\bm{p}_i}W(\bm{p}_i-\bm{p}_j, r), & k=i\\
    \displaystyle \frac{m_j}{\rho_0}\nabla_{\bm{p}_j}W(\bm{p}_i-\bm{p}_j, r), & k=j.
\end{cases}
\end{equation}
The gradient of the kernel function is:
\begin{equation}\label{eq:kernel_gradient}
\nabla_{\bm{p}} W(\bm{p}, r) = 
\begin{cases}
    \frac{48}{\pi r^5}(3q-2)\bm{p}, & 0\leq\frac{\|\bm{p}\|}{r}\leq0.5\\
    -\frac{48}{\pi r^5}\frac{(1-q)^2}{q}\bm{p}, & 0.5<\frac{\|\bm{p}\|}{r}\leq1\\
    0, &{\rm otherwise}
\end{cases}
\end{equation}

GSP also includes a position-based surface tension model~\cite{xing2022pbtf} to better capture the dynamics of the fluid surface. We first detect whether a particle (i.e., a Gaussian kernel) is on the fluid surface based on occlusion estimation. Specifically, we encapsulate a particle with a spherical cover or screen. Each of its neighboring particles generates a projection on the screen (because a particle has a finite volume). The particle is considered on the fluid surface if the total projection area from its neighbors is below a given threshold.

In the original paper~\cite{xing2022pbtf}, surface detection is implemented on the CPU. It is noteworthy that surface detection can be parallelized on the GPU to expedite the simulation, as the calculation of each particle's occluding ratio on the screen is independent of the others.

\begin{wrapfigure}[11]{r}{0.4\linewidth}
    \vspace{-15pt}
    \includegraphics[width=\linewidth]{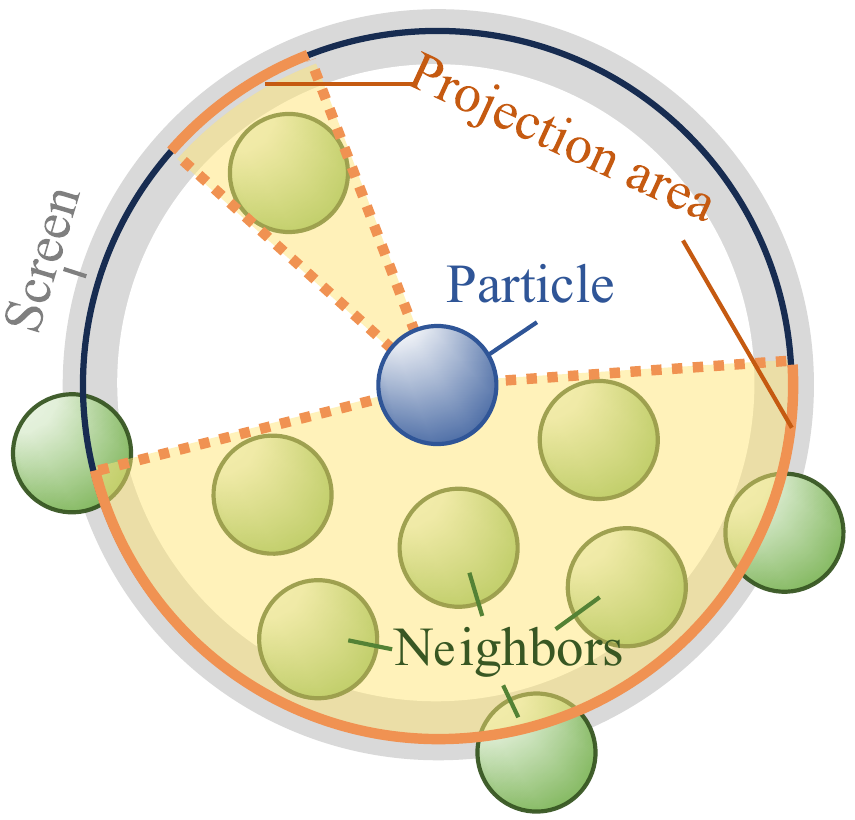}
    \caption{\textbf{Detection of surface particles.}~~An interior particle is detected if its screen is widely shadowed by its neighbors. A boundary particle is detected if at least one part of the particle's screen is not shadowed.}\label{fig:surfacetension}
\end{wrapfigure}

For each neighboring particle, its occluding area on the spherical screen is calculated as follows:
\begin{align}
    \theta &= {\rm tan^{-1}}(\frac{\Delta\bm{p}_y}{\Delta\bm{p}_x+\Delta\bm{p}_z^2})\notag\\
    \phi &= {\rm tan^{-1}}(\frac{\Delta\bm{p}_x}{\Delta\bm{p}_z})\\
    \Delta\theta &= {\rm tan^{-1}}\frac{R}{\|\Delta p\|^2-R^2}\notag\\
    \Delta\phi &= \Delta\theta\notag
\end{align}
where $\Delta\bm{p}$ is the vector from the detection particle to the neighboring particle and $R$ is the particle radius. The shadowed area on the spherical screen is then $[\theta-\Delta\theta,\theta+\Delta\theta]\times[\phi-\Delta\phi,\phi+\Delta\phi]$. We parameterize the screen as an $18\times 36$ environment map, with each column of the environment map corresponding to 18 bits of an integer. We then mask 36 integers and count the mask ratio.

After detecting surface particles on the fluid, we apply tension on the surface. Tensions tends to minimize surface area. Therefore, PBF applies an area constraint to each surface particle to minimize the local surface area nearby. We start by calculating the normal $\bm{n}_i$ of surface particles $i$ as:
\begin{equation}\label{eq:surfacenormal}
    \bm{n}_i = \text{normalize}(-\nabla_{\bm{p}_i}C_{i}^\rho),
\end{equation}
where $C_i^\rho = 0$ indicates the particle is inside the fluid, and $C_i^\rho = -1$ indicates it is outside. After that, we project the neighboring surface particles onto a plane perpendicular to $\bm{n}_i$ and triangularize the plane. The area constraint can then be built as:
\begin{equation}\label{eq:AreaConstraint}
C_i^A=\sum_{t\in T(i)}\frac{1}{2}\left\|(\bm{p}_{t^2}-\bm{p}_{t^1})\times (\bm{p}_{t^3}-\bm{p}_{t^1})\right\|
\end{equation}
where $T(i)$ is the set of neighboring triangles for particle $i$. We use the 2D Delaunay triangulation 
to construct the triangles on the local surface. This process is sequential and cannot be parallelized on the GPU. However, it is sufficiently fast and we translate it from CPU to GPU.

\begin{figure}[H]
    \includegraphics[width=\linewidth]{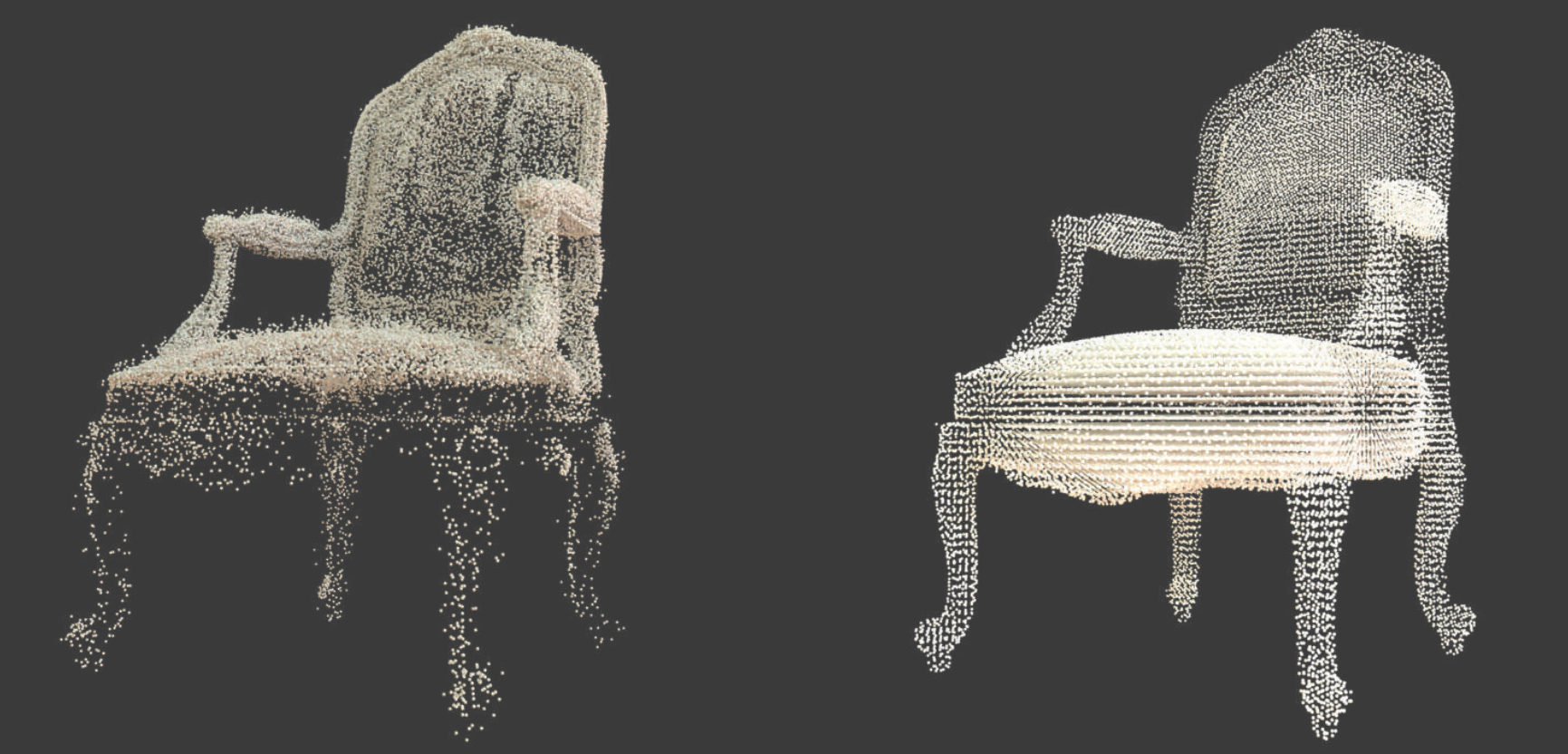}
    \caption{\textbf{Different sampling strategies.}~~We compare the results of different sampling strategies: (left) fill the particle based on the density grid calculated using Gaussian kernels~\cite{xie2023physgaussian}, and (right) uniformly sample within NeuS reconstruction. The point distribution generated by vanilla 3DGS is uneven, which hardly samples the legs or seat of the chair.}\label{fig:boundary_comparison}
\end{figure}

To promote a more uniform particle distribution, additional distance constraints are introduced to push apart particles that are too close to each other:
\begin{equation}\label{eq:DistanceConstraint}
C_{ij}^D={\min}\left\{ 0, \left\|\bm{p}_i-\bm{p}_j\right\|-d_0 \right\},
\end{equation}
where $d_0$ is the distance threshold. The Jacobian of aboved constraints are:
\begin{equation}\label{eq:JacobianTension}
\begin{aligned}
    \nabla_{t^1}C_t^A(\bm{p})&=\frac{(\bm{p}_{t^2}-\bm{p}_{t^1})\times (\bm{p}_{t^3}-\bm{p}_{t^1})\times(\bm{p}_{t^3}-\bm{p}_{t^2})}{2\left\|(\bm{p}_{t^2}-\bm{p}_{t^1})\times (\bm{p}_{t^3}-\bm{p}_{t^1})\right\|},\\
    \nabla_{t^2}C_t^A(\bm{p})&=\frac{(\bm{p}_{t^3}-\bm{p}_{t^2})\times (\bm{p}_{t^1}-\bm{p}_{t^2})\times(\bm{p}_{t^1}-\bm{p}_{t^3})}{2\left\|(\bm{p}_{t^3}-\bm{p}_{t^2})\times (\bm{p}_{t^1}-\bm{p}_{t^2})\right\|},\\
    \nabla_{t^3}C_t^A(\bm{p})&=\frac{(\bm{p}_{t^1}-\bm{p}_{t^3})\times (\bm{p}_{t^2}-\bm{p}_{t^3})\times(\bm{p}_{t^2}-\bm{p}_{t^1})}{2\left\|(\bm{p}_{t^1}-\bm{p}_{t^3})\times (\bm{p}_{t^2}-\bm{p}_{t^3})\right\|},\\
    \nabla_i C_{i j}^D(\mathbf{p})&= 
    \begin{cases}
    \displaystyle \bm{0}, & \left\|\bm{p}_i-\bm{p}_j\right\|>d_0, \\ 
    \displaystyle \frac{\bm{p}_i-\bm{p}_j}{\left\|\bm{p}_i-\bm{p}_j\right\|}, & \text {Others.}\\
    \end{cases}\\
    \nabla_j C_{i j}^D(\mathbf{p})&= 
    \begin{cases}
    \displaystyle \bm{0}, & \left\|\bm{p}_i-\bm{p}_j\right\|>d_0, \\ 
    \displaystyle \frac{\bm{p}_j-\bm{p}_i}{\left\|\bm{p}_i-\bm{p}_j\right\|}, & \text { Others. }\\
    \end{cases}
\end{aligned}
\end{equation}

\section{Sampling and Interpolation Details}

In practice, we found that an uneven distribution of Gaussian kernels results in unstable and inaccurate motion synthesis, while a distribution that is too uniform can detrimentally affect rendering quality. Gaussian kernels tend to distribute primarily unevenly around the surfaces and edges of objects, leading to inaccurate boundary descriptions, which are crucial for interactions between objects in simulation. Conversely, adaptively distributed anisotropic Gaussian kernels are key to representing the spatially varying texture on the object. To address this issue, we maintain two separate sets of points: one sampled from the NeuS mesh surface for simulation, and the other consisting of trained Gaussian kernels for rendering. We compare the results of directly sampling from trained Gaussian kernels to those of sampling from NeuS in Fig.~\ref{fig:boundary_comparison}. The former method can result in sparsely sampled regions, especially for objects with thin parts, potentially affecting simulation quality.

We then discuss how to animate trained Gaussian kernels for rendering dynamics. Denote the set of trained Gaussian kernels for rendering as $S_r$, and the set of sampled points from NeuS for simulation as $S_s$. At time $0$, we initialize GMLS kernels on $S_s$ and find the $k$ nearest neighbors $\{\bm{p}_{s,j}^0: j\in \mathcal{N}(i)\}$ from $S_s$ for each point $\bm{p_{r,i}}$ in $S_r$. Here, $\mathcal{N}(i)$ denotes the set of $k$ nearest neighboring particles' indices of $\bm{p}_{r,i}$ in $S_s$ at time $0$. As the simulation or motion synthesis proceeds, the position $\bm{p}^n_{s,j}$ evolves with time step $n$. We then update $\bm{p}_{r,j}$ by interpolating from $\{\bm{p}_{s,j}^n: j\in \mathcal{N}(i)\}$ with the pre-built GMLS kernel. The interpolation of deformation is achieved in the same way, replacing the physical quantity position $\bm{p}$ to deformation gradient $F$.

\section{Rendering Details}

\paragraph{Shadow} As shadows are crucial to the visual outcomes in dynamic scenes, we re-engineer nearly-soft shadows~\cite{donnelly2006variance} into our system to enhance realism. Following shadow mapping, we splat all Gaussian kernels to a camera positioned at the point light's location, which we denote as light-view. The point light is aligned with the direction of the significantly bright light in the environment map. The resolution of the light-view image is three times that of the original image resolution to address visual discrepancies caused by under-sampling.

We then reproject the points seen in the camera view to the light-view and compare their depths to the previously splatted light-view depth image. A larger depth indicates that the point is occluded by a nearer point and will therefore cast a shadow. A more robust variance method is discussed in \cite{donnelly2006variance}. Softer shadows can be achieved by blurring and averaging the light-view depth image. We compute the shadowing probability using Chebyshev's inequality~\cite{donnelly2006variance} and store the results in a shadow map. Finally, we composite the rendered image with the shadow map to achieve nearly-soft dynamic shadows.

\paragraph{Spray, foam, and bubble} To enhance the realism of fluids, foam, spray and bubble particles are synthesized with \cite{ihmsen2012unified} as a post-processing step. Fluid-air mixtures are generated at the crest of the wave and in the impact zone of the wave.  Spray, foam, and bubble particles are advected by the fluid and dissipate within their predefined lifetimes.

We splat these particles into a foam intensity image using modified additive splatting. Different types of particles use different kernels during splatting~\cite{akinci2013screen}. We preferred a larger overall intensity for surface foam particles to increase their visibility, while we used a comparatively smaller intensity value for spray particles to make them less prominent. Furthermore, we used hollow circle structures for the bubble particles to make their appearance more convincing underwater. The kernel typically has a radius of 2 pixels. However, in practice, we found that the kernel radius should be scaled based on the particle's depth in the view, as particles near the camera occupy more of the view compared to those farther away. Finally, we apply a curve~\cite{akinci2013screen} on the foam intensity image to scale it into $[0, 1]$ and compose it with rendering.


\section{Implementation Details}

We set the simulation time step as $0.005$ seconds throughout the simulations. In our PBD solver, we used $10$ iterations for fluids and $50$ iterations for solids for our experiments, since mass particles on the solid models are more strongly coupled than the ones in the fluid. During the PBF simulation, the surface particles of fluids are updated every two time steps.


\section{More Results}

\begin{figure*}
    \centering

   \begin{minipage}{\textwidth}
        \centering
        \begin{minipage}{\linewidth}
            \centering
            \begin{minipage}{0.196\linewidth}   
                \centering
                \includegraphics[width=\linewidth]{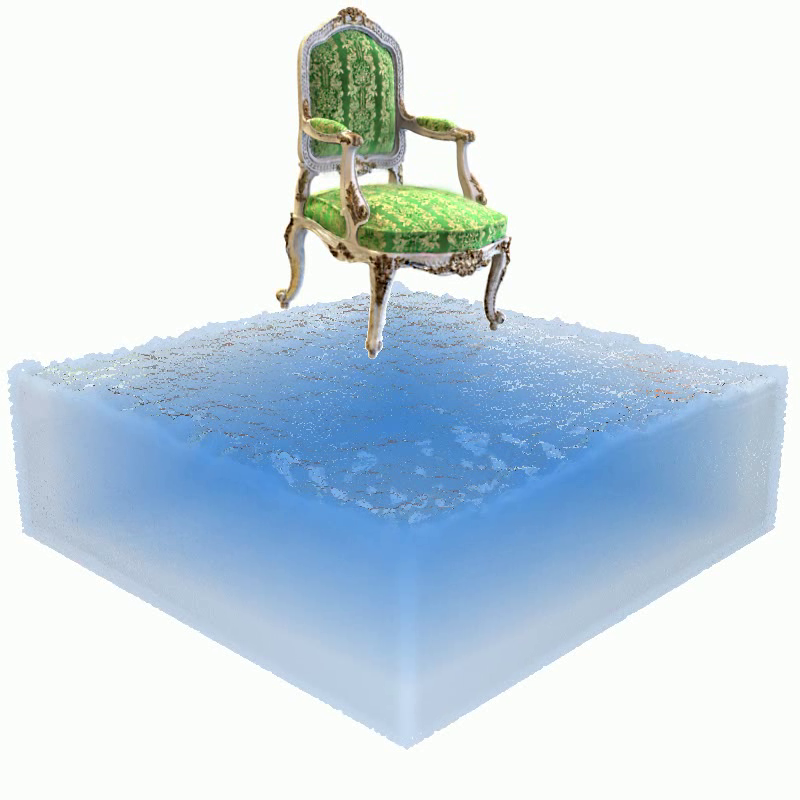}
            \end{minipage}
            \begin{minipage}{0.196\linewidth} 
                \centering
                \includegraphics[width=\linewidth]{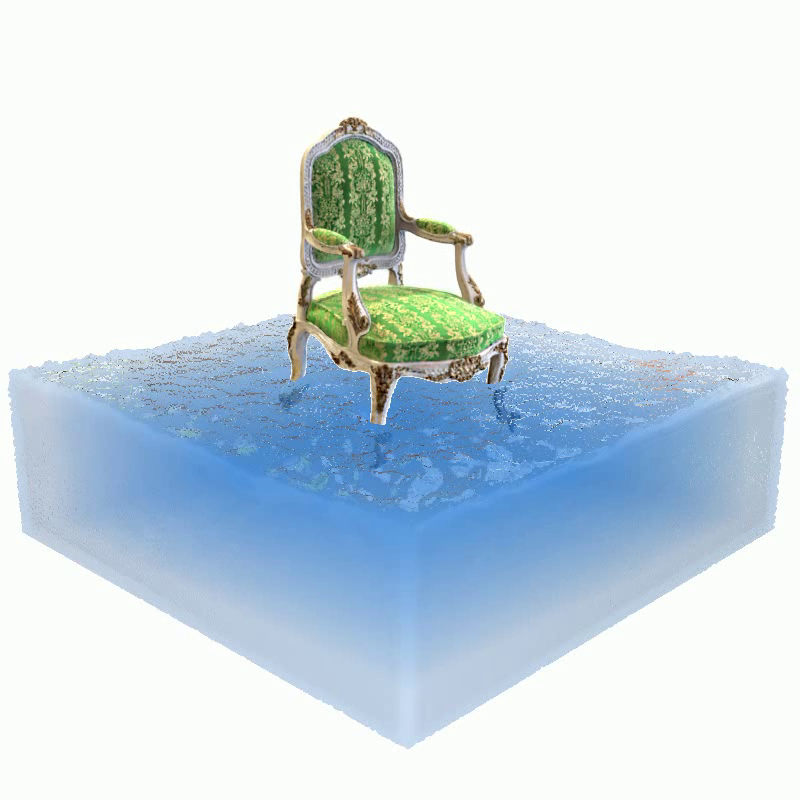}
            \end{minipage}
            \begin{minipage}{0.196\linewidth} 
                \centering
                \includegraphics[width=\linewidth]{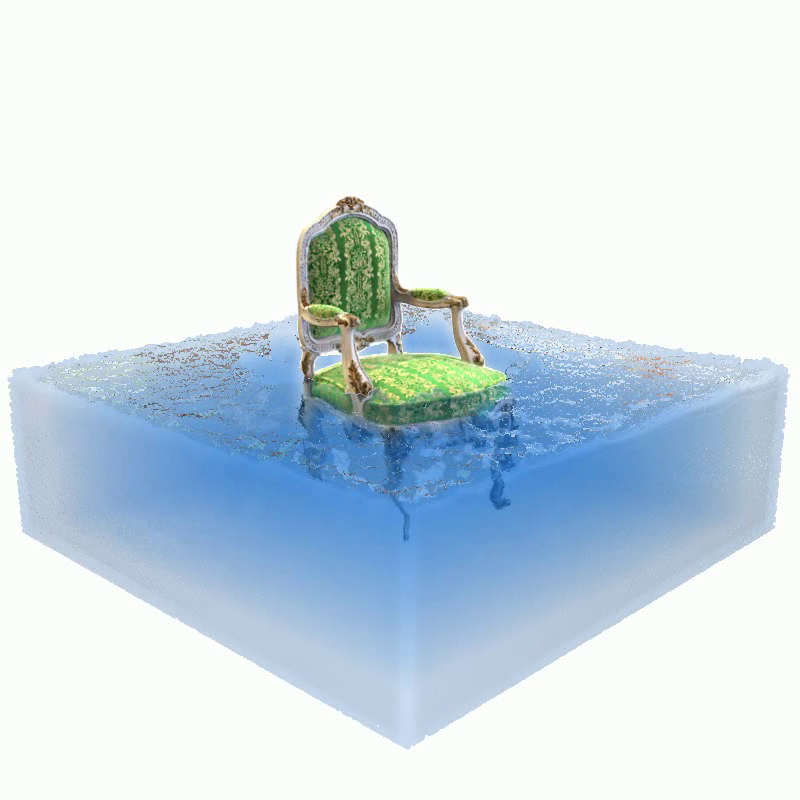}
            \end{minipage}
            \begin{minipage}{0.196\linewidth} 
                \centering
                \includegraphics[width=\linewidth]{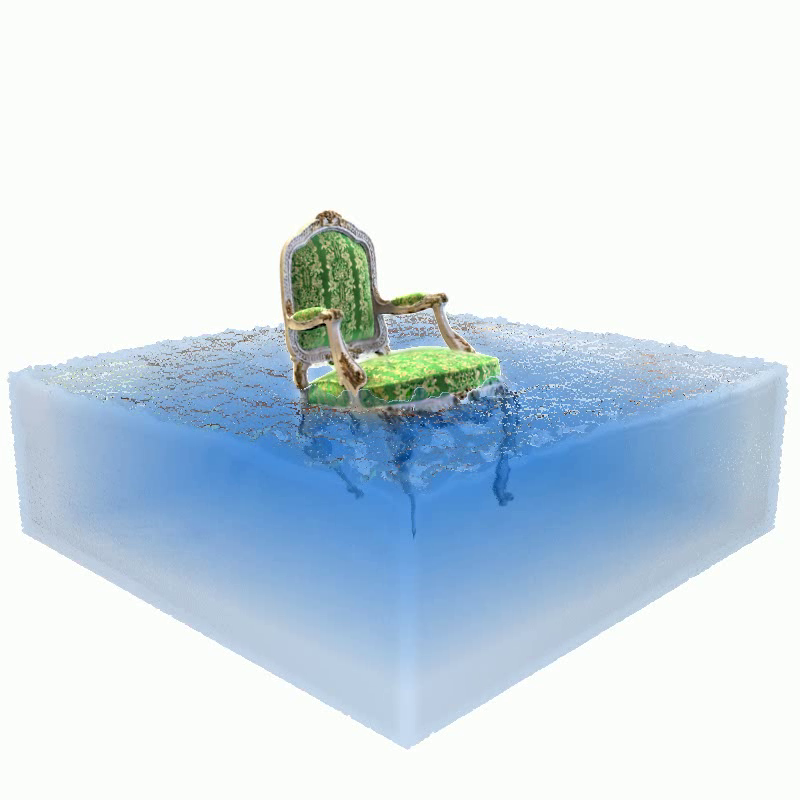}
            \end{minipage}
            \begin{minipage}{0.196\linewidth} 
                \centering
                \includegraphics[width=\linewidth]{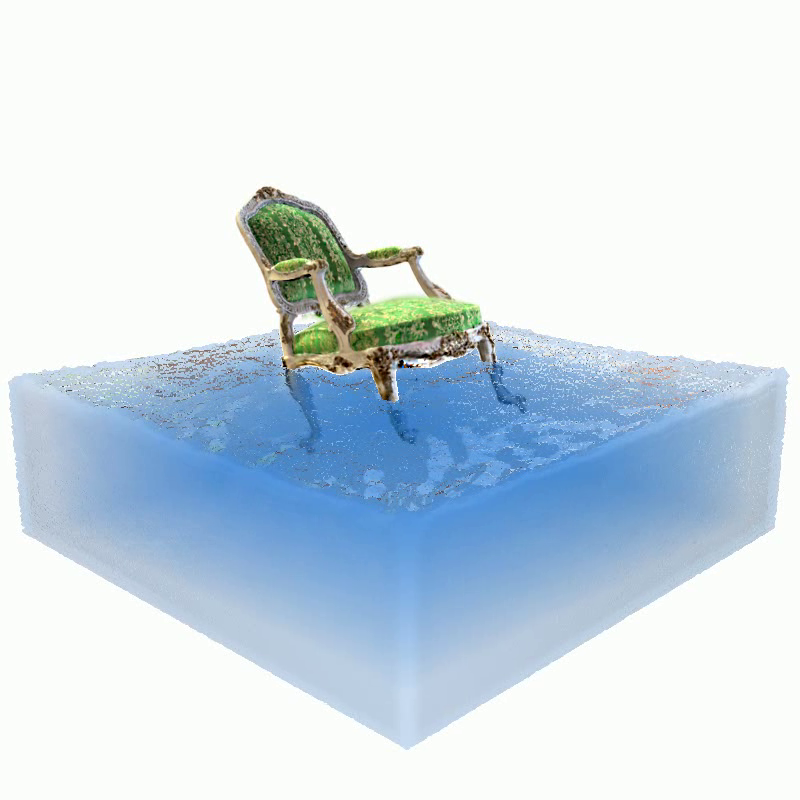}
            \end{minipage}
        \end{minipage}
    \end{minipage}

    \caption{\textbf{Chair.}~~A soft chair falls into the pool, causing deformation and ripples. }
    \label{fig:chair}
\end{figure*}

\begin{figure*}
    \centering

   \begin{minipage}{\textwidth}
        \centering
        \begin{minipage}{\linewidth}
            \centering
            \begin{minipage}{0.196\linewidth}   
                \centering
                \includegraphics[width=\linewidth]{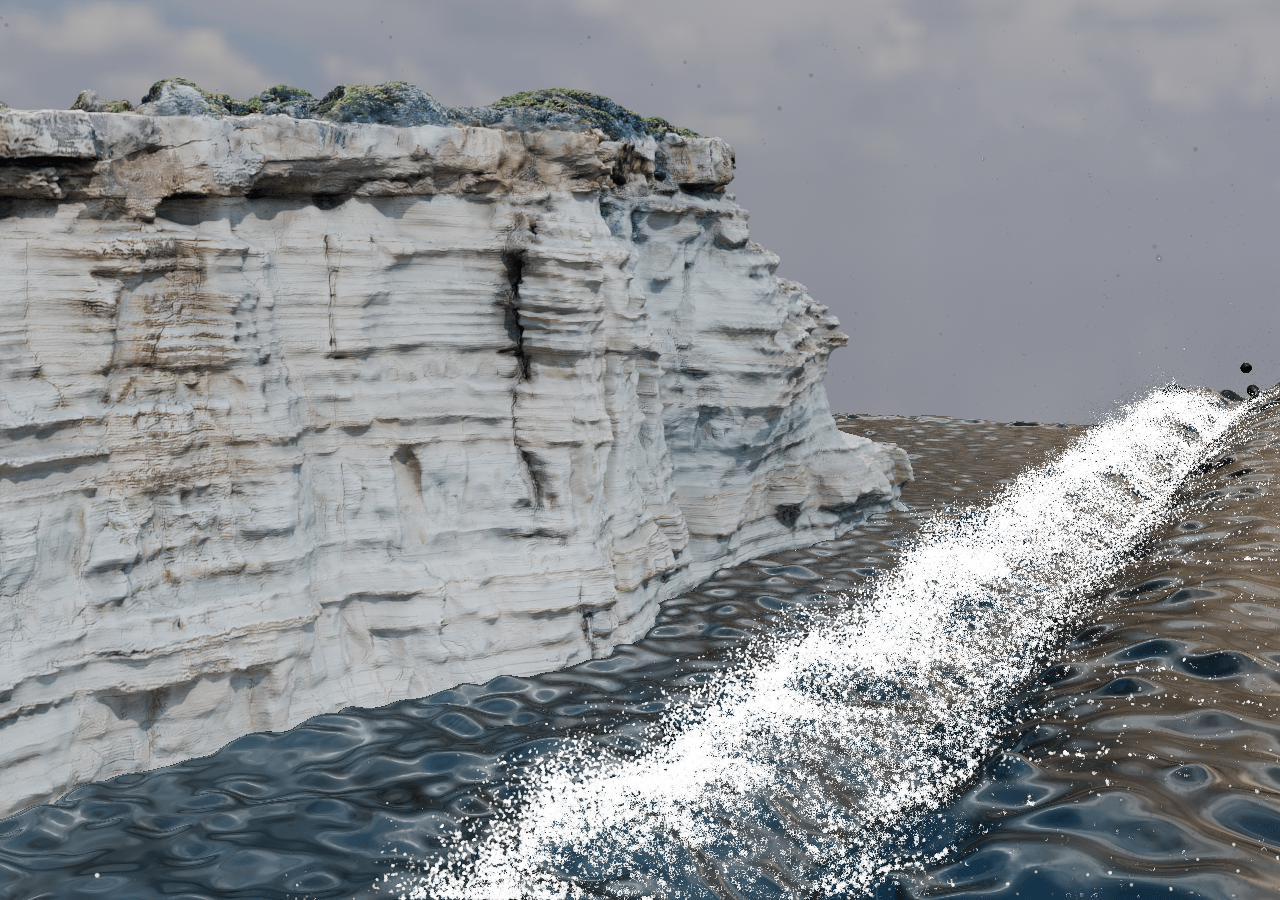}
            \end{minipage}
            \begin{minipage}{0.196\linewidth} 
                \centering
                \includegraphics[width=\linewidth]{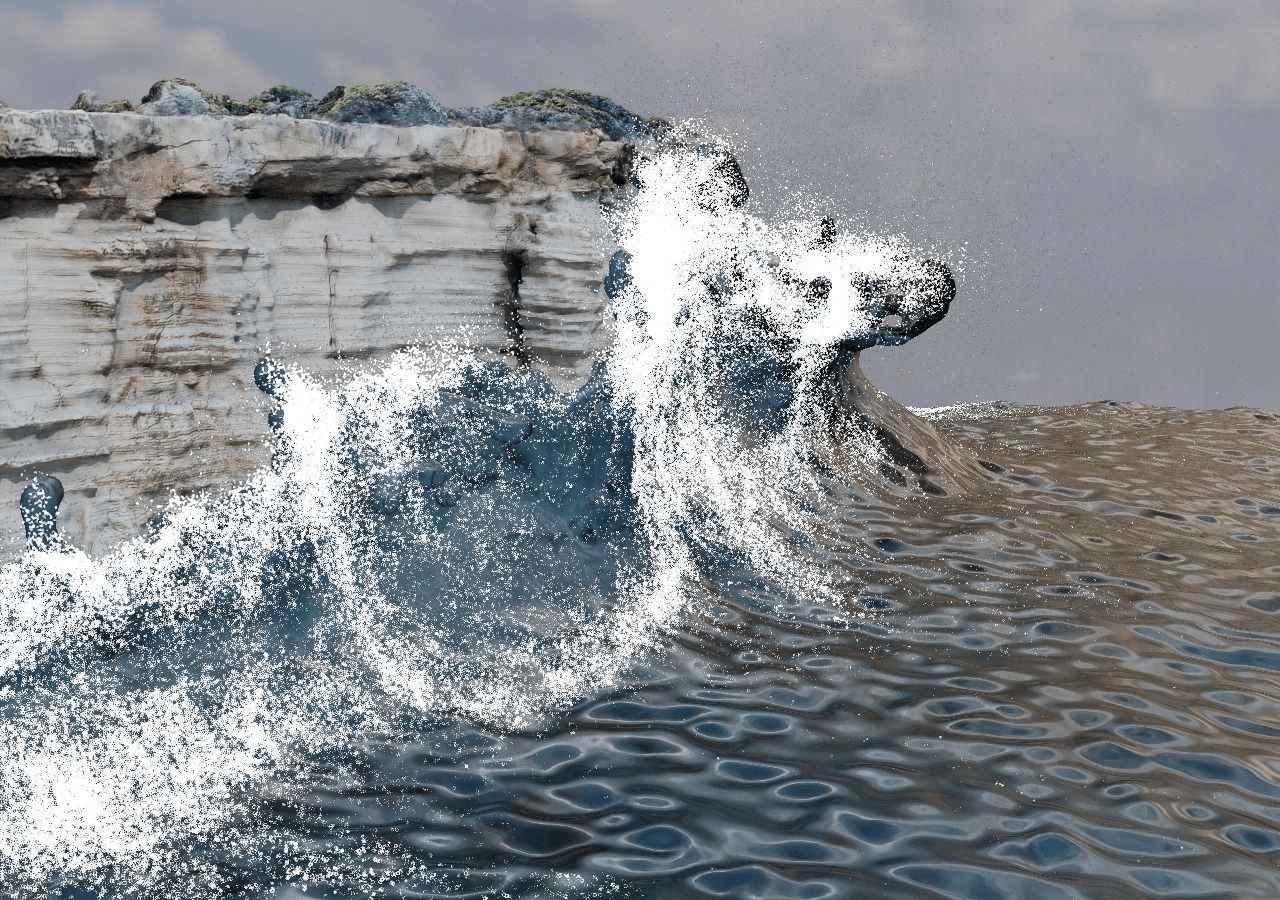}
            \end{minipage}
            \begin{minipage}{0.196\linewidth} 
                \centering
                \includegraphics[width=\linewidth]{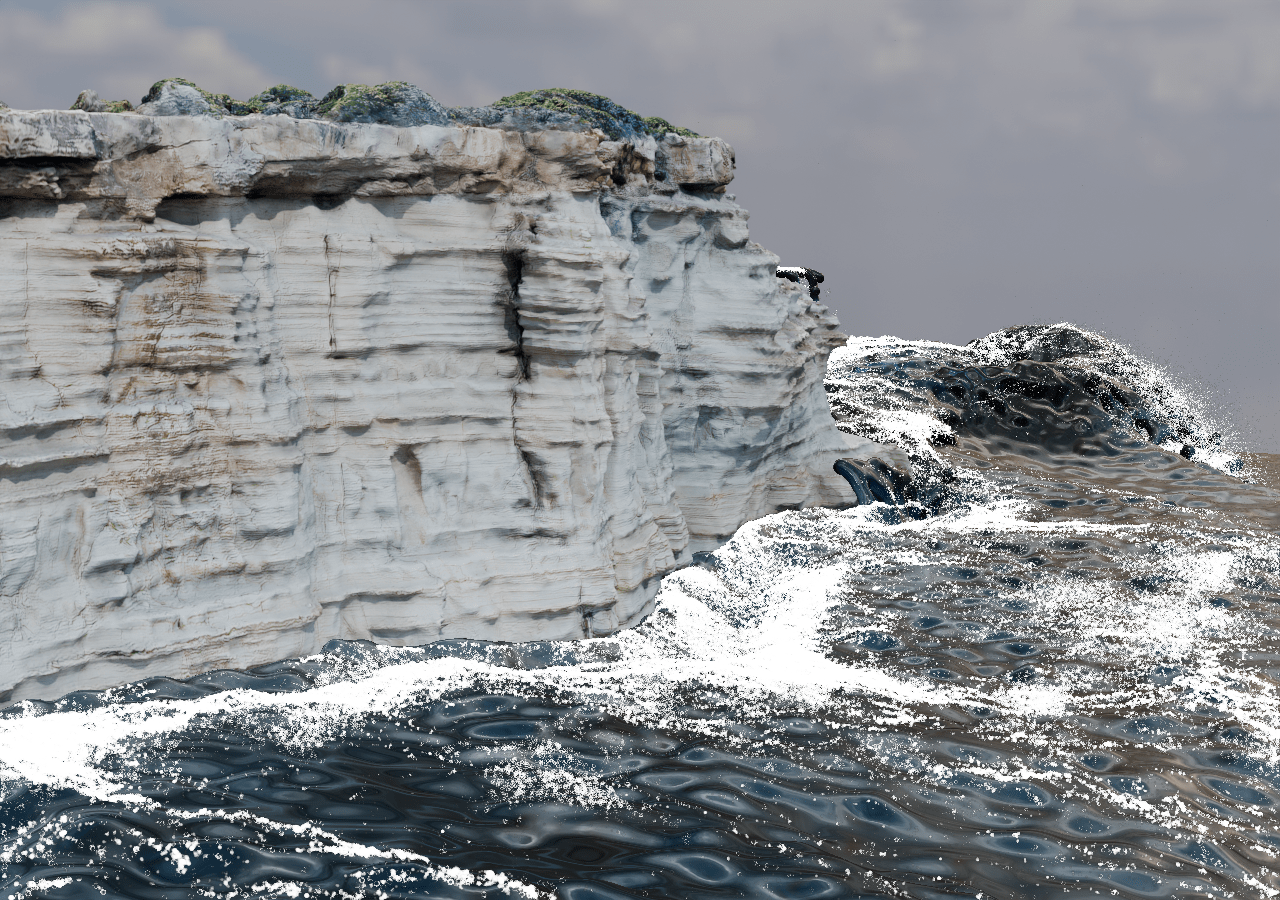}
            \end{minipage}
            \begin{minipage}{0.196\linewidth} 
                \centering
                \includegraphics[width=\linewidth]{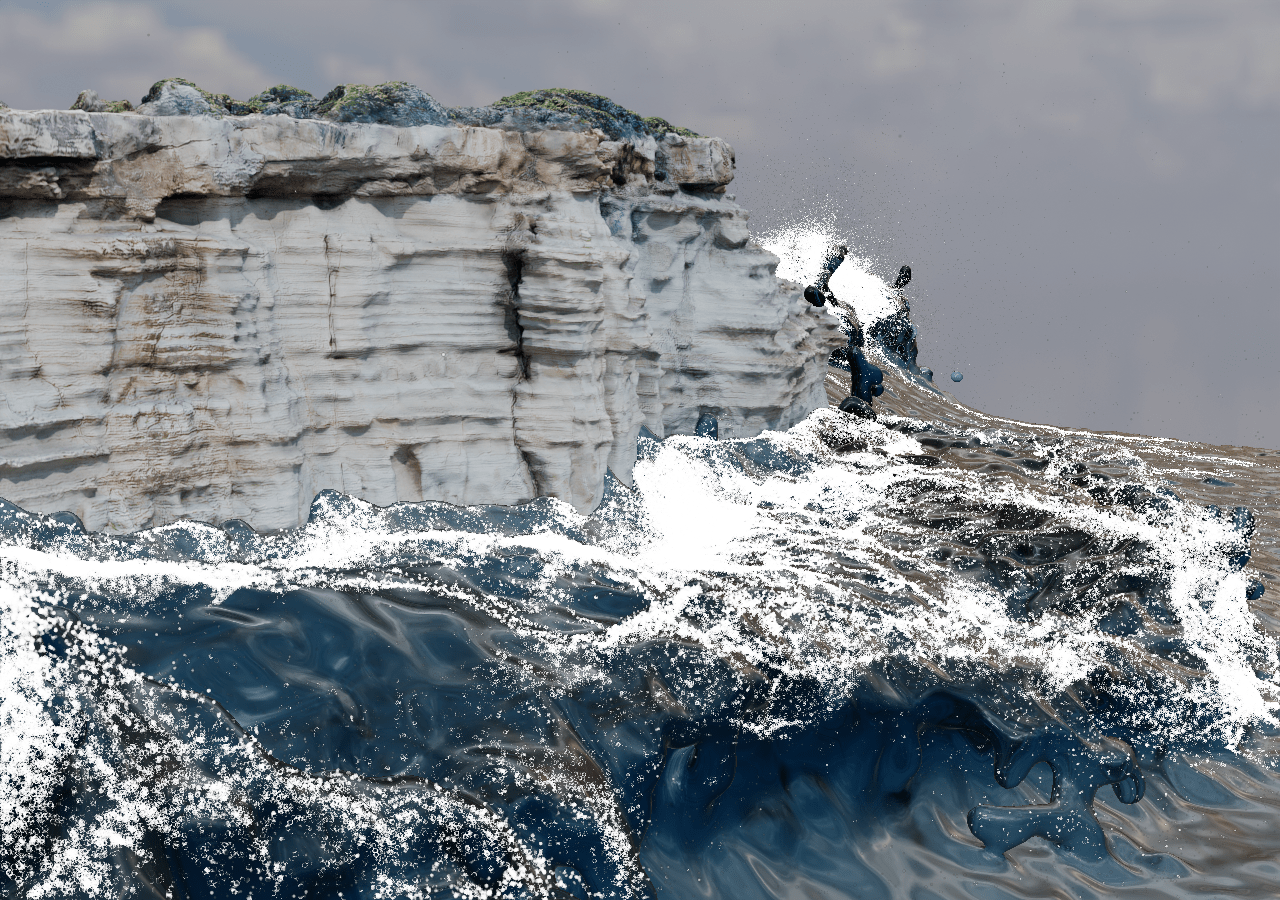}
            \end{minipage}
            \begin{minipage}{0.196\linewidth} 
                \centering
                \includegraphics[width=\linewidth]{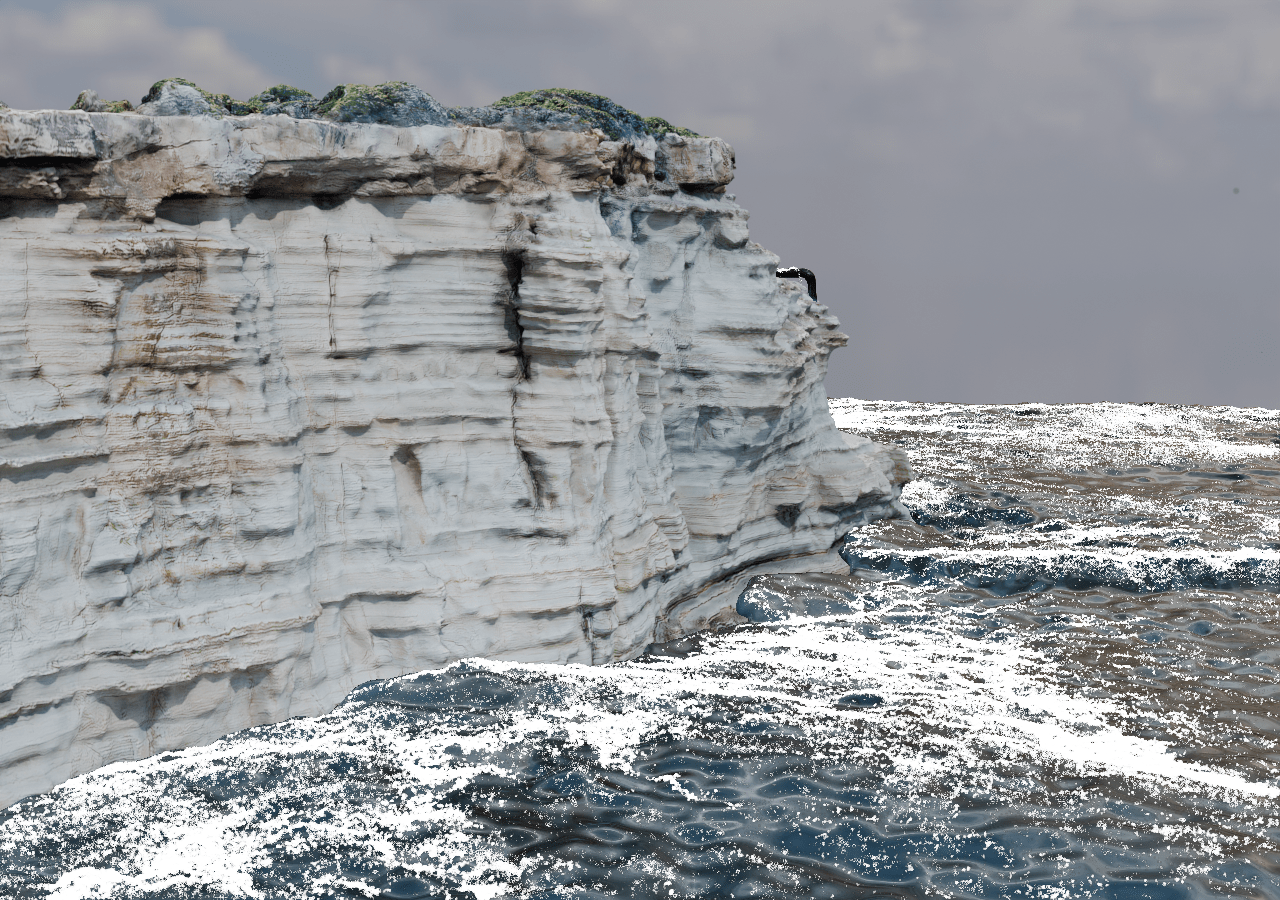}
            \end{minipage}
        \end{minipage}
    \end{minipage}

    \caption{\textbf{Waves crashing on cliff}. A coastal cliff rises from the sandy beach, while the sea waves continuously crash against the rocky surface, generating splashes and foam upon collision.}
    \label{fig:wave}
\end{figure*}

\begin{figure*}
    \centering

   \begin{minipage}{\textwidth}
        \centering
        \begin{minipage}{\linewidth}
            \centering
            \begin{minipage}{0.196\linewidth}   
                \centering
                \includegraphics[width=\linewidth]{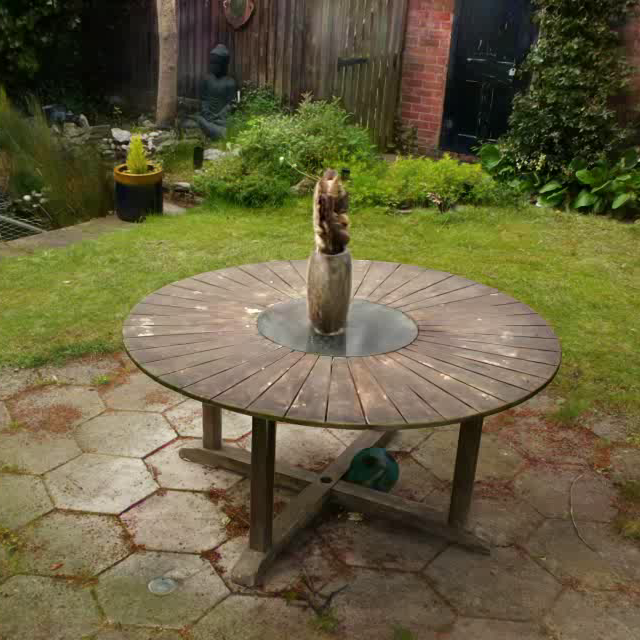}
            \end{minipage}
            \begin{minipage}{0.196\linewidth} 
                \centering
                \includegraphics[width=\linewidth]{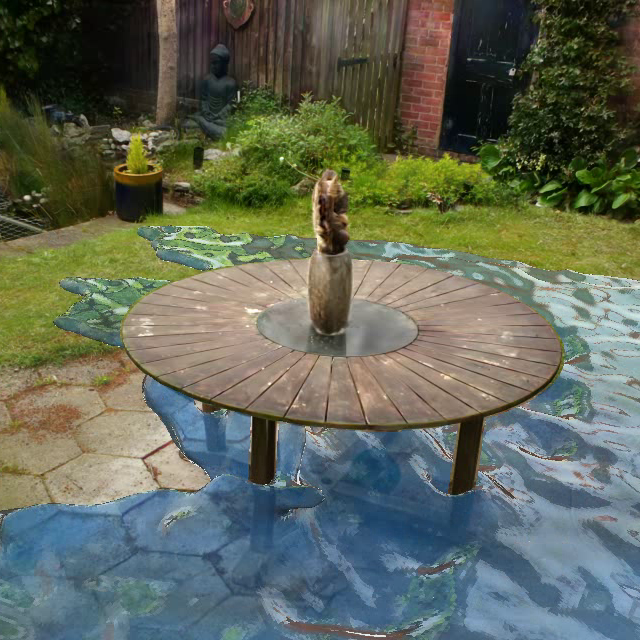}
            \end{minipage}
            \begin{minipage}{0.196\linewidth} 
                \centering
                \includegraphics[width=\linewidth]{samples/figs/main_dynamic/garden/_0003.png}
            \end{minipage}
            \begin{minipage}{0.196\linewidth} 
                \centering
                \includegraphics[width=\linewidth]{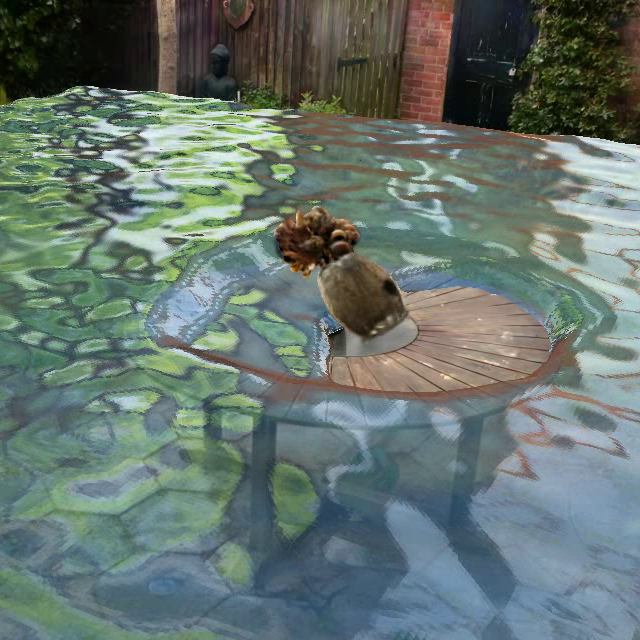}
            \end{minipage}
            \begin{minipage}{0.196\linewidth} 
                \centering
                \includegraphics[width=\linewidth]{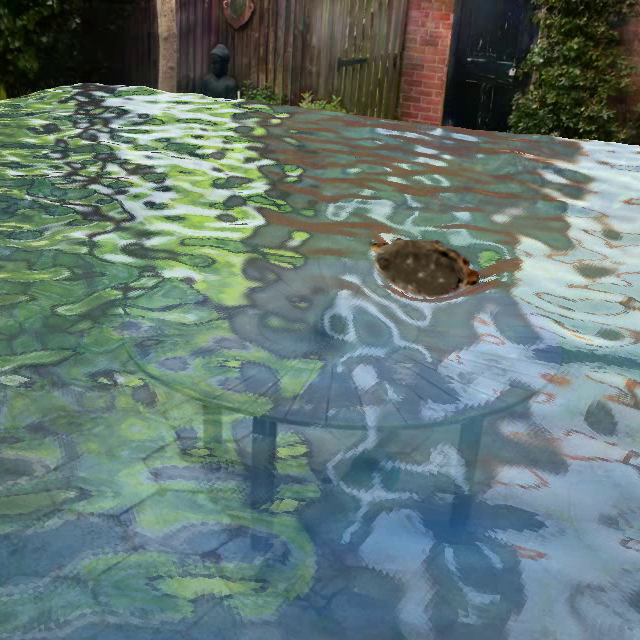}
            \end{minipage}
        \end{minipage}
    \end{minipage}

    \caption{\textbf{Flooding garden.}~~Water leaks into the garden and submerge the table. As the water level goes up, the surface gets more vibrant and washes the potted plant away.}    
    \label{fig:garden}
\end{figure*}

\begin{figure*}
    \centering

   \begin{minipage}{\textwidth}
        \centering
        \begin{minipage}{\linewidth}
            \centering
            \begin{minipage}{0.196\linewidth}   
                \centering
                \includegraphics[width=\linewidth]{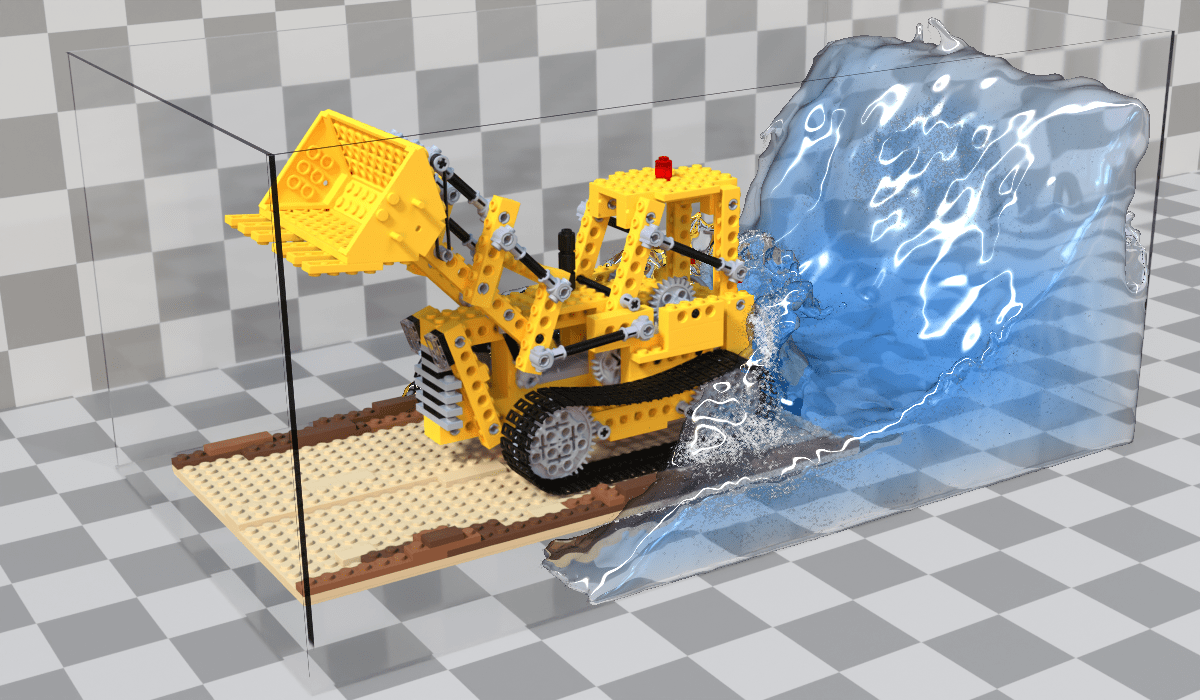}
            \end{minipage}
            \begin{minipage}{0.196\linewidth} 
                \centering
                \includegraphics[width=\linewidth]{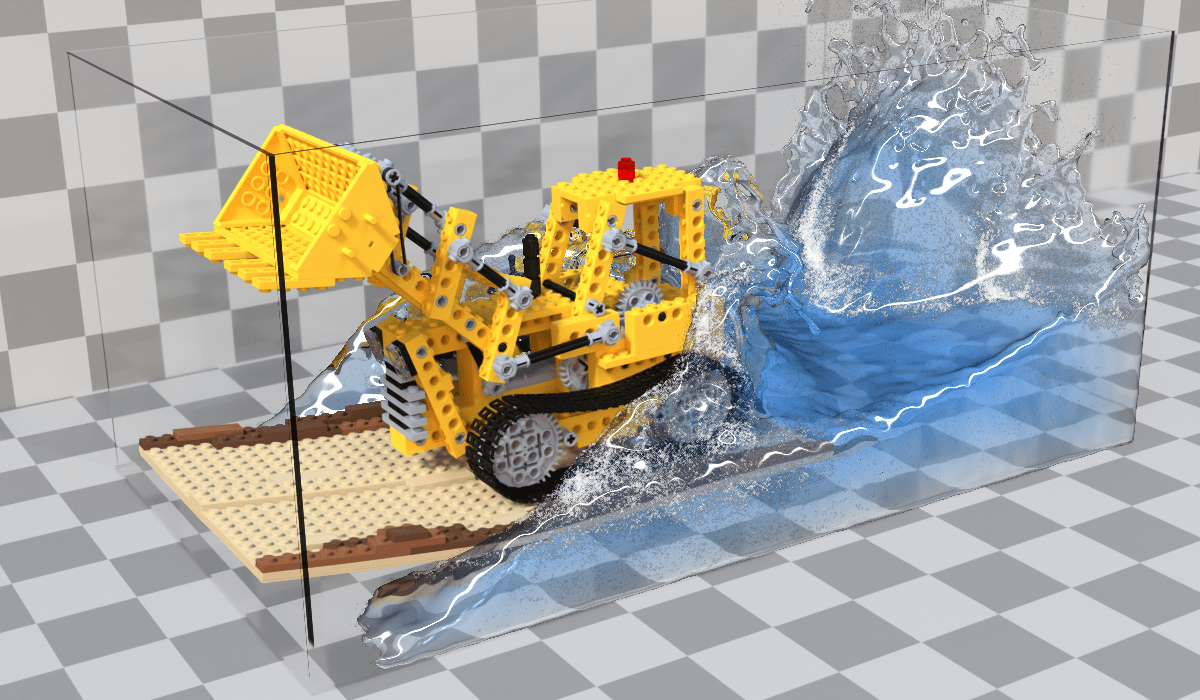}
            \end{minipage}
            \begin{minipage}{0.196\linewidth} 
                \centering
                \includegraphics[width=\linewidth]{samples/figs/main_dynamic/splashing_lego/0420.png}
            \end{minipage}
            \begin{minipage}{0.196\linewidth} 
                \centering
                \includegraphics[width=\linewidth]{samples/figs/main_dynamic/splashing_lego/0710.png}
            \end{minipage}
            \begin{minipage}{0.196\linewidth} 
                \centering
                \includegraphics[width=\linewidth]{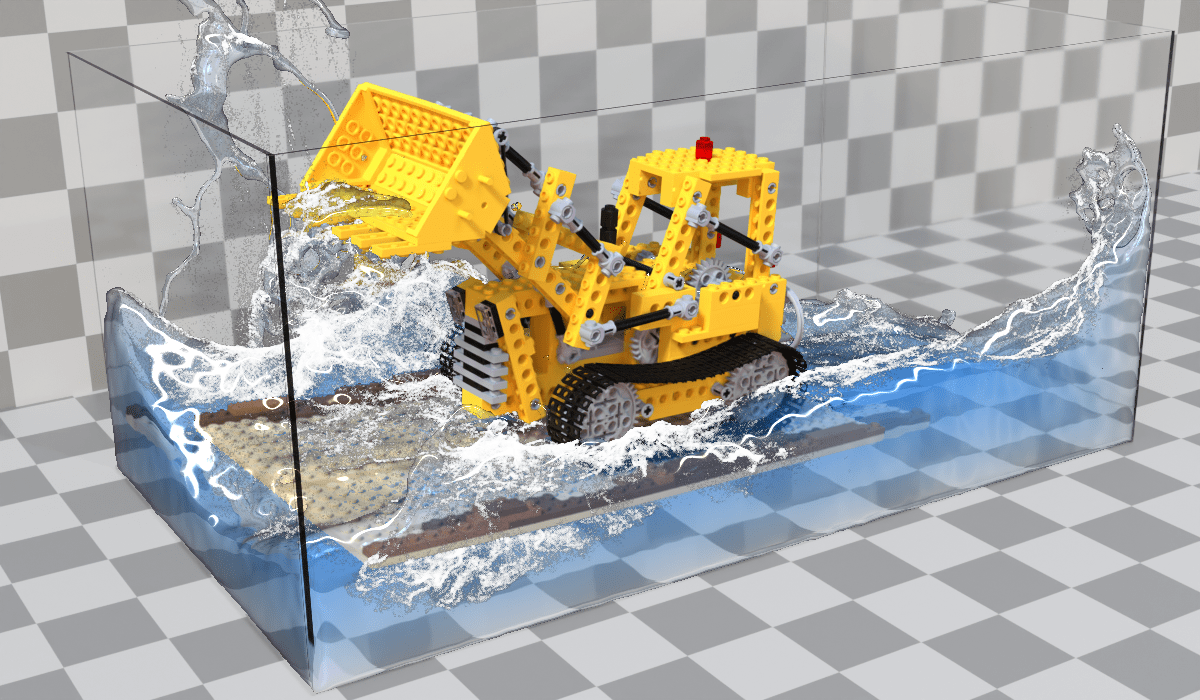}
            \end{minipage}
        \end{minipage}
    \end{minipage}

    \caption{\textbf{Splashing LEGO.}~ Through the two-way coupling dynamics, the LEGO bulldozer is animated to surf on the splashing waves.}
    \label{fig:splashing}
\end{figure*}

\begin{figure*}
    \centering

    \begin{minipage}{\textwidth}
        \centering
        \begin{minipage}{\linewidth}
            \centering
            \begin{minipage}{0.196\linewidth}   
                \centering
                \includegraphics[width=\linewidth,trim={0cm 0cm 0cm 0cm},clip]{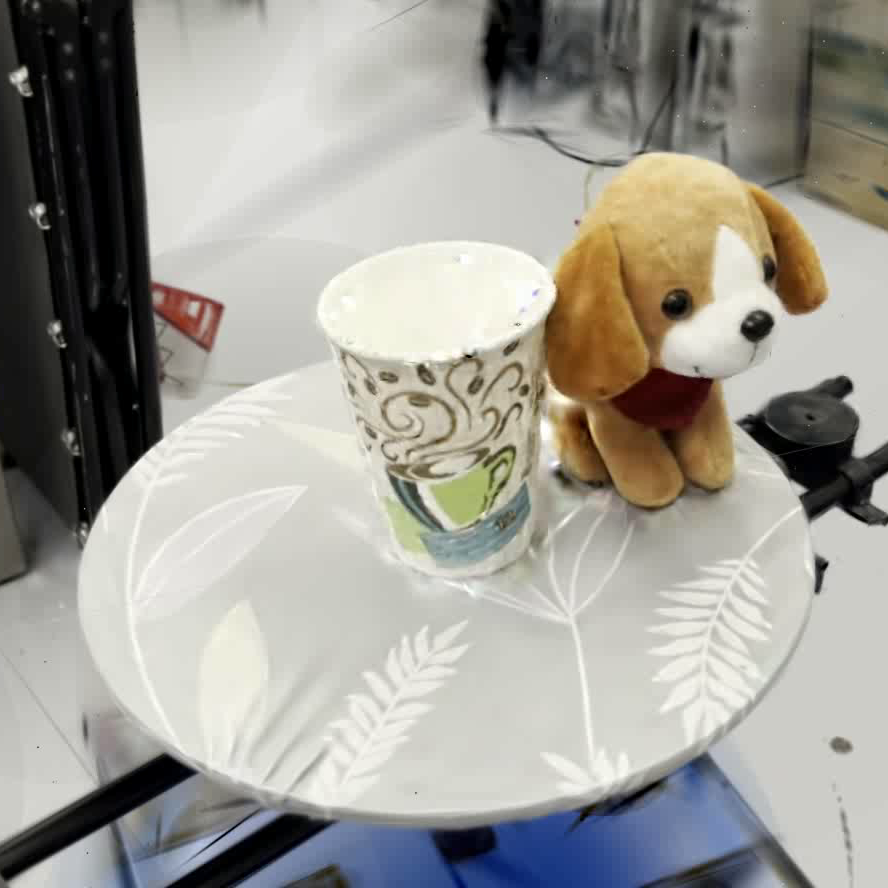}
            \end{minipage}
            \begin{minipage}{0.196\linewidth} 
                \centering
                \includegraphics[width=\linewidth,trim={0cm 0cm 0cm 0cm},clip]{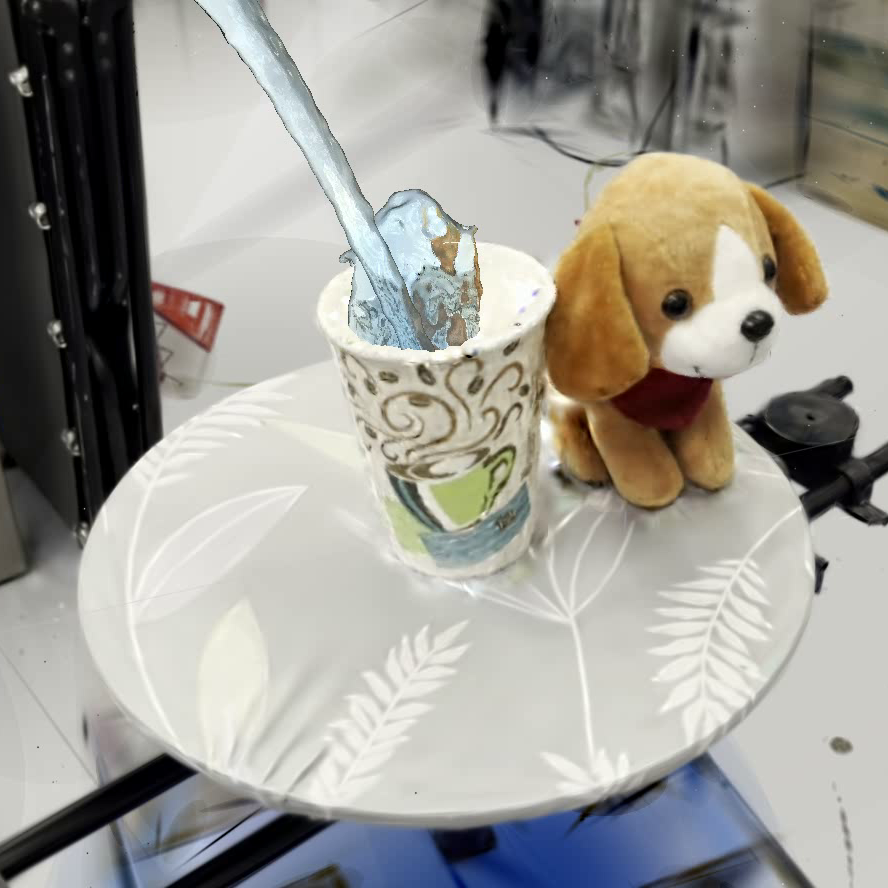}
            \end{minipage}
            \begin{minipage}{0.196\linewidth} 
                \centering
                \includegraphics[width=\linewidth,trim={0cm 0cm 0cm 0cm},clip]{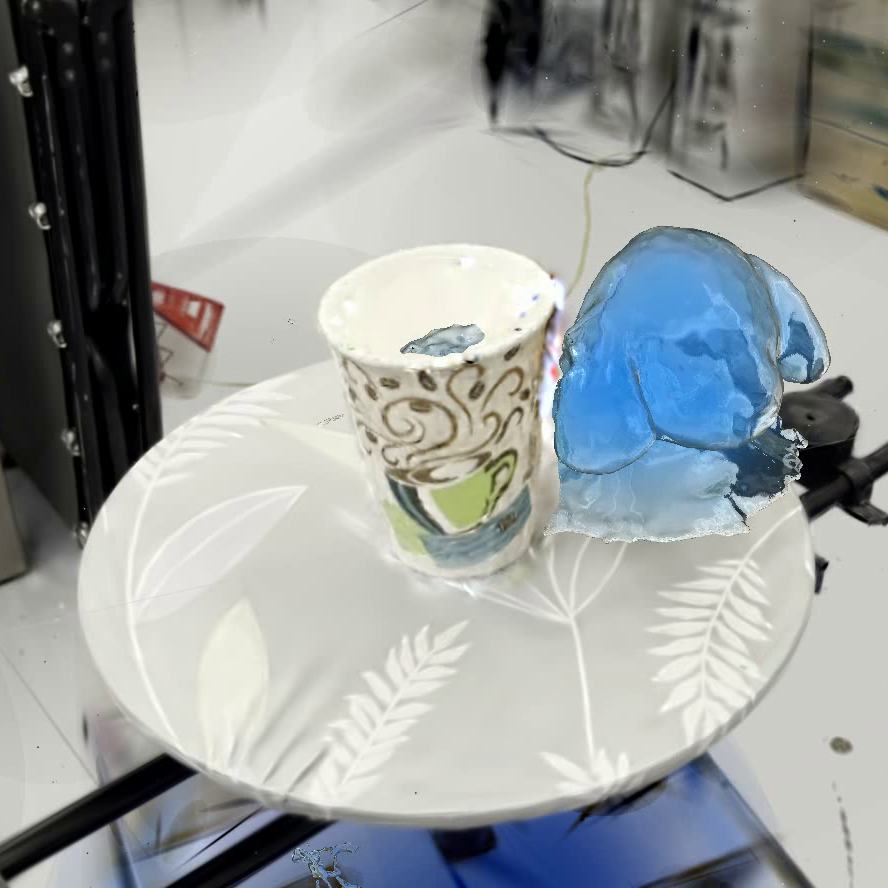}
            \end{minipage}
            \begin{minipage}{0.196\linewidth} 
                \centering
                \includegraphics[width=\linewidth,trim={0cm 0cm 0cm 0cm},clip]{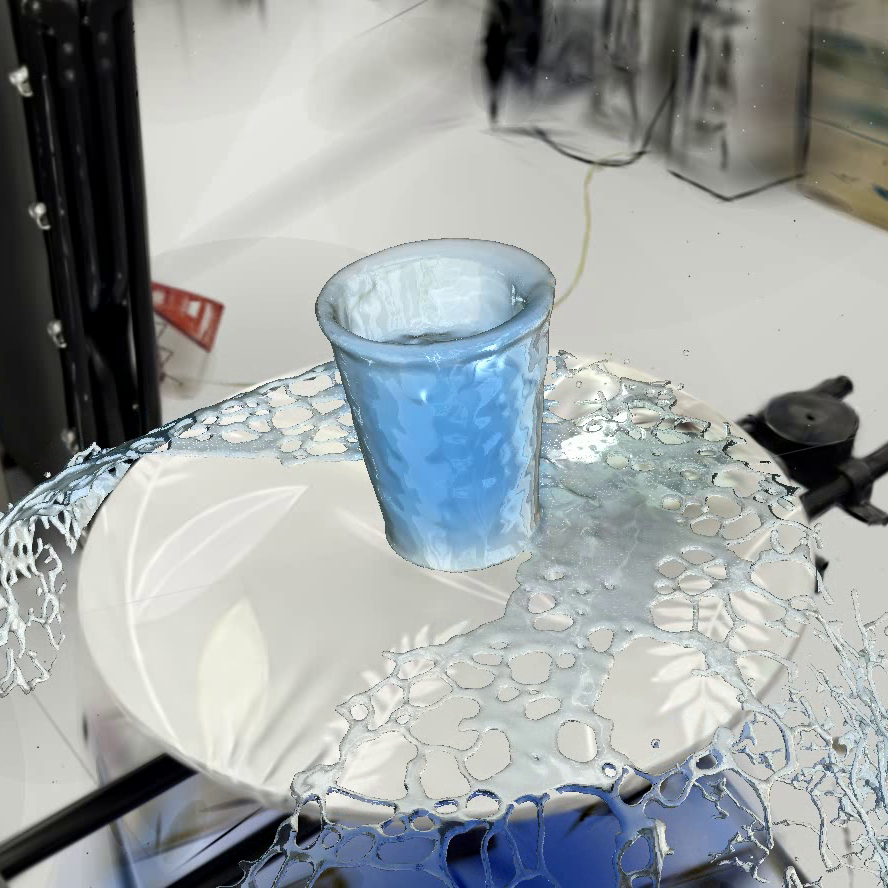}
            \end{minipage}
            \begin{minipage}{0.196\linewidth} 
                \centering
                \includegraphics[width=\linewidth,trim={0cm 0cm 0cm 0cm},clip]{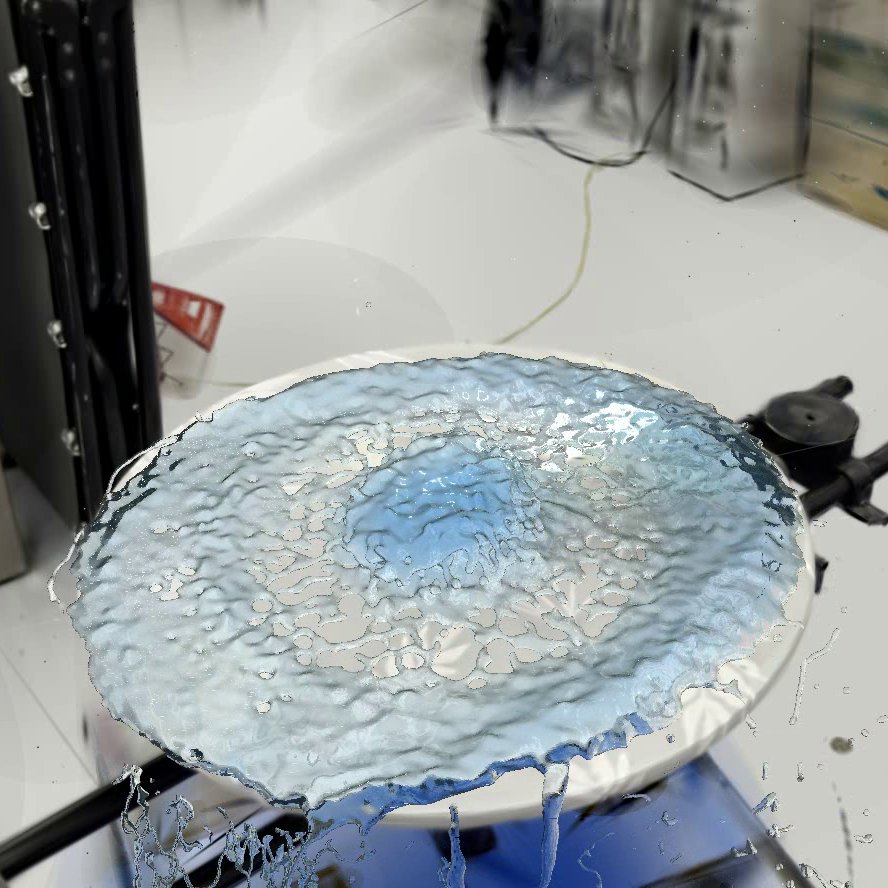}
            \end{minipage}
        \end{minipage}
    \end{minipage}

    \caption{\textbf{``Everything is water''.}~~Pouring water into the paper cup on the table and transforming the cup and the dog toy into water. The water spills out.}
    \label{fig:cup_and_dog}
\end{figure*}

\begin{figure*}
    \centering

   \begin{minipage}{\textwidth}
        \centering
        \begin{minipage}{\linewidth}
            \centering
            \begin{minipage}{0.196\linewidth}   
                \centering
                \includegraphics[width=\linewidth]{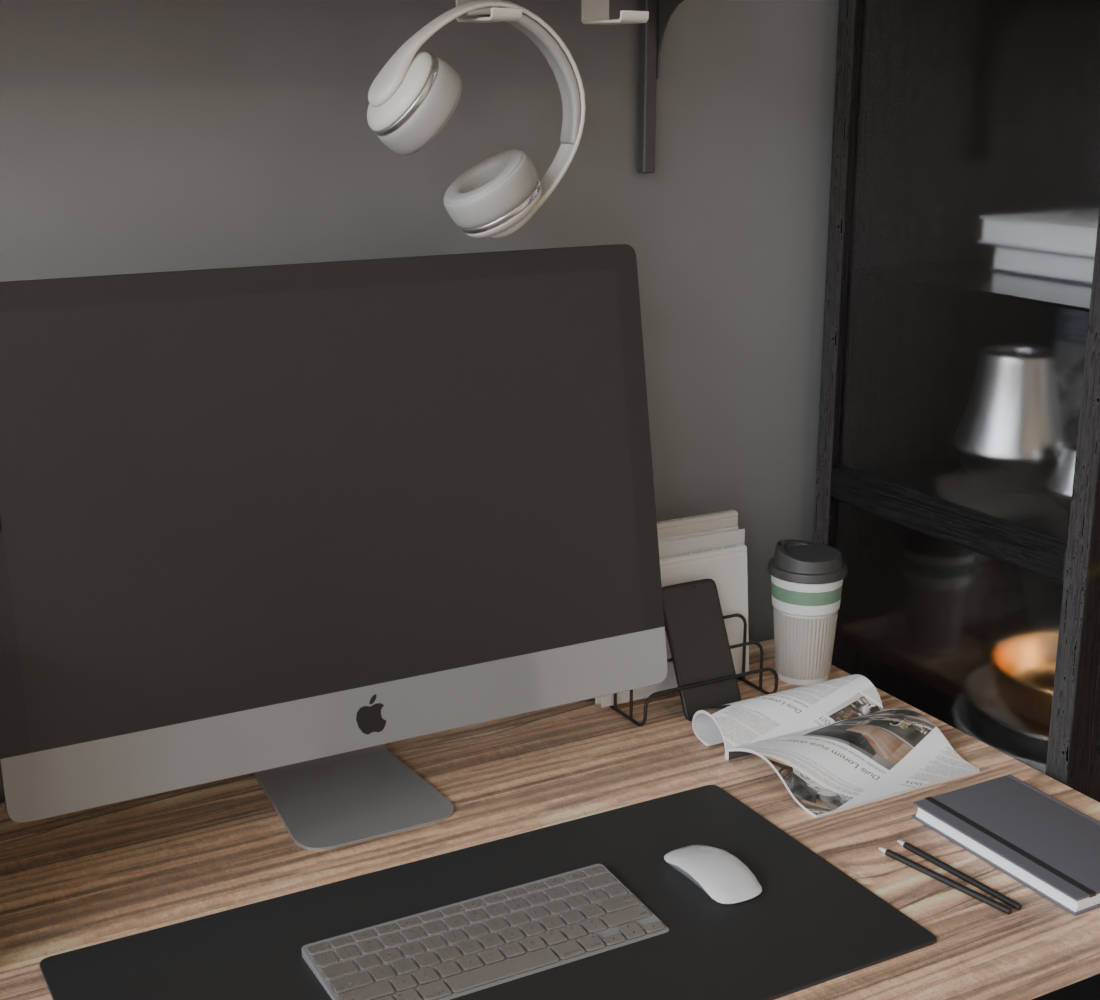}
            \end{minipage}
            \begin{minipage}{0.196\linewidth} 
                \centering
                \includegraphics[width=\linewidth]{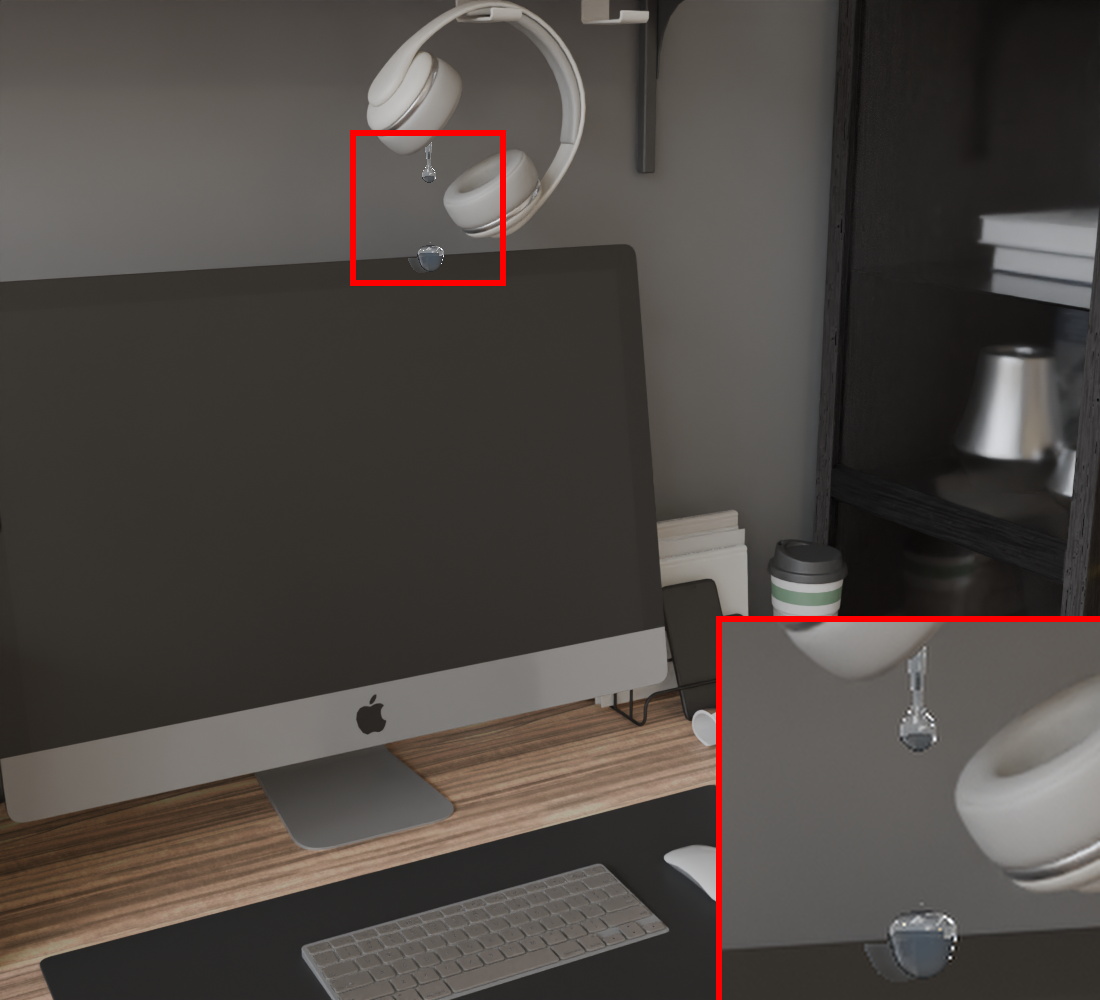}
            \end{minipage}
            \begin{minipage}{0.196\linewidth} 
                \centering
                \includegraphics[width=\linewidth]{samples/figs/main_dynamic/headphone/0122.png.inset.png}
            \end{minipage}
            \begin{minipage}{0.196\linewidth} 
                \centering
                \includegraphics[width=\linewidth]{samples/figs/main_dynamic/headphone/0235.png.inset.png}
            \end{minipage}
            \begin{minipage}{0.196\linewidth} 
                \centering
                \includegraphics[width=\linewidth]{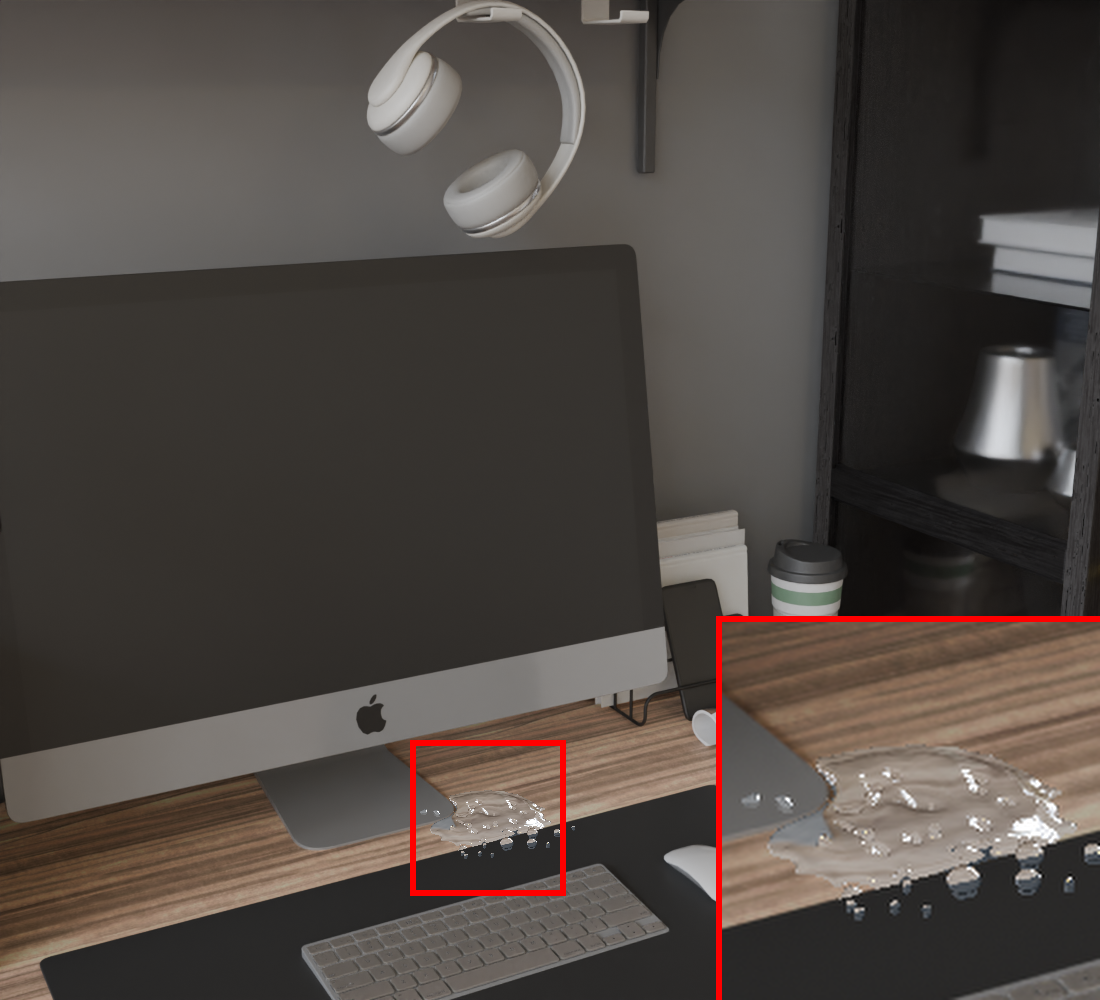}
            \end{minipage}
        \end{minipage}
    \end{minipage}

    \caption{\textbf{Headset waterfall}. Water flows from a headset hanging above an office desk, resembling a faucet. Due to surface tension, the water forms droplets as it falls, sliding down the computer screen and splashing onto the desk, creating a puddle.}
    \label{fig:headset}
\end{figure*}

\begin{figure*}
    \centering

    \begin{minipage}{\textwidth}
        \centering
        \begin{minipage}{\linewidth}
            \centering
            \begin{minipage}{0.196\linewidth}   
                \centering
                \includegraphics[width=\linewidth,trim={0cm 1.5cm 0cm 5.5cm},clip]{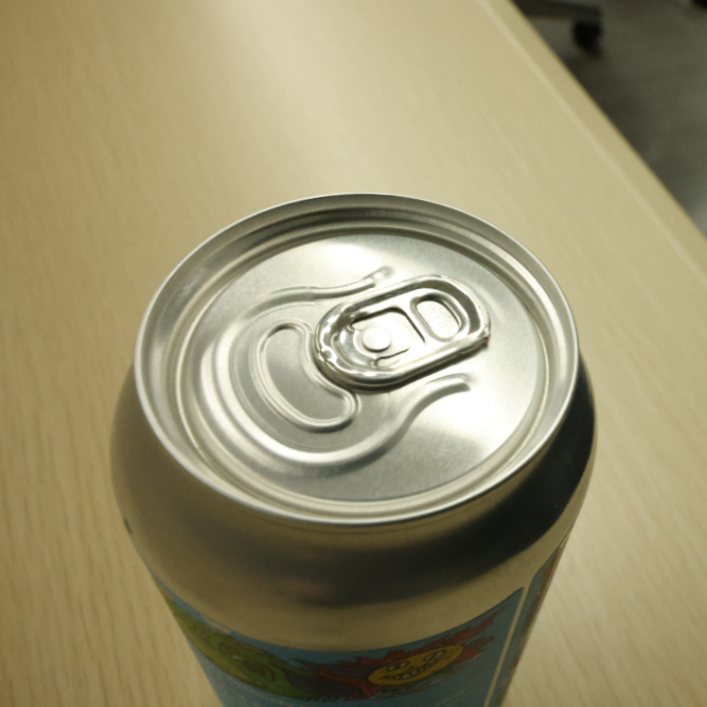}
            \end{minipage}
            \begin{minipage}{0.196\linewidth} 
                \centering
                \includegraphics[width=\linewidth,trim={0cm 1.5cm 0cm 5.5cm},clip]{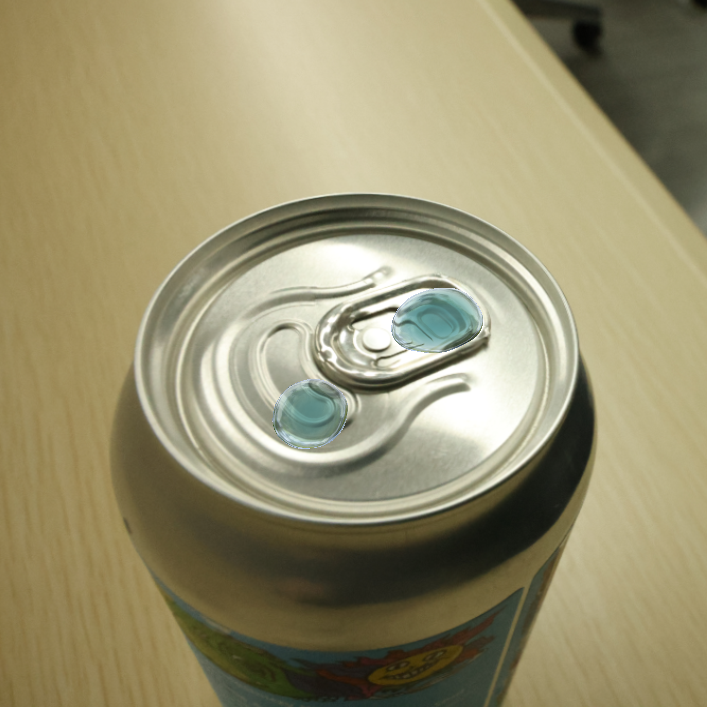}
            \end{minipage}
            \begin{minipage}{0.196\linewidth} 
                \centering
                \includegraphics[width=\linewidth,trim={0cm 1.5cm 0cm 5.5cm},clip]{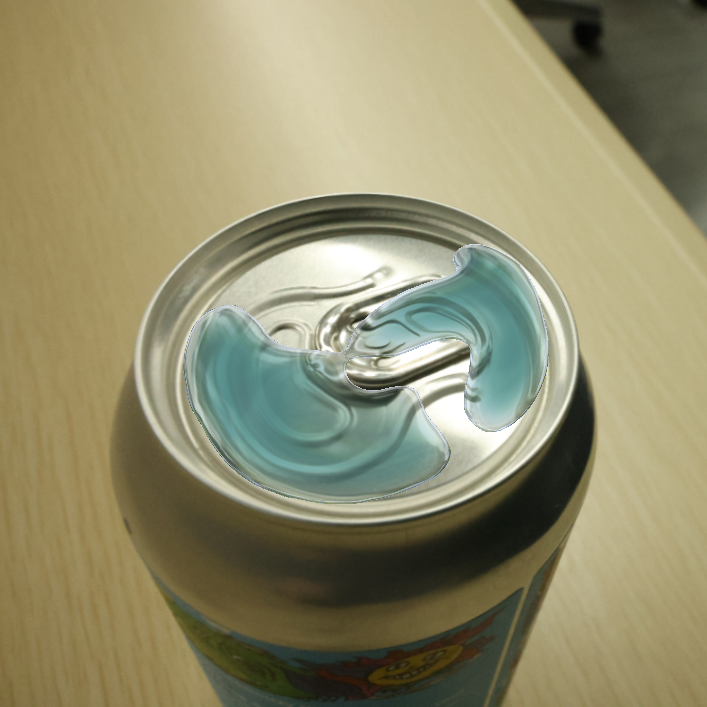}
            \end{minipage}
            \begin{minipage}{0.196\linewidth} 
                \centering
                \includegraphics[width=\linewidth,trim={0cm 1.5cm 0cm 5.5cm},clip]{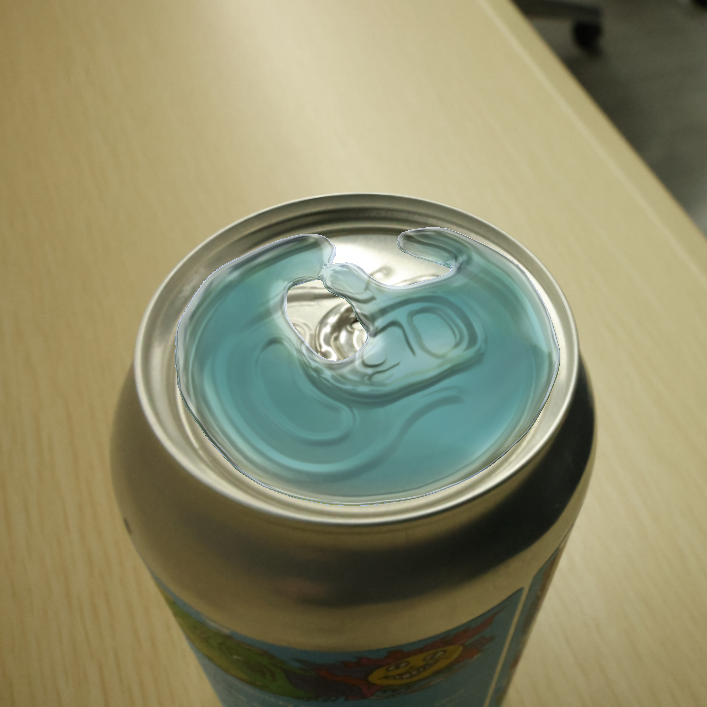}
            \end{minipage}
            \begin{minipage}{0.196\linewidth} 
                \centering
                \includegraphics[width=\linewidth,trim={0cm 1.5cm 0cm 5.5cm},clip]{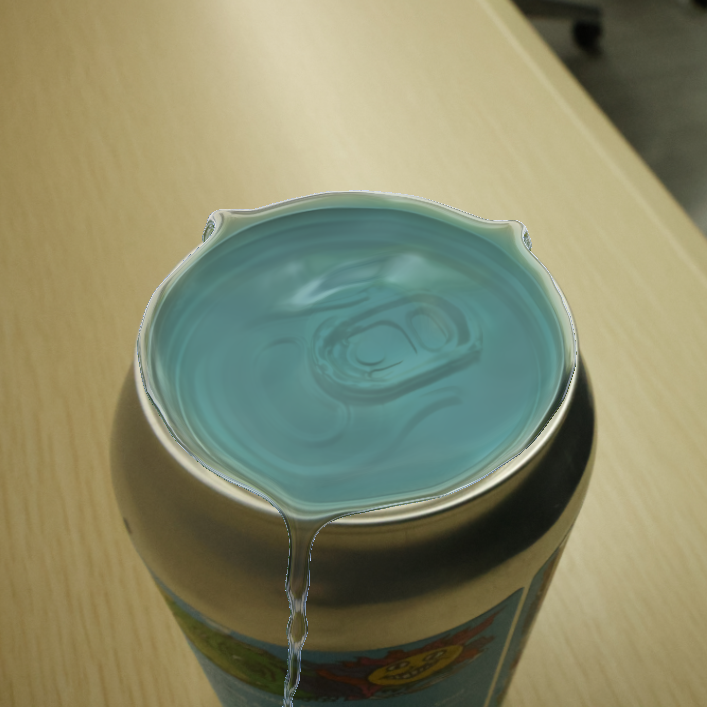}
            \end{minipage}
        \end{minipage}
    \end{minipage}
    \caption{\textbf{Water droplets on can.}~~Droplets of water fall onto the surface of a soda can, coalesce due to surface tension and gradually overflow.}
    \label{fig:can}
\end{figure*}

\begin{figure*}
    \centering

   \begin{minipage}{\textwidth}
        \centering
        \begin{minipage}{\linewidth}
            \centering
            \begin{minipage}{0.196\linewidth}   
                \centering
                \includegraphics[width=\linewidth]{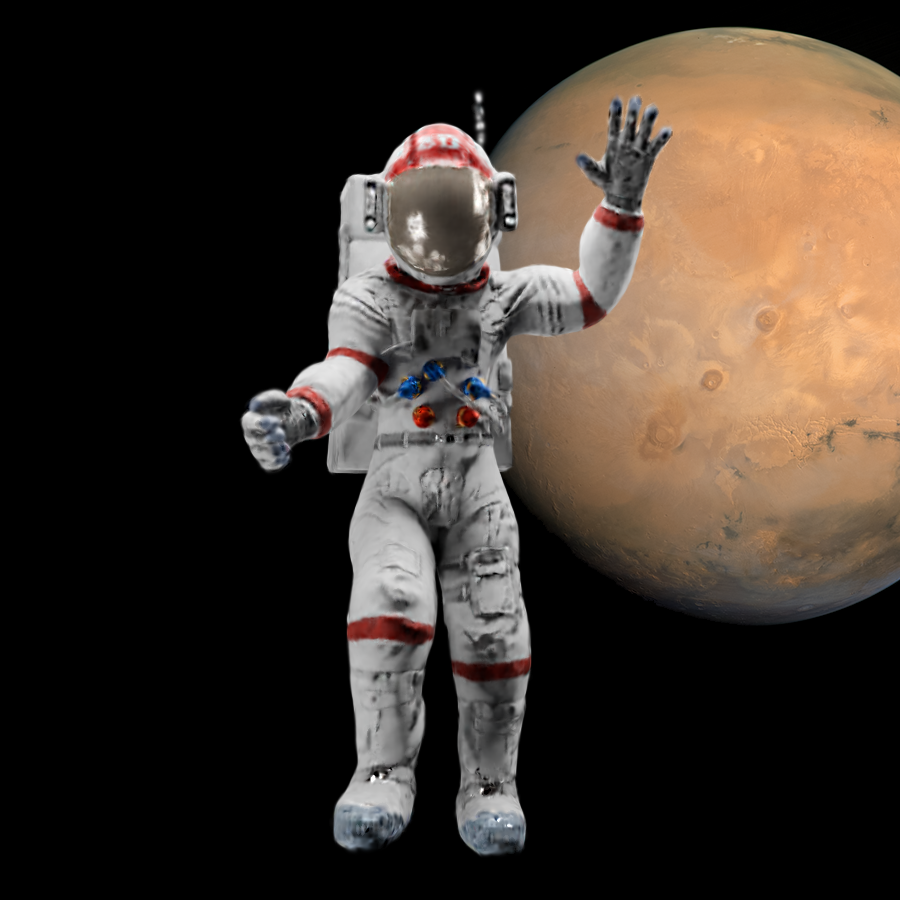}
            \end{minipage}
            \begin{minipage}{0.196\linewidth} 
                \centering
                \includegraphics[width=\linewidth]{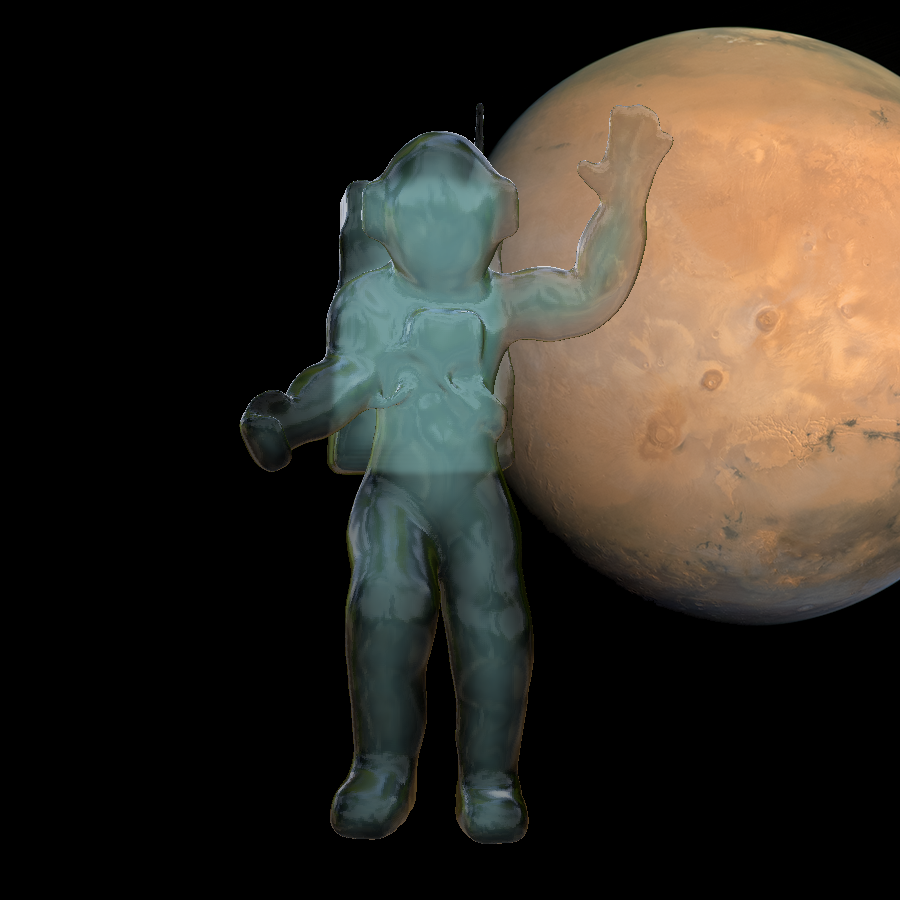}
            \end{minipage}
            \begin{minipage}{0.196\linewidth} 
                \centering
                \includegraphics[width=\linewidth]{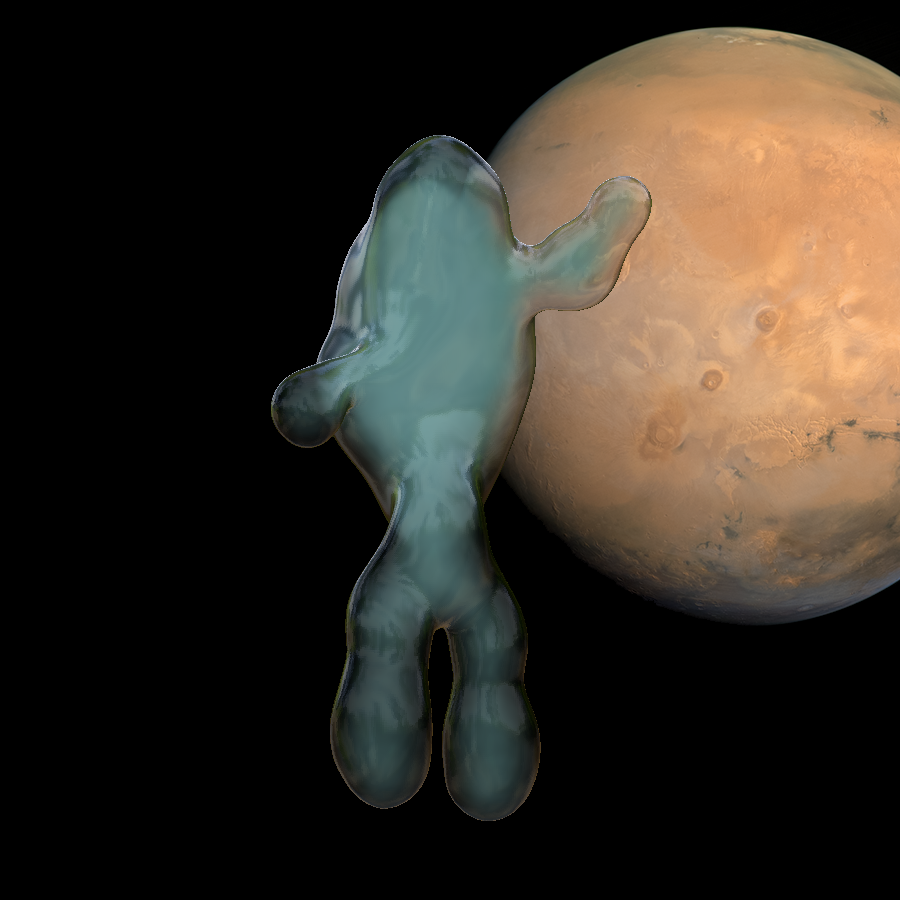}
            \end{minipage}
            \begin{minipage}{0.196\linewidth} 
                \centering
                \includegraphics[width=\linewidth]{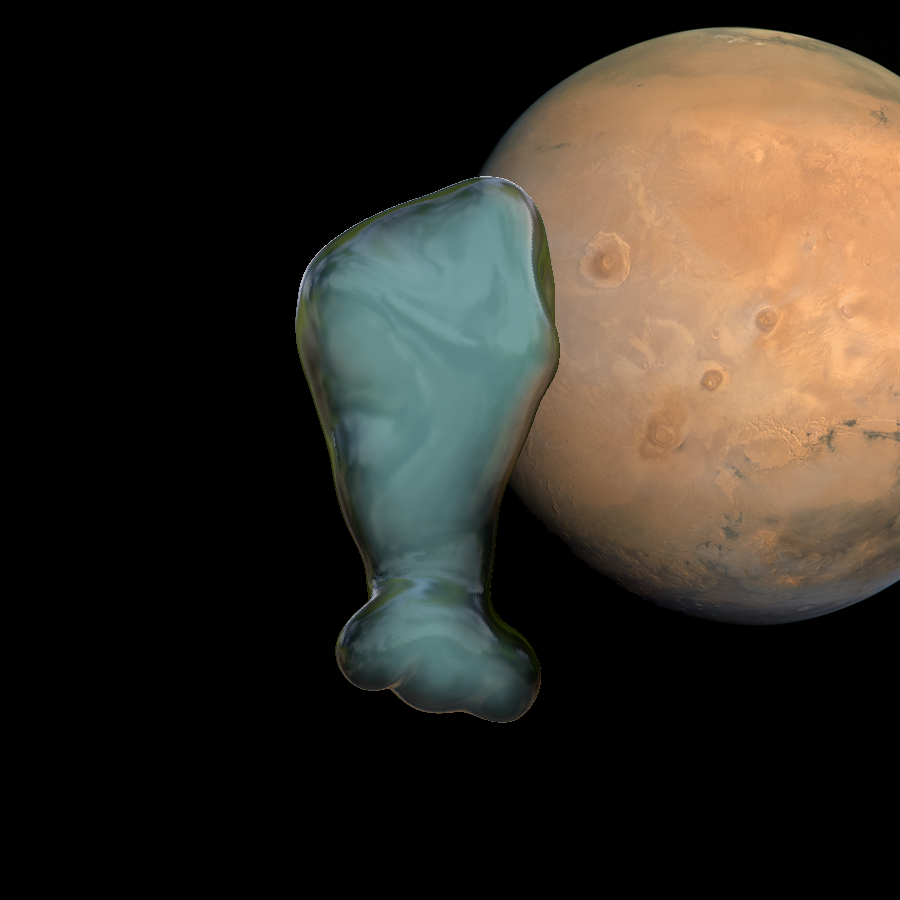}
            \end{minipage}
            \begin{minipage}{0.196\linewidth} 
                \centering
                \includegraphics[width=\linewidth]{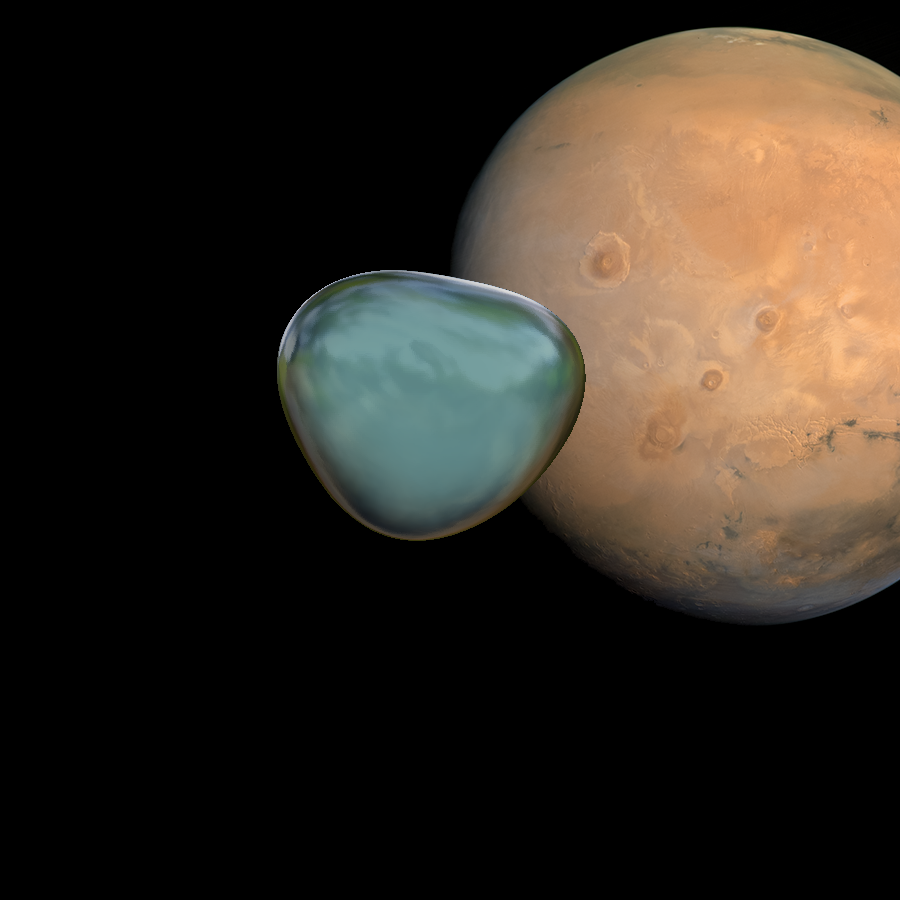}
            \end{minipage}
        \end{minipage}
    \end{minipage}

    \caption{\textbf{Black magic.}~~An astronaut in space strucked by the black magic of the Trisolarans, and get transformed into a water sphere.}
    \label{fig:astronaut}
\end{figure*}

\begin{figure*}
    \centering

   \begin{minipage}{\textwidth}
        \centering
        \begin{minipage}{\linewidth}
            \centering
            \begin{minipage}{0.196\linewidth}   
                \centering
                \includegraphics[width=\linewidth,trim={3.5cm 5cm 3.5cm 4cm},clip]{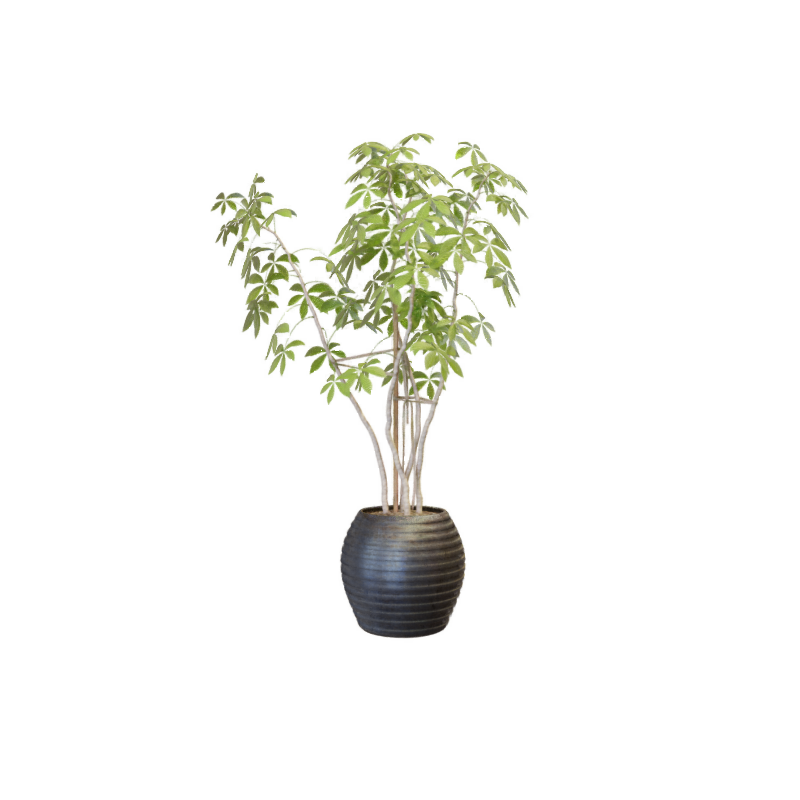}
            \end{minipage}
            \begin{minipage}{0.196\linewidth} 
                \centering
                \includegraphics[width=\linewidth,trim={3.5cm 5cm 3.5cm 4cm},clip]{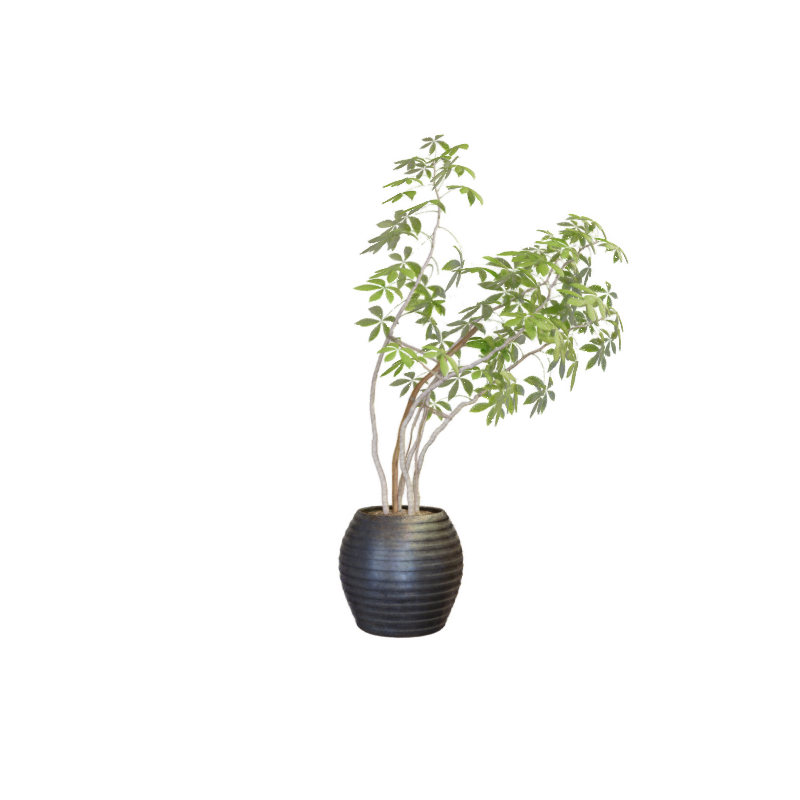}
            \end{minipage}
            \begin{minipage}{0.196\linewidth} 
                \centering
                \includegraphics[width=\linewidth,trim={3.5cm 5cm 3.5cm 4cm},clip]{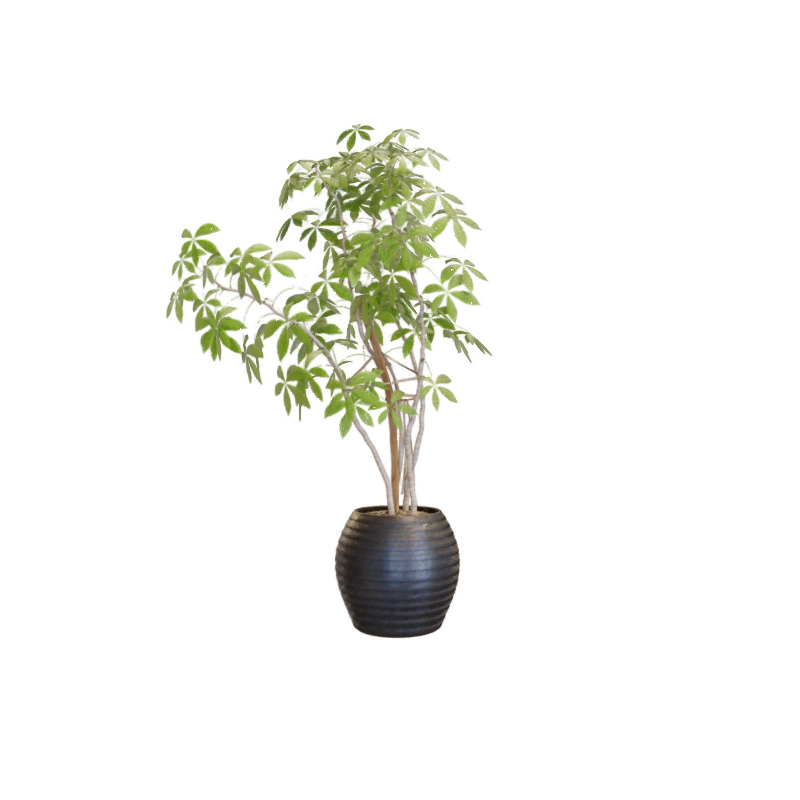}
            \end{minipage}
            \begin{minipage}{0.196\linewidth} 
                \centering
                \includegraphics[width=\linewidth,trim={3.5cm 5cm 3.5cm 4cm},clip]{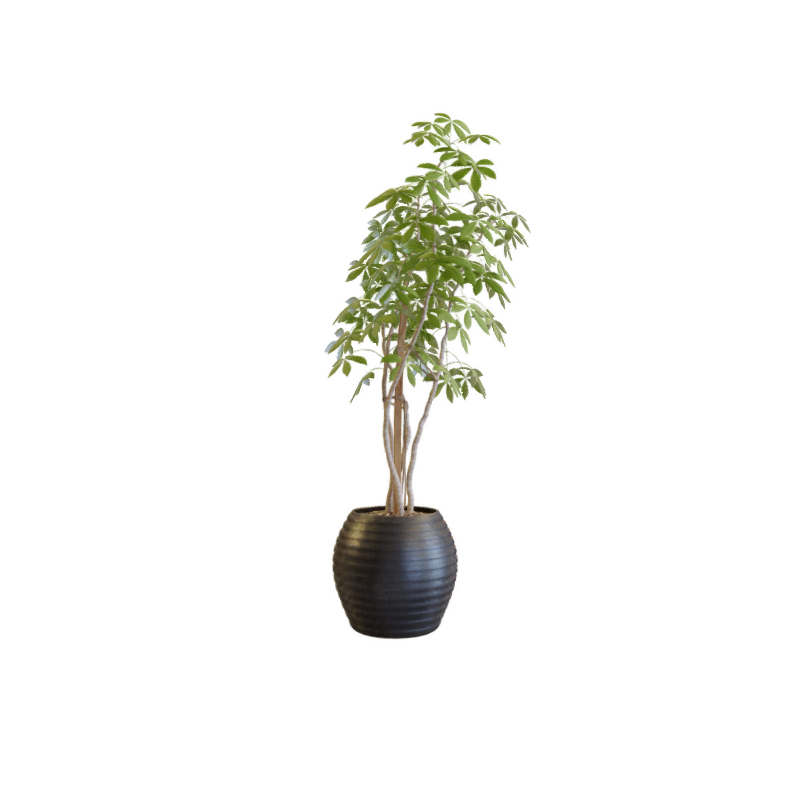}
            \end{minipage}
            \begin{minipage}{0.196\linewidth} 
                \centering
                \includegraphics[width=\linewidth,trim={3.5cm 5.5cm 3.5cm 4cm},clip]{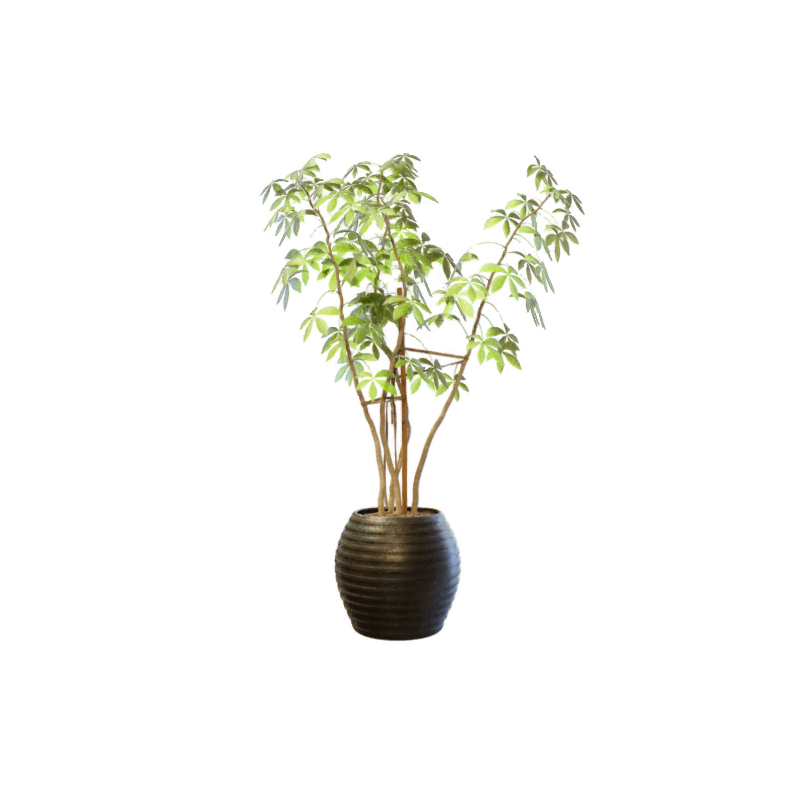}
            \end{minipage}
        \end{minipage}
    \end{minipage}

    \caption{\textbf{Deformable ficus}. A deformable ficus plant undergoes continuous shape changes as it is dragged and manipulated by external forces.}
    \label{fig:ficus}
\end{figure*}

\begin{figure*}
    \centering

   \begin{minipage}{\textwidth}
        \centering
        \begin{minipage}{\linewidth}
            \centering
            \begin{minipage}{0.196\linewidth}   
                \centering
                \includegraphics[width=\linewidth,trim={4cm 0cm 4cm 0cm},clip]{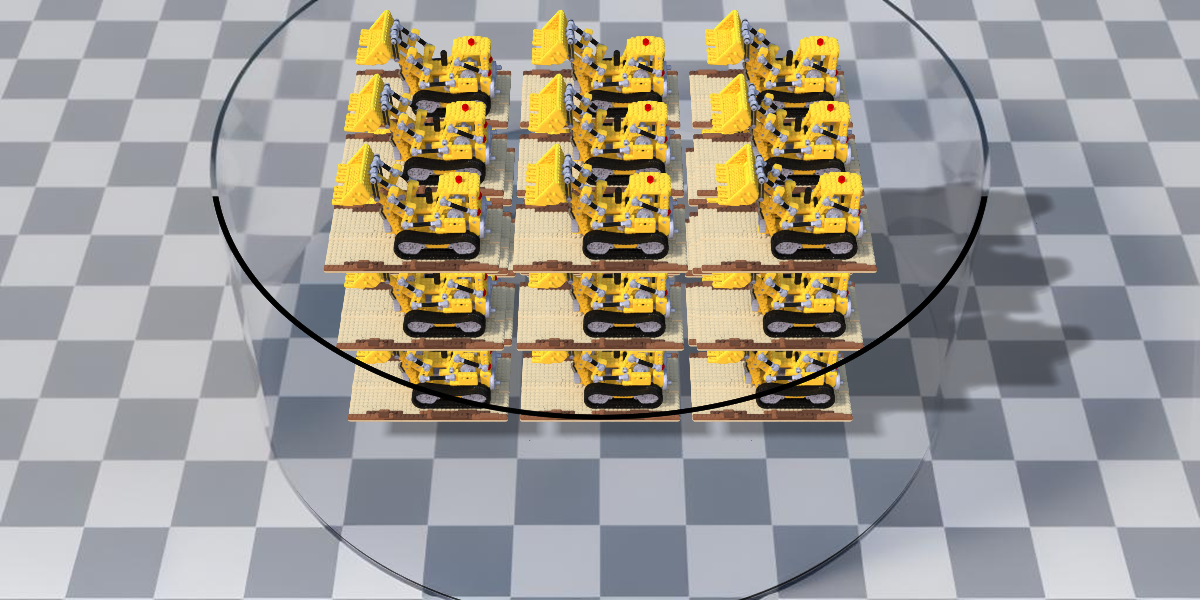}
            \end{minipage}
            \begin{minipage}{0.196\linewidth} 
                \centering
                \includegraphics[width=\linewidth,trim={4cm 0cm 4cm 0cm},clip]{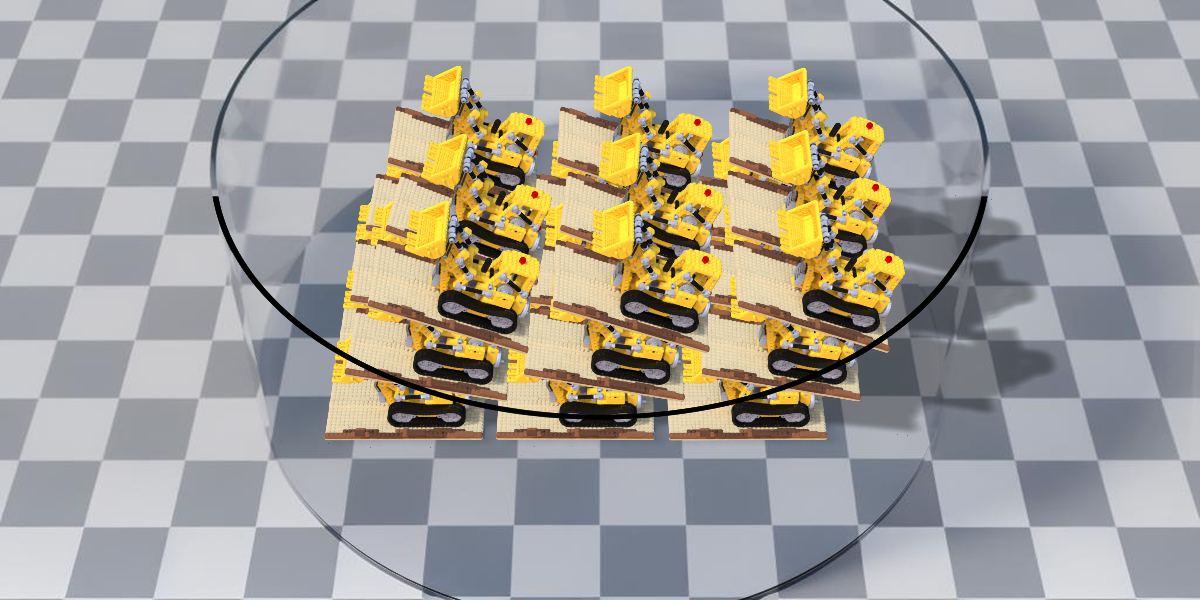}
            \end{minipage}
            \begin{minipage}{0.196\linewidth} 
                \centering
                \includegraphics[width=\linewidth,trim={4cm 0cm 4cm 0cm},clip]{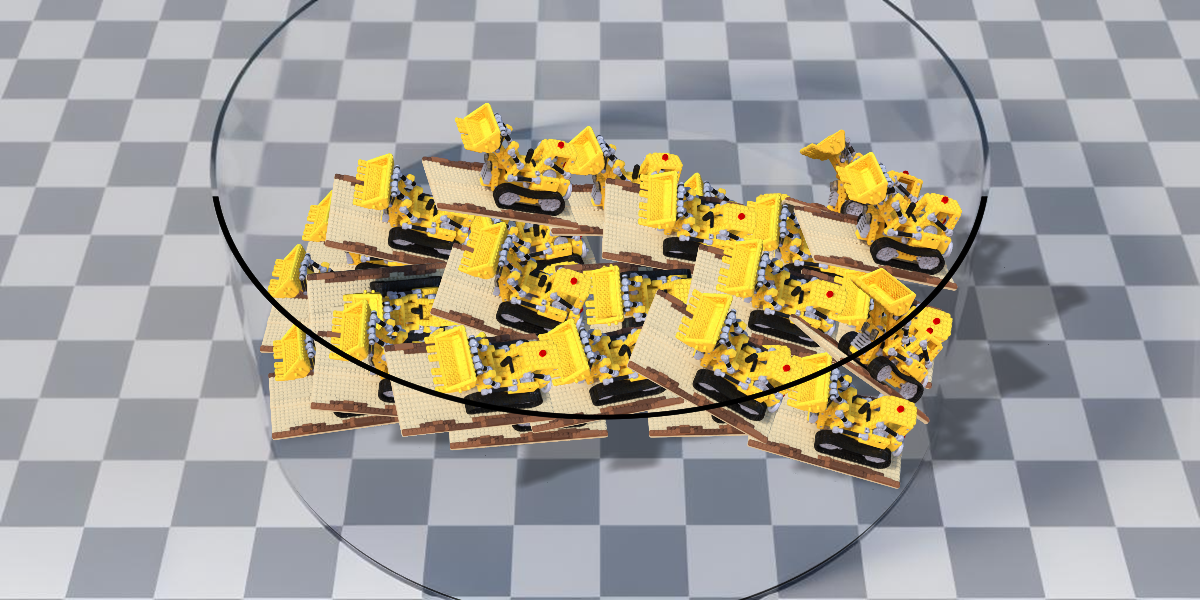}
            \end{minipage}
            \begin{minipage}{0.196\linewidth} 
                \centering
                \includegraphics[width=\linewidth,trim={4cm 0cm 4cm 0cm},clip]{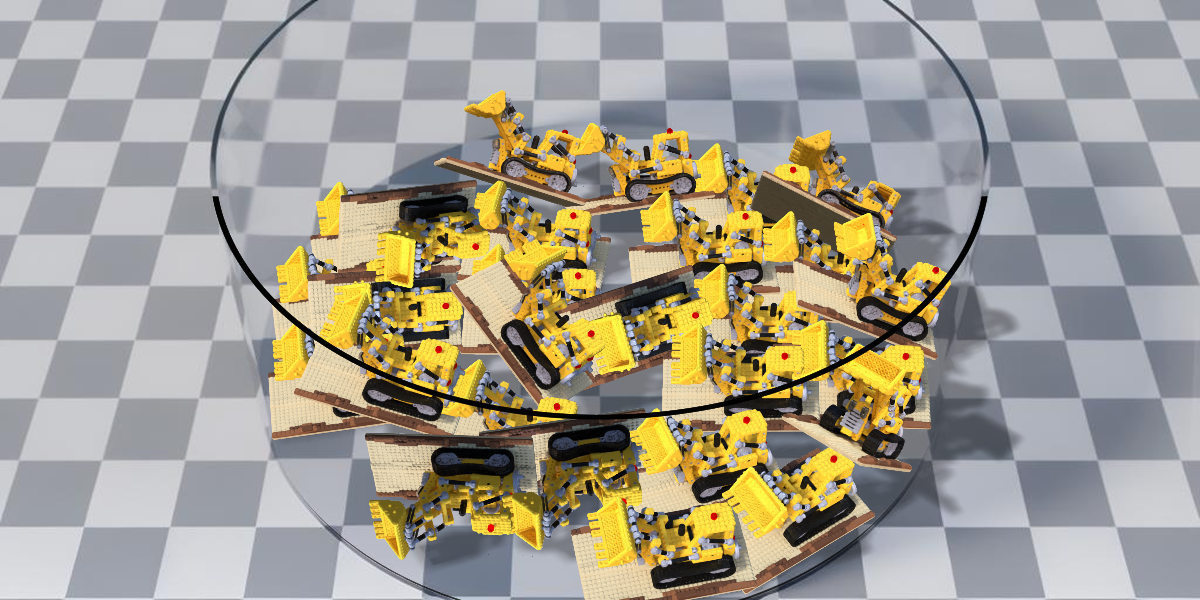}
            \end{minipage}
            \begin{minipage}{0.196\linewidth} 
                \centering
                \includegraphics[width=\linewidth,trim={4cm 0cm 4cm 0cm},clip]{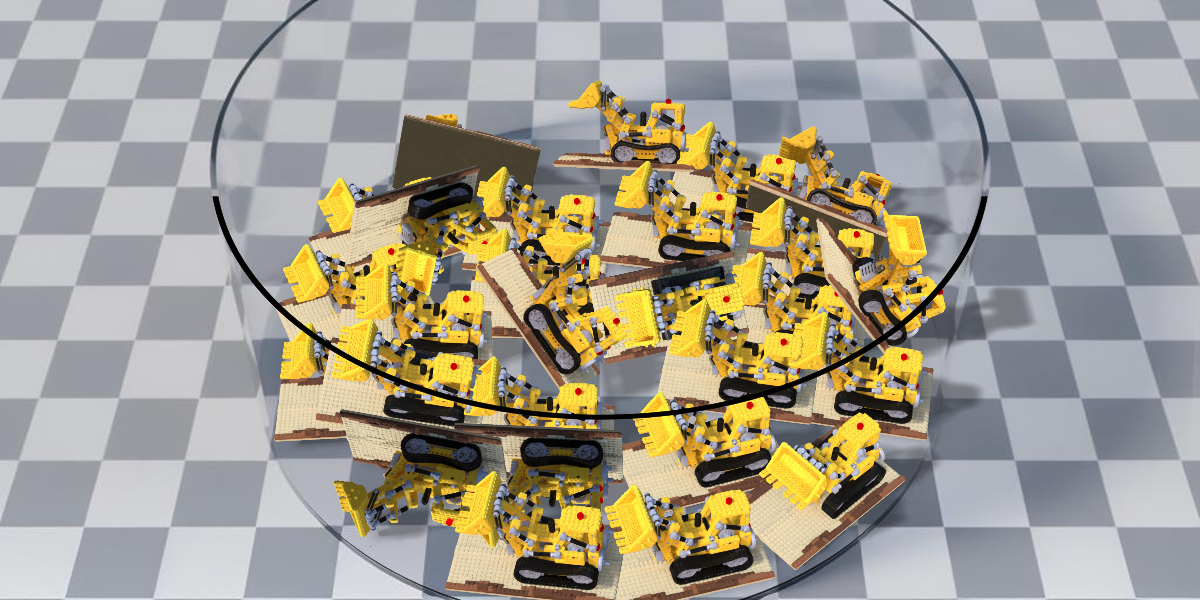}
            \end{minipage}
        \end{minipage}
    \end{minipage}

    \caption{\textbf{LEGO bulldozers in glass bowl.} A collection of LEGO bulldozer rigid bodies fall into a round glass box, colliding with each other. They cast shadow on the ground and eventually stack and scatter throughout the box.}
    \label{fig:piled_bulldozers}
\end{figure*}

We show the results of
\textbf{Chair} (Fig.~\ref{fig:chair}),
\textbf{Waves} (Fig.~\ref{fig:wave}),
\textbf{Garden} (Fig.~\ref{fig:garden}),
\textbf{Lego} (Fig.~\ref{fig:splashing}),
\textbf{Cup \& dog} (Fig.~\ref{fig:cup_and_dog}),
\textbf{Headset} (Fig.~\ref{fig:headset}),
\textbf{Can} (Fig.~\ref{fig:can}),
\textbf{Astronaut} (Fig.~\ref{fig:astronaut}),
\textbf{Ficus} (Fig.~\ref{fig:ficus}),
and \textbf{Bulldozers} (Fig.~\ref{fig:piled_bulldozers}).
For better visualization, please refer to the supplementary video.

\end{document}